\documentclass{aa}  

\usepackage{amssymb}
\usepackage{graphicx}
\usepackage{rotating}
\usepackage{float}
\usepackage{txfonts}
\usepackage{hyperref}
\usepackage{afterpage}
\hypersetup{draft}
\newcommand\vv{{\mathrm v}  }
\begin{document}

   \title{Structure and evolution of a tidally heated star}


   \author{D. Estrella-Trujillo
           \inst{1}
        \and   
            S. J. Arthur
            \inst{2}
        \and
            G. Koenigsberger\thanks{Visiting scholar, Astronomy Department, Indiana University}      
            \inst{3} 
         \and
            E. Moreno
           \inst{4}
          }

   \institute{ Instituto de Ciencias F\'{\i}sicas, Universidad Nacional Aut\'onoma de M\'exico,
              Ave. Universidad S/N, Chamilpa, Cuernavaca, M\'exico \\
              \email{dtrujillo@icf.unam.mx}
         \and
              Instituto de Radioastronom\'{\i}a y Astrof\'{\i}sica, Universidad Nacional Aut\'onoma de México, Antigua Carretera a P\'atzcuaro \#8701, Ex-Hda.\
San Jos\'e de la Huerta, Morelia, Michoac\'an, M\'exico C.P. 58089,
              \email{j.arthur@irya.unam.mx}
         \and
              Instituto de Ciencias F\'{\i}sicas, Universidad Nacional Aut\'onoma de M\'exico,
              Ave. Universidad S/N, Chamilpa, Cuernavaca, M\'exico \\
              \email{gloria@astro.unam.mx}
         \and
              Instituto de Astronom\'{\i}a, Universidad Nacional Aut\'onoma de M\'exico, Apdo. Postal 70-264,
              Ciudad de M\'exico, 04510 M\'exico \\
             }

   \date{Full Draft Version \#1 (2022 May 27)           }

 
 \abstract
 {The shearing motion of tidal flows that are excited in non-equilibrium binary stars transform kinetic energy into heat via  a process referred to as tidal heating. 
        }
{We aim to explore the way tidal heating affects the stellar structure.
        }
{We used the TIDES code, which solves the equations of motion of the three-dimensional (3D) grid of volume elements that conform  multiple layers of a rotating binary star to obtain an instantaneous value for the angular velocity,  $\omega''$, as a function of position in the presence of gravitational, centrifugal, Coriolis, gas  pressure, and  viscous forces.  The released energy,  $\dot{E,}$ was computed using a prescription for turbulent viscosity that depends on the instantaneous velocity gradients. The $\dot{E}$ values  for each radius were injected into a MESA stellar structure  calculation.  The method is illustrated for a  1.0+0.8 M$_\odot$ binary system, with an orbital period of  $P$=1.44\,d and departures  from synchronous rotation of 5\% and 10\%.
         }
{Heated models have a larger radius and surface luminosity, a smaller surface convection zone, and lower nuclear reaction rates than the equivalent standard stellar models, and their evolutionary tracks extend to higher temperatures.  The magnitude of these effects depends  on the amount of injected energy, which, for a fixed set of stellar, rotation and orbital parameters, depends on the perturbed star's density structure and turbulent viscosity. 
         }
{Tidal heating offers a possible alternative for describing 
phenomena such as bloated or overluminous binary components, age discrepancies, and  aspherical mass ejection, as well as the extended main sequence turnoff in clusters.  However, establishing its actual role requires 3D stellar structure models commensurate with the nonspherically symmetric properties of tidal perturbations. 
 }
 
 
   \maketitle

\section{Introduction}

Binary stars are generally assumed to be in equilibrium, thus allowing them to be modeled as single stars 
during most of their lifetime. In an equilibrium state, the rotation and orbital rates are equal, the orbit is circular and the orbital and rotation axes are parallel. There are, however, many binary stars that are not in 
equilibrium, such as those in eccentric orbits as well as those in circular orbits that have either
not yet attained equilibrium or have departed from this state due to evolutionary changes. In such cases,
the tidal interaction excites shearing flows where kinetic energy is converted into heat.

\citet{1968Ap&SS...1..411K} addressed the question of whether the energy dissipation rates, $\dot{E,}$ due to such tidal interactions could affect the internal stellar structure, concluding that if only molecular and radiative viscosities are active in the stellar material, then the energy dissipation rates due to shearing motions are  too small to have any effect on the star's  structure \cite[see also][]{Kundo1990}. Hence, $\dot{E}$ is generally neglected in the stellar structure calculations. However, \citet{1968Ap&SS...1..411K} also noted that if the so-called "turbulent viscosity," $\nu_\mathrm{turb}$,  is present, then the internal stellar structure could be affected.  

Turbulent viscosity is associated with the idea that turbulence in a fluid dynamical system induces an effective viscosity associated with the size of the eddies \citep{1987flme.book.....L, 1988AdSpR...8b.135S}.  Numerical and laboratory experiments have been used to estimate its value in stars 
\citep{1999A&A...347..734R,2004A&A...425..243M,2007ApJ...655.1166P,2018A&A...620A..22M}. In binary stars with external convective layers, $\nu_\mathrm{turb}$ is assumed to be related to $\nu_\mathrm{cv}$, which is associated with the convective eddies predicted by the mixing-length theory  \citep{1971CoASP...3...53S}. The scaling between the tidally induced $\nu_\mathrm{turb}$ and $\nu_\mathrm{cv}$  depends on the largest eddy turnover time and the tidal frequency \citep{1966AnAp...29..313Z, 1989A&A...220..112Z,1977ApJ...212..243G, 1977Icar...30..301G, 2020MNRAS.497.3400D}. Values of $\nu_\mathrm{cv}$ typically lie in the range between 10$^{10}$ - 10$^{13}$ cm$^{2}$s$^{-1}$ \citep{2007ApJ...655.1166P}. However, determining the actual values of $\nu_\mathrm{turb}$, in addition to  determining the conditions under which the shear instability is triggered remain challenging issues \citep[see, e.g.,][and references therein]{2004A&A...425..243M,  2016ApJ...821...49G, 2016A&A...592A..33M}.

The value of $\nu_\mathrm{turb}$ enters into the expression for calculating  $\dot{E}$ and determines  the amount of energy released by the  motions of the shearing layers. Substantial ongoing efforts are being devoted to determining the impact of  $\nu_\mathrm{turb}$ and $\dot{E}$ on the rotation and orbital element evolution \citep[see, e.g.,][and references therein]{2014ARA&A..52..171O, 2021MNRAS.503.5789T, 2022MNRAS.512.3651P}. Much less consideration has been given to evaluating the potential effects of turbulence-induced heating on the stellar structure, although  the dissipative heating rate could exceed the luminosity carried by convection and  significantly alter the internal dynamics of stars and planets \citep{2017ApJ...845L..17C}.  Also, the role of tidal heating is now well accepted as a mechanism responsible for the Jupiter moon Io's volcanism \citep{1988Icar...75..187S} and the subsurface liquid oceans in the  icy moons of the Solar System \citep{1998Icar..135...64G,1999JGR...10424015P, 2002Icar..156..143H}.

\citet[][hereinafter KM2016]{2016RMxAA..52..113K} explored the possibility that tidally induced energy dissipation could cause a star to become more extended than a single-star counterpart. Focusing on the post-main sequence phases, they found that  $\dot{E}$ could act to rapidly expand the outer layers, resulting in a runaway process with each radius increase leading to a larger value of $\dot{E}$ and, therefore, an even larger radius increase. However, these results were based on three  approximations.  The first is that the tidal perturbation amplitudes were computed only for a surface layer, thus neglecting the contribution from deeper stellar regions and relied on a coarse estimate of the stellar density. The second is the use of a constant $\nu_\mathrm{turb}$ value, while in reality it depends strongly on position and is time-variable.  The third simplifying approximation is the assumption that the tidal shear energy dissipation leads only to a radius increase, with no other effect on the stellar structure. 

In this paper we perform more realistic calculations for which we have implemented an n-layer calculation of the tidal perturbations, including a prescription for the turbulent viscosity allowing it to be computed  as a function of the time-varying and location-dependent velocity fields within the perturbed star.  In addition, the structure and evolution of the perturbed star are determined by injecting the tidal energy into a standard structure and evolution model.   

The question of the degree to which tidal shear energy dissipation may affect the internal structure of asynchronously rotating binary stars  and their evolution has implications for several important phenomena, including the onset of mass-transfer and common envelope processes, the subsequent evolutionary path followed by the remnant star and the ejection of circumstellar material.  In addition, it could have a bearing on the study of stellar populations if the observable properties of tidally heated stars differ significantly from those of their nonperturbed counterparts.    

In Section 2, we describe the tidal shear energy dissipation method and the grid of models.  In Section 3, we describe the stellar structure models that were constructed and the effects of injecting tidal shear energy dissipation into such models. In Section 4, we discuss the results and in Section 5, we present our conclusions.


\section{Tidal shear energy dissipation \label{sec:model_viscosity}}

\subsection{Method}

The response of the  stellar layers to the external gravitational field of a companion is computed 
with the n-layer {\it TIDES} code\footnote{The {\it Tidal Interactions with Dissipation of Energy due 
to Shear TIDES} code in all its versions is available upon request and is easily implemented in any 
operating system running a Fortran or GNU Fortran compiler} \citep{2021A&A...653A.127K}, which is based on the numerical method introduced in \citet{1999RMxAA..35..157M} and \citet{2007A&A...461.1057T}, and  upgraded in  \citet{Moreno:2011jq}.  

The TIDES code can be described as a quasi-hydrodynamic Lagrangian scheme which  simultaneously solves the orbital motion of the companion and the equations of motion of a 3D
grid of volume elements covering the inner, rigidly rotating "core"  of the tidally perturbed primary star.  The core is defined as the interior region that is rotating as a solid body and does not necessarily coincide with the nuclear burning region.  The equations of motion include the gravitational acceleration of both stars as well as centrifugal, Coriolis, and gas pressure accelerations.  The motions of individual elements are coupled to those of neighboring elements and to the core through viscous stresses.  The method is fully described in \citet{Moreno:2011jq} and \citet{2021A&A...653A.127K}.

The initial dimensions of the volume elements are ($\ell^0_r$, $\ell^0_\varphi$, $\ell^0_\theta$)  
determined by the selected 3D grid size. The mass contained in each volume element remains
constant over time and is determined by the polytropic stellar structure that is selected.  Once the calculation is initiated, the solution of the equations of motion determines the new location of the center of mass of each volume element and its distance from neighboring elements determines the
new dimensions ($\ell_r$, $\ell_\varphi$, $\ell_\theta$).  These, in turn, are used to compute the new gas pressure within each volume element and the corresponding acceleration term, which is then included in the new acceleration to compute the centers of mass locations  in the next integration time step.

The typical depth of the volume elements is $\Delta R/R_1<$0.1, where $R_1$ is the primary star
radius. The equations are solved in the frame of reference with origin in the center of the
primary star and that rotates with the binary orbit, using  a seventh order Runge-Kutta integrator.
The secondary is considered to be a point-mass source and its orbital plane is coplanar
with the primary star's equator.


The TIDES calculation captures the effects due to the oscillatory nature of the tidal flows, in addition to those that are due to a differential rotation structure. However, it neglects  buoyancy effects and heat and radiation transfer,  as well as  detailed microphysical processes involving the diffusion and advection of chemical elements which, in most cases, occur on different timescales than the tidal forcing.  Hence, the TIDES calculation  provides the 3D internal rotation and energy dissipation structures that result from the particular instantaneous dynamical conditions  in the system, but the impact of the above processes  on the subsequent dynamical evolution cannot be assessed. 

The time-marching algorithm is applicable for binary stars with arbitrary rotation velocity and eccentricity, as long as neighboring grid elements retain contact over at least $\sim$80\% of their surface and the centers of mass of two adjoining grid elements do not overlap. Perturbations that
depart from these conditions halt the computation.  For these very strong perturbations, a full SPH calculation with a different scheme is required, which goes beyond the scope of our current investigation.

The rate of energy dissipation per unit volume is as given in \citet{Moreno:2011jq}:

\begin{equation} \dot{E}_{V} \simeq {\nu \rho} \left \{ \frac{4}{3} \left (
\frac{\partial \omega{'}}{\partial {\varphi{'}}} \right )^2 +
\left [ r'^{2} \left ( \frac{\partial \omega{'}}{\partial r{'}} \right )^2 +
\left ( \frac{\partial \omega{'}}{\partial \theta{'}} \right )^2
\right ]{\sin}^2{\theta{'}} \right \},
\label{dis1} \end{equation}
 
\noindent with $\nu$  as the kinematical viscosity, $\rho$ as the mass density, $\omega'$  as the angular velocity, and $r'$, $\theta'$, $\varphi'$ are, respectively, the radius, latitude, and longitude coordinates. The primes indicate that the variables are measured in the noninertial frame of reference that rotates with the companion star's orbital velocity and with its origin at the center of the perturbed star. 

For the new version of TIDES used in this paper, we implemented a self-consistent calculation of turbulent viscosity instead of providing it as a fixed input parameter as in previous versions. Then we chose the simplest formulation \citep{1987flme.book.....L}:

\begin{equation}
 \nu_\mathrm{turb} =\lambda \ell_\mathrm{t} \Delta u_\mathrm{t},         \label{eq_Landau_Lifshitz}
\end{equation}

\noindent where $\ell_\mathrm{t}$ is the characteristic length of the largest eddies that are associated
with the turbulence,  $\Delta u_\mathrm{t}$ is the typical average velocity variation of the flow over
the length, $\ell_\mathrm{t}$, and $\lambda$ is the proportionality parameter that is analogous to 
$\alpha$   introduced by \citet{1988AdSpR...8b.135S} in the $\alpha$-disk model. 




The largest amplitude perturbation caused by the tidal interaction is in the azimuthal 
direction \citep{1981ApJ...246..292S,2009ApJ...704..813H}. Our n-layer  numerical simulation considers only the 
azimuthal motions and their radial gradients.  Thus,  we set $\ell_\mathrm{t}$=$\ell_r$ and took $\Delta u_\mathrm{t}$ as  the typical average velocity variation between a given volume element and the medium surrounding it within this distance, $\ell_r$. This corresponds to velocity variations between any individual volume element and  the elements above and below it.  

In the numerical scheme, the value of $\Delta u_\mathrm{t}$ is obtained as follows.
Let $r$ be the radius of a particular element which we call $e_\mathrm{i}$ and 
$\vv'_\mathrm{i}$=$\omega'_\mathrm{i} r \sin \theta$ its linear velocity. Here, $\omega'_\mathrm{i}$ is 
the angular velocity  in the frame of reference, $S'$, that rotates with the binary orbit,
and $\theta$ is the polar angle. The distance from the center of $e_\mathrm{i}$ to its top and 
bottom surfaces is, respectively, $r+\ell_r/2$ and $r-\ell_r/2$.  The algorithm averages 
the angular velocity of the volume elements that lie above and below $e_\mathrm{i}$ and then multiplies 
this average by ($r+\ell_r/2)\sin\theta$ (above) and ($r-\ell_r/2)\sin\theta$ (below).   
This gives the typical velocities above and below $e_\mathrm{i}$, which we call $\vv_\mathrm{top}$ and 
$\vv_\mathrm{bottom}$, respectively. Hence, the average velocity of the elements surrounding $e_\mathrm{i}$  
is $\vv_\mathrm{ave}$= ($\vv_\mathrm{top}$ + $\vv_\mathrm{bottom}$)/2 and the typical average velocity variation of the flow over the length scale $\ell_r$ is $\Delta u_t$=$\vv_\mathrm{}$-$\vv_\mathrm{ave}$. 

The values of $\ell_r$ and $\Delta u_t$ obtained as described above for each volume element in the grid are inserted in Eq.~\ref{eq_Landau_Lifshitz}.  The total viscosity is computed as $\nu$=$\nu_\mathrm{turb}$+$\nu_\mathrm{molec}$, where $\nu_\mathrm{molec}$ is the  molecular viscosity and it is given as input. This value
of $\nu$ at each volume element is now used by the TIDES algorithm to solve the equations of  motion in the next time step of the calculation.


\begin{table}
\caption{Description of TIDES input parameters.}
\label{table1}
\centering
\begin{tabular}{l l l }     
\hline\hline
Param      & Value  & Description         \\        
\hline
$P_\mathrm{orb}$  & 1.44   &Orbital period (d)              \\
$e$        & 0      &Orbital eccentricity              \\
$m_1$      & 1.0    &Primary mass (perturbed star) (${\rm M_\odot}$)  \\
$m_2$      & 0.8    &Secondary mass (point source) (${\rm M_\odot}$)  \\
$R_1$      & Tab.\ref{table_TIDES_runs} &Primary unperturbed radius (${\rm R_\odot}$)  \\  
$\omega_0$ &....    &Rotation angular velocity of rigid core \\
$\Omega_0$ &....    &Orbital angular velocity in a circular orbit \\
$\beta_0^0$&Tab.\ref{table_TIDES_runs}&Synchronicity parameter $\beta_0^0$=$\omega_0/\Omega_0$   \\
$\nu_\mathrm{molec}$ & 10$^{-16}$  &Molecular viscosity (${\rm R_\odot^2\, d^{-1}}$)   \\
$\lambda$  &Tab.\ref{table_TIDES_runs}&Turbulent viscosity coefficient\\
$n$        &Tab.\ref{table_TIDES_runs}&Polytropic index   \\   
$\Delta R/R_1$& 0.06   &Layer thickness   \\
$N_r$      &Tab.\ref{table_TIDES_runs}, \ref{10layer_TIDES_runs}  &Number of layers \\
$N_\varphi$ & 200   & Number of partitions in longitude \\
$N_\theta$& 20      & Number of partitions in latitude \\
$\mathit{Tol}$      &10$^{-7}$& Tolerance for the Runge-Kutta integration \\
\hline
\hline
\end{tabular}
\end{table}

\begin{table*}
\caption{TIDES five-layer {\bf models}.}
\label{table_TIDES_runs}
\centering
\begin{tabular}{c c c c c c c c c c}    
\hline\hline
{\bf Model} &$R_1$&$n$&$\nu_\mathrm{const}$&$\lambda$& $V_\mathrm{rot}$&$\nu_\mathrm{max}$&$\dot{E}_\mathrm{k=1}$ &$\dot{E}_\mathrm{k=4}$ &$\dot{E}_\mathrm{tot}$\\  
       & (R$_\odot$)&  ....  & (R$_\odot^2/d$) &  ....    & (km/s)    & (R$_\odot^2/d$) & (ergs/s)    &  (ergs/s) & (ergs/s)\\
\hline
&&&&&&&&    \\
&&& Block 1:  $\beta_{0}=$0.95&&&&&  \\
\hline
2 &0.99 & 3.0 &9.1$\times$10$^{-4}$&....&32.6&....         &2.0$\times$10$^{30}$&4.1$\times$10$^{30}$&1.0$\times$10$^{31}$\\
4 &0.99 & 3.0 &....         &1.0 &32.6&6.3$\times$10$^{-4}$&1.8$\times$10$^{29}$&6.9$\times$10$^{29}$&1.9$\times$10$^{30}$\\
6 &0.99 & 3.0 &....         &0.1 &32.6&6.5$\times$10$^{-5}$&3.1$\times$10$^{28}$&7.3$\times$10$^{28}$&2.0$\times$10$^{29}$\\
7 &1.20  & 3.0 &9.1$\times$10$^{-4}$&....&39.6&....         &6.5$\times$10$^{30}$&1.3$\times$10$^{31}$&3.3$\times$10$^{31}$\\
8 &1.20  & 3.0 &....         &1.0 &39.6&1.4$\times$10$^{-3}$&5.5$\times$10$^{29}$&4.4$\times$10$^{30}$&9.1$\times$10$^{30}$\\
10&1.20  & 3.0 &....         &0.1 &39.6&1.4$\times$10$^{-4}$&5.9$\times$10$^{28}$&4.5$\times$10$^{29}$&9.1$\times$10$^{29}$\\
11&1.40  & 3.0 &3.2$\times$10$^{-3}$&....&46.2&....         &5.8$\times$10$^{31}$&1.2$\times$10$^{32}$&3.0$\times$10$^{32}$\\
12&1.40  & 3.0 &....         &1.0 &46.2&2.9$\times$10$^{-3}$&1.2$\times$10$^{30}$&2.1$\times$10$^{31}$&3.9$\times$10$^{31}$\\
14&1.40  & 3.0 &....         &0.1 &46.2&2.9$\times$10$^{-4}$&1.2$\times$10$^{29}$&2.2$\times$10$^{30}$&4.0$\times$10$^{30}$\\
23&2.25 & 3.0 &9.0$\times$10$^{-2}$&....&74.2&....         &3.0$\times$10$^{34}$&6.0$\times$10$^{34}$&1.5$\times$10$^{35}$\\
24&2.25 & 3.0 &....         &1.0 &74.2&3.1$\times$10$^{-2}$&1.2$\times$10$^{32}$&2.9$\times$10$^{33}$&5.0$\times$10$^{33}$\\
26&2.25 & 3.0 &....         &0.1 &74.2&3.2$\times$10$^{-3}$&9.3$\times$10$^{30}$&3.1$\times$10$^{32}$&5.5$\times$10$^{32}$\\
41&1.648&3.8&....         &1.0 &54.4&7.5$\times$10$^{-3}$&8.4$\times$10$^{29}$&6.0$\times$10$^{30}$&1.1$\times$10$^{31}$\\
\hline
&&&&&&&&  \\
&&& Block 2:  $\beta_{0}=$1.05&&&&&  \\
\hline
61&0.97 &1.5&....&1.0&35.4&4.9$\times$10$^{-4}$&6.9$\times$10$^{29}$&9.2$\times$10$^{30}$&2.5$\times$10$^{31}$\\
65&1.20 &1.5&....&1.0&43.8&9.7$\times$10$^{-4}$&4.0$\times$10$^{30}$&7.8$\times$10$^{31}$&2.1$\times$10$^{32}$\\
66&1.20 &3.0&....&1.0&43.8&1.4$\times$10$^{-3}$&4.2$\times$10$^{29}$&4.3$\times$10$^{30}$&9.2$\times$10$^{30}$\\
67&1.20 &3.8&....&1.0&43.8&1.6$\times$10$^{-3}$&3.7$\times$10$^{28}$&2.2$\times$10$^{29}$&5.2$\times$10$^{29}$\\
69&1.648&3.8&....&1.0&60.1&7.6$\times$10$^{-3}$&6.0$\times$10$^{29}$&6.1$\times$10$^{30}$&1.1$\times$10$^{31}$\\
\hline
&&&&&&&&  \\
&&&Block 3:  $\beta_{0}=$1.10&&&&&  \\
\hline
73&1.20 &3.0&....&1.0&45.8&2.7$\times$10$^{-3}$&5.0$\times$10$^{30}$&3.4$\times$10$^{31}$&7.5$\times$10$^{31}$\\
75$^{(a)}$&1.648&3.8&....&1.0&62.9&1.5$\times$10$^{-2}$&6.5$\times$10$^{30}$&4.9$\times$10$^{31}$&8.8$\times$10$^{31}$\\
75$^{(b)}$&1.648&3.8&....&1.0&62.9&1.5$\times$10$^{-2}$&7.9$\times$10$^{30}$&4.9$\times$10$^{31}$&8.9$\times$10$^{31}$\\
75$^{(c)}$&1.648&3.8&....&1.0&62.9&1.5$\times$10$^{-2}$&8.9$\times$10$^{30}$&4.8$\times$10$^{31}$&8.9$\times$10$^{31}$\\
\hline
\hline
\end{tabular}
\tablefoot{Column descriptions: (1) model number; (2) stellar radius; (3) polytropic index; (4) fixed value in the constant viscosity models; (5) $\lambda$ value; (6)  rotation velocity at the equator; (7) maximum viscosity value in the variable viscosity runs; (8) and (9),   dissipation rate in, respectively,  the layer closest to the core (layer 1) and in the layer below the surface (layer 4); (10) total energy dissipation rate
over the entire 3D structure.  Model 75 is a case in which the inner layers are still evolving toward the stationary state so values for different orbital cycles are listed: $^a$Cycle 100; $^b$Cycle 150;  $^c$Cycle 200.} 
\end{table*}
\begin{table*}
\caption{TIDES ten-layer models.}
\label{10layer_TIDES_runs}
\centering
\begin{tabular}{c c c c c c c c c c}    
\hline\hline
Model&$R_1$&$n$&$\beta_0$&$\lambda$& $V_\mathrm{rot}$&$\nu_\mathrm{max}$&$\dot{E}_\mathrm{k=1}$ &$\dot{E}_\mathrm{k=9}$ &$\dot{E}_\mathrm{tot}$\\  
       & (R$_\odot$)&  ....  & (R$_\odot^2/d$) &  ....    & (km/s)    & (R$_\odot^2/d$) & (ergs/s)    &  (ergs/s) & (ergs/s)\\
\hline
62&0.97 &1.5&1.05&1.0&35.4&4.9$\times$10$^{-4}$&1.8$\times$10$^{27}$&9.2$\times$10$^{30}$&2.6$\times$10$^{31}$\\
63&0.97 &2.2&1.05&1.0&35.4&5.1$\times$10$^{-4}$&2.2$\times$10$^{27}$&4.4$\times$10$^{30}$&1.0$\times$10$^{31}$\\
64&0.97 & 2.2 & 1.10& 1.0&37.1& 1.0$\times$10$^{-3}$&7.5$\times$10$^{27}$&1.2$\times$10$^{31}$ &8.6$\times$10$^{31}$ \\
72 & 0.99 & 3.0 & 0.95 & 1.0& 32.6 &6.2$\times$10$^{-4}$&1.0$\times$10$^{27}$&6.8$\times$10$^{29}$&2.2$\times$10$^{30}$\\
68&1.294&1.5&1.05&1.0&47.2&5.0$\times$10$^{-3}$&6.1$\times$10$^{30}$&1.2$\times$10$^{34}$&2.9$\times$10$^{34}$\\
70&1.648&3.8&1.05&1.0&60.1&7.5$\times$10$^{-3}$&6.0$\times$10$^{27}$&5.9$\times$10$^{30}$&1.1$\times$10$^{31}$\\
71 &1.648 &3.8 & 0.20 &1.0&11.4&1.6$\times$10$^{-1}$ &6.8$\times$10$^{32}$ &4.0$\times$10$^{34}$ &1.0$\times$10$^{35}$\\
   
\hline
\hline
\end{tabular}
\end{table*}

\begin{table*}
\caption{Energy dissipation rate radial profiles from the TIDES ten-layer models. }
\label{tab_input_dotE}
\centering
\begin{tabular}{llllllll}
\hline\hline
\multicolumn{3}{c}{Model 62} & \multicolumn{1}{c}{Model 63} &\multicolumn{2}{c}{Model 70} & \multicolumn{2}{c}{Model 68}    \\
\multicolumn{3}{c}{$R_1$=0.97, $n$=1.5} &\multicolumn{1}{c}{$R_1$=0.97, $n$=2.2} &
    \multicolumn{2}{c}{$R_1$=1.648, $n$=3.8} & \multicolumn{2}{c}{$R_1$=1.294, $n$=1.5}     \\
\hline
$k$ & r\,(R$_\odot$)&$\dot{E}_{r}$ & $\dot{E}_{r}$ &r\,(R$_\odot$)&$\dot{E}_{r}$  &r\,(R$_\odot$)&$\dot{E}_{r}$  \\
\hline
 1 &0.39& 1.8$\times$10$^{27}$&2.2$\times$10$^{27}$& 0.66 &6.0$\times$10$^{27}$&0.52 &6.1$\times$10$^{30}$         \\
 2 &0.45& 3.4$\times$10$^{27}$&3.8$\times$10$^{27}$& 0.76 &1.8$\times$10$^{28}$&0.60 &3.5$\times$10$^{31}$        \\
 3 &0.50& 1.1$\times$10$^{28}$&1.1$\times$10$^{28}$& 0.86 &7.3$\times$10$^{28}$&0.67 &1.2$\times$10$^{32}$       \\
 4 &0.56& 5.2$\times$10$^{28}$&4.7$\times$10$^{28}$& 0.96 &2.5$\times$10$^{29}$&0.75 &3.1$\times$10$^{32}$        \\
 5 &0.62& 2.5$\times$10$^{29}$&2.3$\times$10$^{29}$& 1.05 &5.2$\times$10$^{29}$&0.83 &6.0$\times$10$^{32}$        \\
 6 &0.68& 7.9$\times$10$^{29}$&7.3$\times$10$^{29}$& 1.15 &7.6$\times$10$^{29}$&0.90 &9.6$\times$10$^{32}$       \\
 7 &0.74& 1.4$\times$10$^{30}$&1.4$\times$10$^{30}$& 1.25 &8.9$\times$10$^{29}$&0.98 &1.2$\times$10$^{33}$       \\
 8 &0.80& 2.0$\times$10$^{30}$&1.9$\times$10$^{30}$& 1.35 &1.6$\times$10$^{30}$&1.06 &2.1$\times$10$^{33}$       \\
 9 &0.85& 9.2$\times$10$^{30}$&4.4$\times$10$^{30}$& 1.45 &5.9$\times$10$^{30}$&1.14 &1.2$\times$10$^{34}$       \\
 10&0.91& 1.2$\times$10$^{31}$&1.5$\times$10$^{30}$& 1.55 &1.5$\times$10$^{30}$&1.22 &1.2$\times$10$^{34}$       \\
\hline
\hline
\end{tabular}
\tablefoot{Column 1 lists the layer number; cols. 2, 5, and 7 list the distance of the layer from the star's center; cols. 3, 4, 6, and 8 list the energy dissipation rate for each layer in units of ergs/s.  Models 62 and 63 have the same radial grid.
}
\end{table*}

\subsection{TIDES input parameters and model runs}

The modeled binary star is called the "primary." Its  mass is $m_1$ and it has an initial, unperturbed radius, $R_1$. Its initial rotation condition is one of rigid rotation with its rotation velocity given in terms of the synchronicity parameter, $\beta_0$=$\omega_0$/$\Omega_0$, where $\omega_0$ is the angular velocity of the inner core (assumed throughout to rotate as a rigid body) and $\Omega_0$ is the orbital angular velocity at a reference point in the orbit, for example, periastron in the case of nonzero eccentricity. The TIDES input parameters are described in Table~\ref{table1}, where we also list the values that were kept constant. The stellar masses and orbital period were chosen to be the same as in KM2016.

In this paper, we mainly consider rotation rates that are close to corotation.  Specifically, we analyzed cases with $\beta_0$=0.95, 1.05 and 1.10. We also probed the effects for stellar radii in the range $R_1$=0.97--2.25\,R$_\odot$, which correspond to those of a 1\,M$_\odot$ star from the time it reaches the main sequence (MS) until shortly after the terminal age main sequence (TAMS). These radii are smaller than the Roche Lobe radius of the primary star ($R_\mathrm{RL}$\,=2.5 R$_\odot$). The radii that were probed and the corresponding surface equatorial velocities $V_\mathrm{rot}$ in the unperturbed star are listed in columns 2 and 6, respectively, of Tables~\ref{table_TIDES_runs} and  \ref{10layer_TIDES_runs}. 


The tidal shear energy dissipation rates depend on the tidal velocity gradient, the viscosity, and the density, with the latter depending on the assumed polytropic structure. The outer convective layers of low-mass stars are generally represented with a $n$=1.5 polytrope.  However, the  inner layers correspond to polytropes with larger $n$ values, depending on the evolutionary stage.  For the early main sequence models,  we  chose $n$=3, which provides a closer match to the values of the density in the outer layers. 
For later evolutionary states, we considered polytropic indices that provide a best approximation to the density  structure of the MESA model in the outer layers.  


\begin{figure}
\centering
\includegraphics[width=0.75\columnwidth]{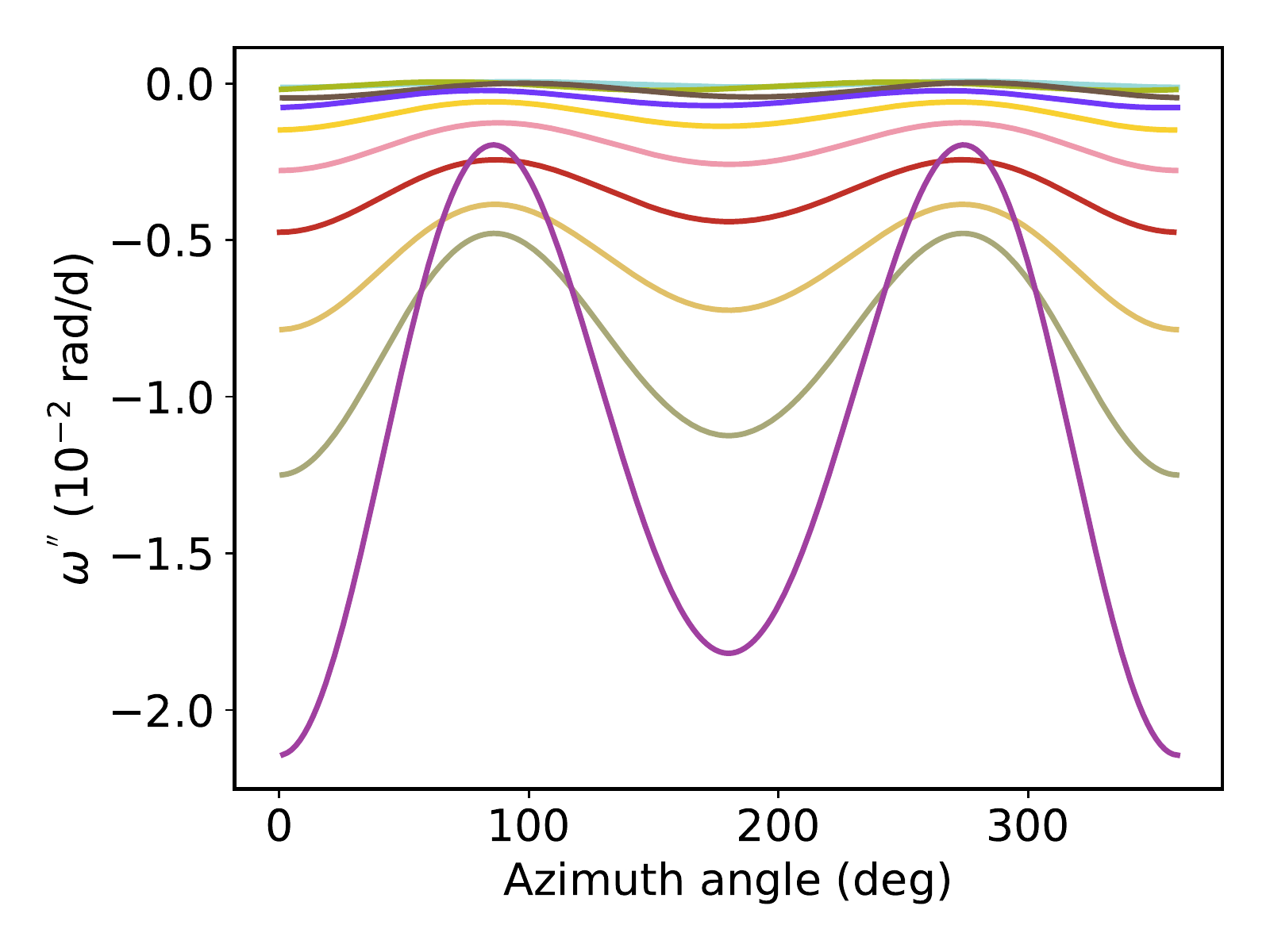}
\includegraphics[width=0.75\columnwidth]{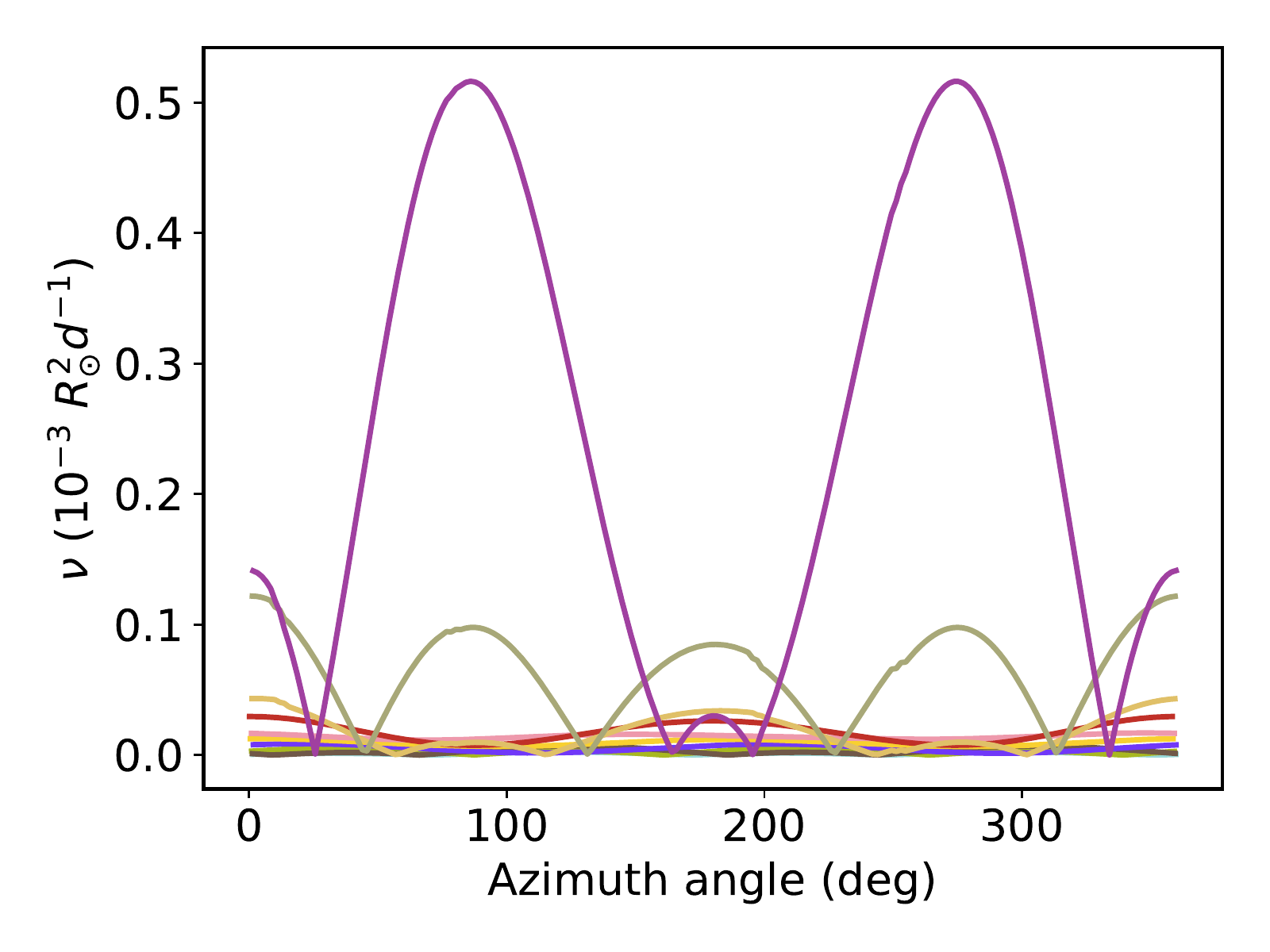}
\includegraphics[width=0.75\columnwidth]{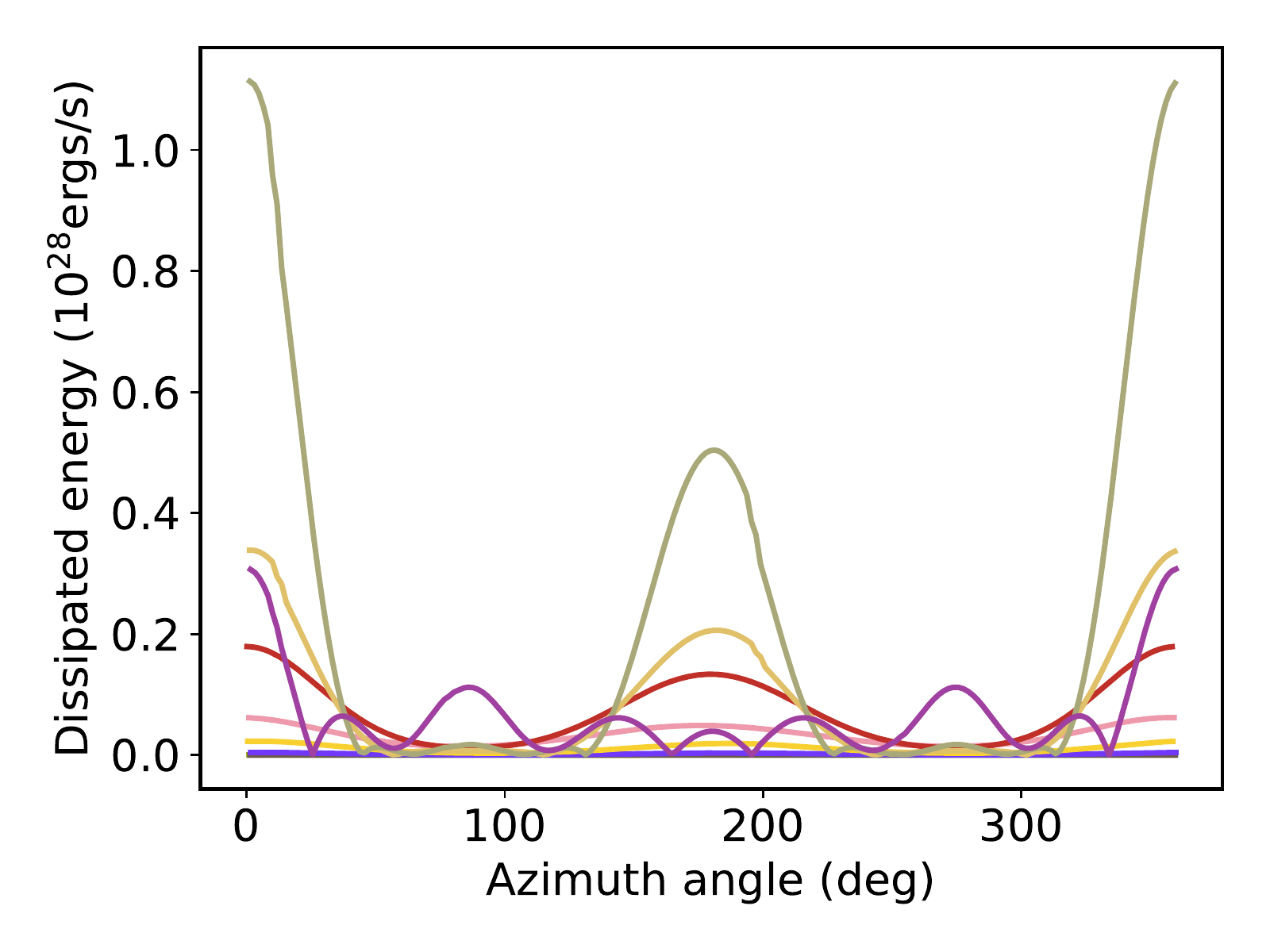}
\caption{Angular velocity, viscosity and dissipated energy as a function of azimuth angle in the
ten layers of model 63.
The azimuth angle is measured in the direction of the companion's orbital motion and $\varphi$=0 corresponds to the sub-binary longitude. {\it Top}: Angular velocity $\omega''$ at the equator in the perturbed star rest frame.
The surface layers (largest amplitude curves) are more strongly coupled to the tidal field than the inner layers and thus are forced to  lag behind the supersynchronously rotating core.  {\it Middle}: Turbulent viscosity values computed  by the model. {\it Bottom}: Tidal shear energy dissipation rate. 
}
\label{fig_model63_omdp_visc}
\end{figure}

The numerical experiments to determine the tidal shear energy dissipation rates are listed in Table~\ref{table_TIDES_runs} and are divided into three blocks.  The aim of each block is to examine the dependence  on particular input parameters. 

In the first block, we probe the dependence on the stellar radius and $\lambda$ (the turbulent viscosity coefficient; see Eq.~\ref{eq_Landau_Lifshitz}) while holding $\beta_0$=0.95. Calculations were performed for a range in stellar radii $R_{1}$/R$_\odot$=[0.99, 2.25].  For each radius, three viscosity representations were tested:  a constant value as described in  \citet{2021A&A...653A.127K}, with the same values that were  used in KM2016 and listed here in column 4 of Table~\ref{table_TIDES_runs}; values computed with $\lambda$=0.1; and value  computed with $\lambda$=1. The value of $\lambda$ is listed in column 5 of Table~\ref{table_TIDES_runs}. 
The second block of numerical experiments listed in Table~\ref{table_TIDES_runs} was performed with
$\beta_0$=1.05.  This means that the primary star's rotation is slightly super-synchronous.  Here, we fixed $\lambda$=1 and tested different polytropic indices as well as different stellar radii within the range tested in block 1. 

The third block of numerical computations was performed with a slightly greater departure from synchronicity than those in block 2, namely $\beta_0$=1.10.  All other input parameters were kept the same. Table~\ref{10layer_TIDES_runs} shows results of numerical computations with ten layers, instead of the five that are used in the models listed in Table~\ref{table_TIDES_runs}.  These models  probe the behavior of layers that lie closer to the nuclear burning core. Since we retained the same layer thickness for all computations, adding more layers allows us to probe regions that lie deeper in the star.  For example, the models  in Table 2 include layers down to 0.7 $R_1$, where $R_1$ is the unperturbed equilibrium radius of the star, while the corresponding model in Table 3 reaches down to 0.4 $R_1$. We have also performed experiments with 20 layers, arriving at nearly the same results as with ten layers because the perturbed velocities decline very rapidly at smaller radii.  Increasing the grid size beyond ten layers significantly increases the processing time and is deemed unnecessary for our current purposes. For problems where the main concern is modeling the layers near the surface, we find that a reduced number of layers yields results that are comparable to a larger radial grid computation.  This is illustrated in Table~\ref{table_compare_layers},  where we list energy dissipation rates in each layer obtained from a five-layer and a ten-layer computation. 

The TIDES computation provides the angular velocity, viscosity, and energy dissipation rates as a function of  azimuth angle $\varphi$ for each radius, $r,$ and colatitude, $\theta,$ of the computational grid and as a function of time, $t$. Thus, the energy dissipation rates are represented as $\dot{E}_{r,\theta,\varphi,t}$. The temporal dependence cannot be neglected because  asynchronous binaries undergo orbital-phase dependent variability. Therefore, for each model run, we chose to output the data at 40 orbital phases within each of five orbital cycles.  An inspection of the five cycles allows for an assessment of long-term variability patterns and also whether the transitory state of the numerical integration has transpired.   The latter has usually occurred within  $\sim$200 orbital cycles.  As the nature of the phase-dependent variability is not the subject of this paper, $\dot{E}_{r,\theta,\varphi,t}$ was averaged over the 40 orbital phases within an orbital cycle in which the calculation has reached the stationary state.  This yields $\dot{E}_{r,\theta,\varphi}$.  The radial profile  $\dot{E}_{r}$ is obtained by integrating over $\theta$ and $\varphi$. 

For a general characterization of the models in Tables 2 and 3, we opted to list the energy dissipation rates in the deepest layer of the calculation and also in the layer that neighbors the surface. The deepest layer is represented by  $\dot{E}_{\mathrm{k=1}}$, to indicate that it refers to the first layer, which in Table 2 corresponds to $r\sim$0.7\,$R_1$ and in Table 3, it corresponds to $r\sim$0.4\,$R_1$.  The layer that is contiguous to the surface layer is represented by $\dot{E}_{\mathrm{k=4}}$ in Table 2 and $\dot{E}_{\mathrm{k=9}}$ in Table 3, both of which correspond to $r\sim$0.9\,$R_1$. We note that the $k$ number index denotes the layer, with $k$=1 corresponding always to the layer that lies closest to the core.



\subsection{Results}

An example of the results is illustrated in Fig.~\ref{fig_model63_omdp_visc}, where we plot the data obtained in model 63.  The top panel illustrates the angular velocity,   $\omega''$, in the equatorial latitude ($\theta$=90$^\circ$), measured in the rotating frame of the primary star\footnote{The double-prime notation indicates that it is measured with respect to the reference frame $S''$, which rotates at the same constant rate as the primary star core, as opposed to the $S'$ reference frame that rotates with the orbital motion of the companion which for eccentric orbits is not constant \citep[see ][]{2021A&A...653A.127K}.} 
as a function of the azimuth angle. The sinusoidal shape corresponds to the equilibrium tide. The peak-to-peak amplitude of each curve decreases with decreasing distance from the stellar center, as expected from the tidal force. These characteristics are shared by all the models that were run for this paper, as shown in Appendix~\ref{sec:plots_block1}. Another general feature is that the overall shape and the  peak-to-peak amplitude do not depend significantly on the viscosity  for the range of viscosity values that were explored, a result that was already found in the one-layer calculations that were performed by KM2016.  

The corresponding behavior of the viscosity in model 63 is illustrated in the middle panel of Fig.\ref{fig_model63_omdp_visc}.  Its value in the layers closest to the surface is a few orders of magnitude larger than near the core.   Thus, the inner layers are significantly less coupled to the outer layers compared to the case in which a constant viscosity is used (models 2, 7, 11, and 23). Thus, the outer layers approach synchronous rotation more rapidly than the inner layers.  This results in a radial gradient in the average velocity of each layer, a result that is consistent with the conclusion of  \citet{1989ApJ...342.1079G}, asserting that a star synchronizes from the surface inward.
The differential rotation structure that is obtained with a variable viscosity is in contrast with the uniform average rotation structure in models 2, 7, 11, and 23, where the viscosity is kept constant,  illustrating the important role that is played by this parameter. 


We list in column 7 of Table~\ref{table_TIDES_runs} the maximum viscosity, $\nu_\mathrm{max}$.  It appears in the calculation  always near the surface and around the equatorial latitude. For stars with radii between 0.99\,R$_\odot$ -- 2.25\,R$_\odot$, we find  $\nu_\mathrm{max}$=4$\times$10$^{13}$\,cm$^2$\,s$^{-1} - $ 2 $\times$10$^{15}$\,cm$^2$\,s$^{-1}$ ($\lambda$=1).

The bottom panel of  Fig.\ref{fig_model63_omdp_visc} shows  $\dot{E}_{r,\varphi}$, the energy dissipation rate for each radius as a function of azimuth. Its behavior  mimics the $\varphi$-dependence of the viscosity as might be expected given that the density remains relatively constant within each layer and only the velocity gradients change as a function of azimuth angle.  

We give the energy dissipation rates in the deepest layer, $\dot{E}_\mathrm{k=1}$, and the layer immediately below the surface layer, $\dot{E}_\mathrm{k=4}$,    in cols. 8 and 9, respectively, of Table~\ref{table_TIDES_runs}. The last column of this table lists $\dot{E}_\mathrm{tot}$, the total energy dissipation rate obtained by integrating $\dot{E}_{r}$  over all layers.  The relatively small difference  between $\dot{E}_\mathrm{tot}$ and  $\dot{E}_\mathrm{k=4}$ is due to the fact that the layers near the surface are responsible for  the largest share of the dissipated energy. In fact,  the greatest contribution to the total energy dissipation rate for models with a polytropic index $>$1.5 arises in the layer that is adjacent to the  surface (in this case, $\dot{E}_\mathrm{k=4}$),  not in the surface layer. A priori, it would appear curious that  the energy dissipation rate in the surface layer is not the greatest since it is the one subject to the strongest tidal perturbations.
There are two explanations for this apparent contradiction.  The first resides in the fact that only the inner boundary of the surface interacts with a neighbor, whereas all other layers interact with both a layer above and one below.  The second factor is based on the interplay between the three terms that enter into the energy dissipation calculation (see Eq.~\ref{dis1}).  When the polytropic index $>$1.5, the density decrease near the surface is more pronounced  than the increase  in viscosity and in the velocity gradients.  This effects is  illustrated by comparing models 65-67, where $\dot{E}_\mathrm{tot}$ decreases by over two orders of magnitudes due to the decreasing surface density that results from changing the polytropic index from 1.5 to 3.8. 

The depth dependence of $\dot{E}_r$ for different density structures  is most clearly illustrated with models that are computed with layers that are deeper, such as those listed in Table~\ref{10layer_TIDES_runs}. These are ten-layer runs for which $\dot{E}_\mathrm{k=1}$ correspond to a layer that lies at $\sim$0.4\,R$_\odot$.  A  layer-by-layer comparison down to this depth  clearly illustrates the density-dependence of $\dot{E}_r$, as shown  in Table~\ref{tab_input_dotE} as a comparison models 62 and 63.

The energy dissipation rate is very sensitive to the value of the synchronicity parameter, $\beta_0$. For example, models 66 and 73 have identical input parameters except for $\beta_0$:  the latter model is slightly more supersynchronous than the former ($\beta_0$=1.10 versus $\beta_0$=1.05). The maximum viscosity value of model 73 is approximately twice as large as that of model 66, and $\dot{E}_{\mathrm{tot}}$ is one order of magnitude larger. Model 71 illustrates the case in which the inner core rotates significantly slower than the other cases studied ($\beta_0$=0.20), but because of the large departure from synchronicity, it has one of the largest values of $\dot{E}$ in all layers. Thus, increasing departures from synchronicity while holding other parameters constant does lead to increasing energy dissipation rates.

Finally, the near-zero-values displayed in the viscosity plots merit a comment. The  value of $\nu_\mathrm{turb}$ is calculated using the instantaneous velocity  gradients.  There are longitudes in our calculations at which these gradients vanish and thus the minimum  $\nu_\mathrm{turb} \rightarrow$0  (see Fig.\ref{fig_model63_omdp_visc}). However, in  reality, the transfer of the  kinetic energy of the flow into eddies that then act as the viscosity source is not instantaneous, nor are the eddies expected to disappear instantaneously when the velocity gradient vanishes.  Thus, there may be a minimum viscosity that is larger than the molecular viscosity even when the velocity gradients momentarily vanish. Furthermore, in convective envelopes, a base-level viscosity  associated with the convective eddies are expected to always be present.  

\begin{table}
\caption{Heat injection in MESA models. \label{tab_MESA_models}}
\centering
\begin{tabular}{l l l l }
\hline\hline
Name           &  Added heat & $\dot{E}$ profile & Note    \\
\hline
 h0       & no           & none     & .... \\
Set2 h1   & yes          & Model 63    & constant    \\
Set2 h10  & yes          & Model 63$\times$10   & constant \\
Set3 h1   & yes          & Model 63    & gradual  \\
Set3 h10  & yes          & Model 63$\times$10   & gradual \\
\hline
\hline
\end{tabular}
\tablefoot{The name of each MESA model is listed in col. 1; col. 2 indicates whether it includes injected heat.
The injected heating profile is indicated in col. 3 and the manner in which it was injected in
col. 4.}
\end{table}

\begin{table}
\caption{Ages and events in MESA models. \label{tab_MESA_models2}}
\centering
\begin{tabular}{l l}
\hline\hline
{ Age }            & Event                                                  \\
  (Gyr)\\
\hline
              & {\bf Standard model h0:}                   \\
9.207        & Core H-abundance [N(H)]$\simeq$0.0001    \\  
10.758       & $R_1 \simeq$1.5 R$_{\odot}$ \\ 
11.644        & end of calculation; $R_1$=2.25 R$_{\odot}$ \\    
              & {\bf Set 2:}                   \\
9.247       &N(H) $\simeq$0.0001, model h1 \\
9.381       &N(H)$\simeq$0.0001, model h10   \\    
              & {\bf Set 3:}                   \\
9.237        &N(H)$\simeq$0.0001, model h1        \\   
9.3591        &N(H)$\simeq$0.0001, model h10              \\ 
10.708       & $R_1 \simeq$1.5 R$_{\odot}$, model h1       \\ 
9.765        & $R_1 \simeq$1.5 R$_{\odot}$, model h10       \\
11.673        & $R_1$=2.25 R$_{\odot}$ end of the run  h1        \\    
11.728        & $R_1$=2.25 R$_{\odot}$ end of the run  h10        \\    
\hline
\hline
\end{tabular}

\end{table}
\begin{figure}
\centering
\includegraphics[width=0.45\columnwidth]{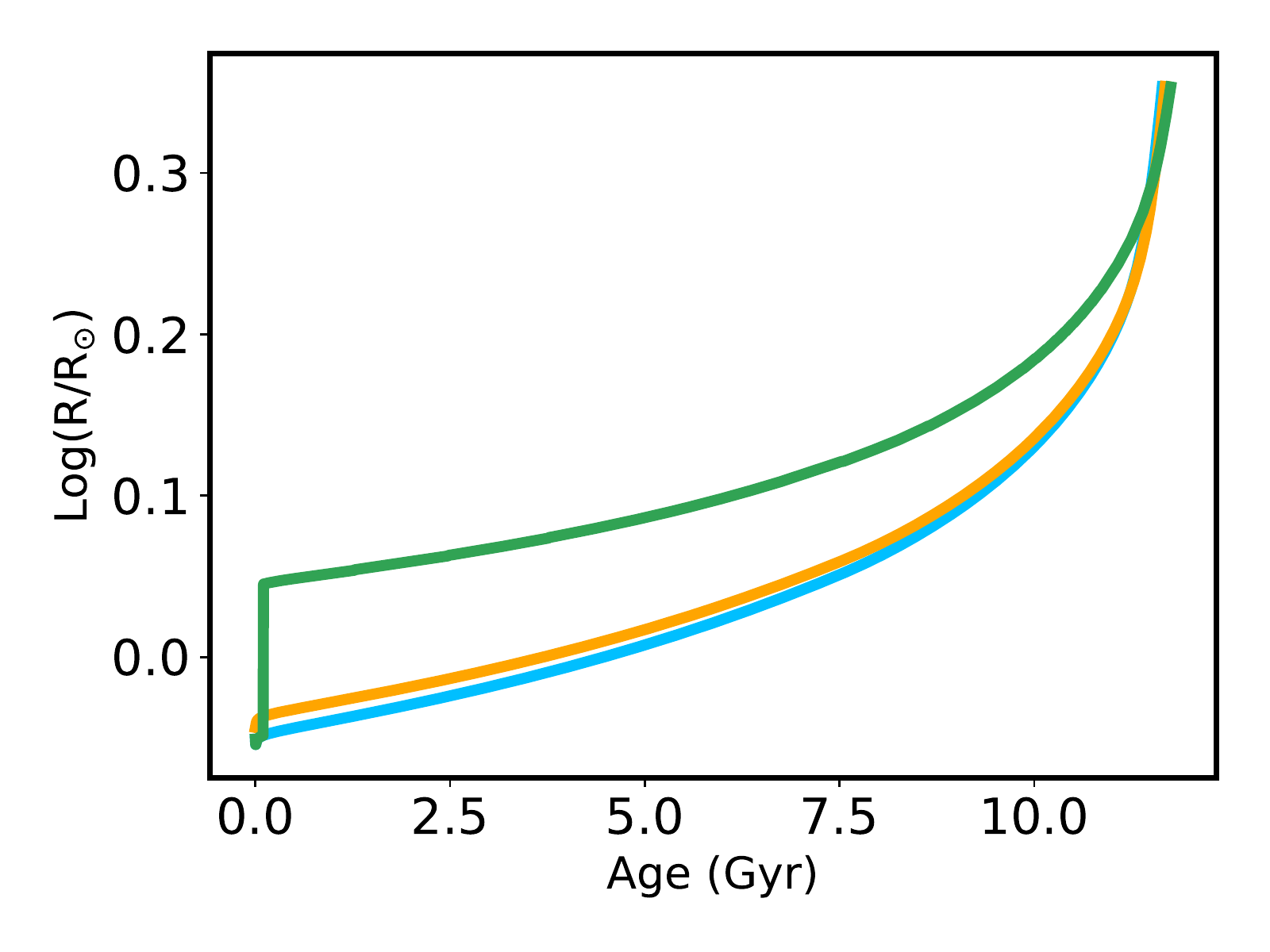}
\includegraphics[width=0.45\columnwidth]{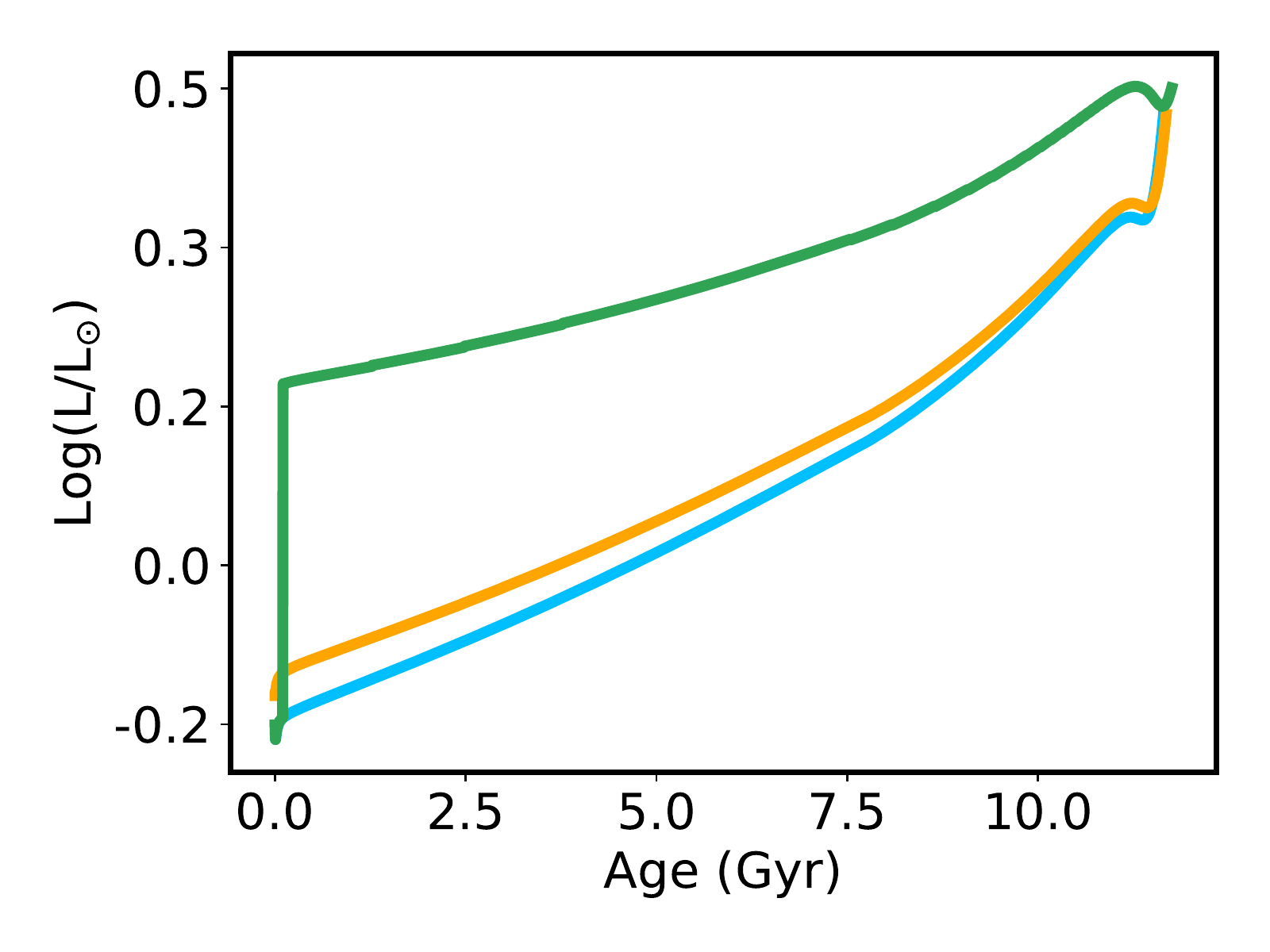}
\includegraphics[width=0.45\columnwidth]{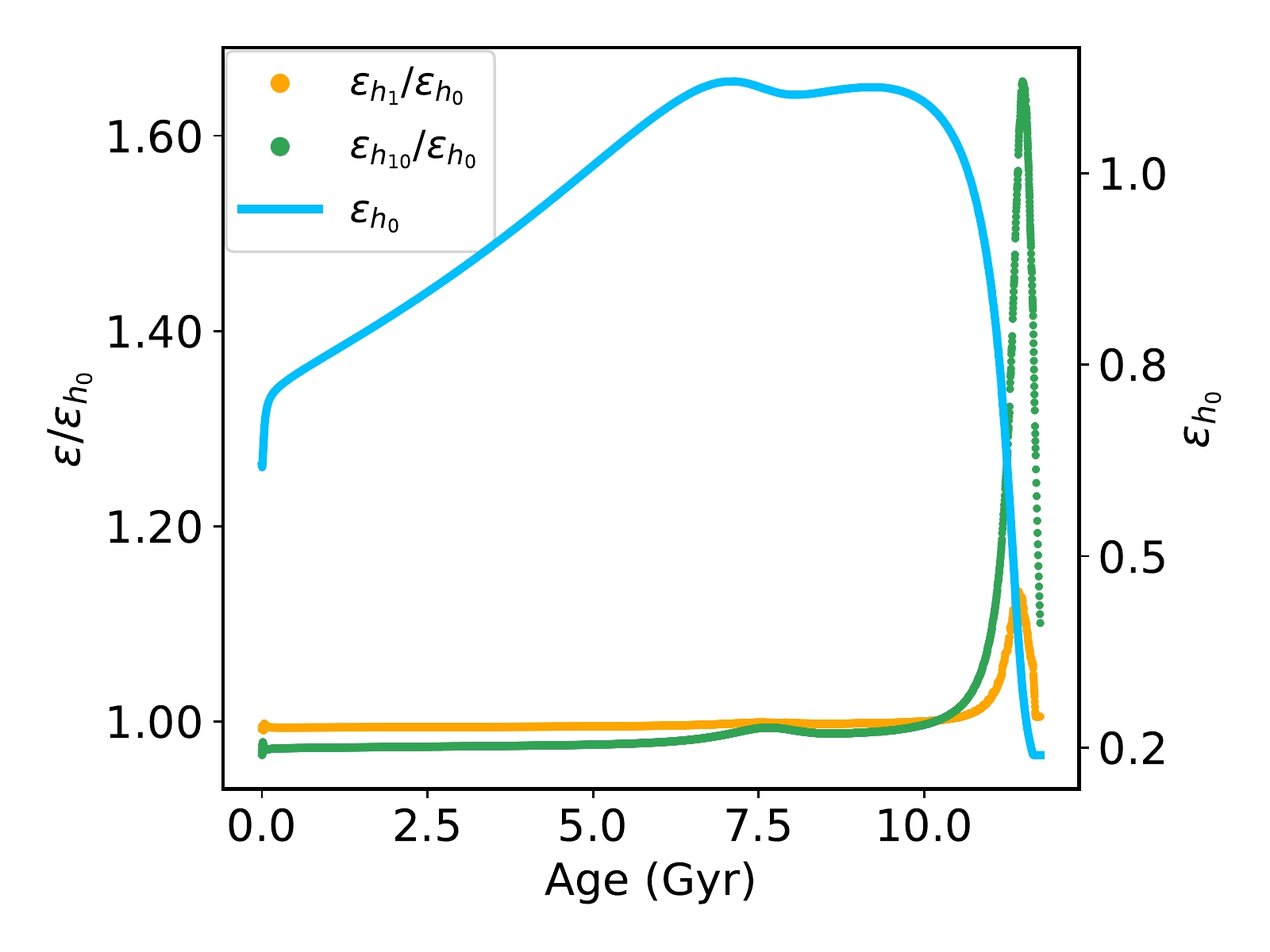}
\includegraphics[width=0.45\columnwidth]{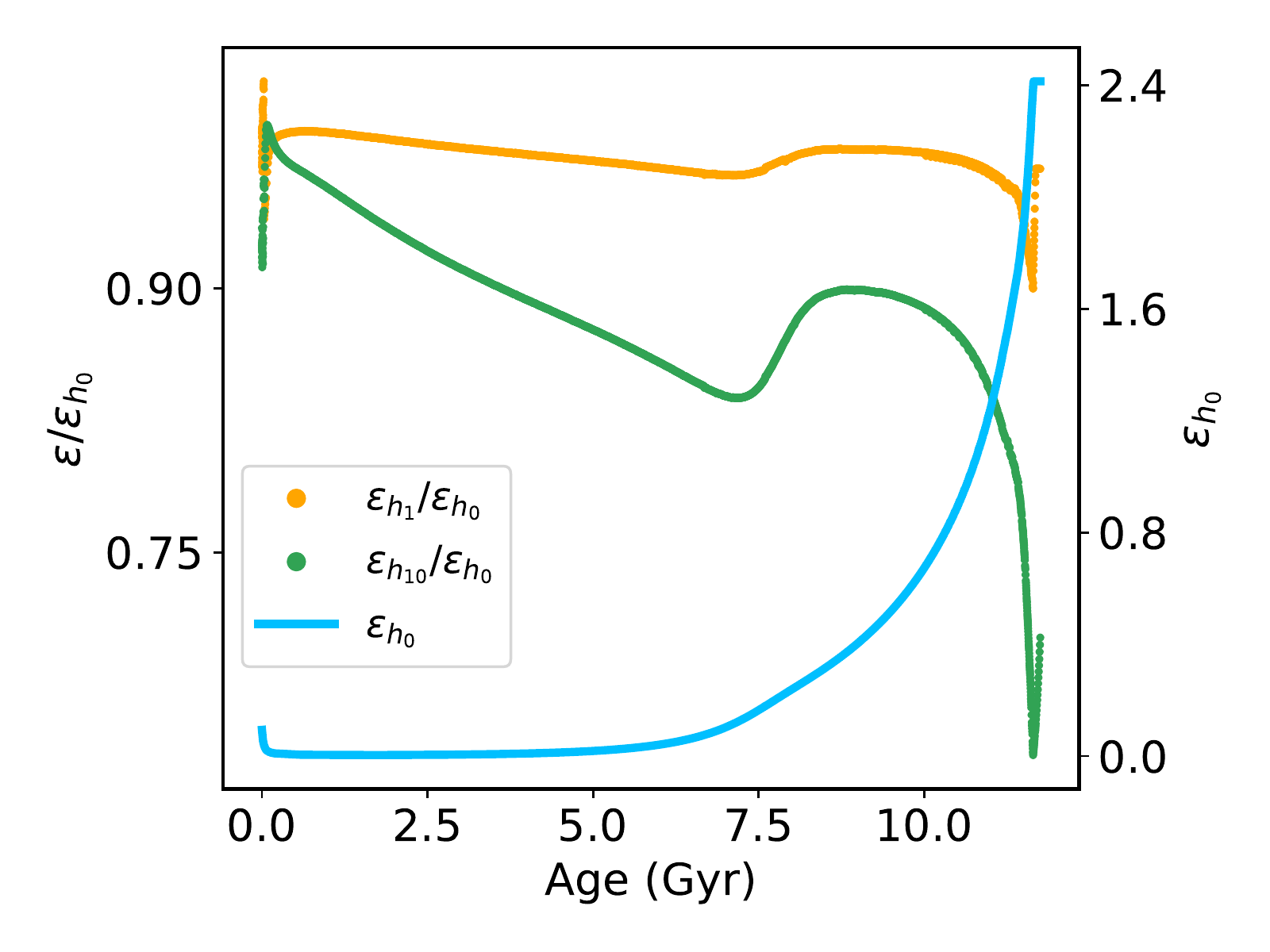}
\caption{ Properties of the MESA models of set 2, in which the heating profile is introduced continuously throughout the evolutionary trajectory. {\it Top}: Stellar radius (left) and luminosity (right). The blue curve corresponds to the nonheated models.  The orange and green curves correspond, respectively, to the heating profile of model 63 and a heating profile that is ten times larger. {\it Bottom}: Proton-proton reaction rate power (left) and CNO reaction rate power (right). The blue curve corresponds to the nonheated models and its ordinate is on the right side of the plot and given in units of log(L$_\odot$). The orange and green curves correspond to the ratio of rates in the heated and nonheated models, with colors the same as in the top panels.
}
\label{fig_MESA_Set2}
\end{figure}

\section{Stellar structure models}

The stellar structure models were computed with version 15140 of the MESA open-source 1D stellar evolution code\footnote{\url{https://docs.mesastar.org/en/r15140/}} \citep{{2011ApJS..192....3P},{2013ApJS..208....4P},{2015ApJS..220...15P},{2018ApJS..234...34P},{2019ApJS..243...10P}}. We ran models for a 1~M$_\odot$ star with solar abundances ($Z=0.02$) and an initial surface rotation velocity of 40~km~s$^{-1}$ on the zero age main sequence (ZAMS). All the models were run until the stellar radius reached 2.25~R$_\odot$, which occurs shortly after the central H fraction falls below $10^{-4}$. By this point, the surface rotation velocity has fallen to about 10~km~s$^{-1}$.

We include the effect of energy dissipation on the stellar envelope using the \verb;other_energy; hook provided in the MESA code. The extra heat we inject into the envelope is taken from the  energy dissipation rate as a function of radius, $\dot{E}_r$, of a TIDES binary model. In particular, we chose the energy dissipation profile of model~63, which is listed in col.~4 of Table~\ref{tab_input_dotE}. The parameters of this model are given in Table~\ref{10layer_TIDES_runs}. The  TIDES code is not an evolution code and the radius of a given model is fixed, whereas the MESA simulations show that the radius of the star changes during the main sequence. We account for this in our models by normalizing the radius of the TIDES energy dissipation profile and apply the profile to the normalized stellar radii of the MESA models. In addition, MESA has many more interior radial points than the TIDES model and we find the energy dissipation rate at each of these by linear interpolation between the nearest TIDES model points.

We run three sets of models: (i) a standard model (h0) which evolves the star from the ZAMS to the stopping point with no added heat; (ii) set 2, in which heat is injected steadily into the stellar envelope with the TIDES energy dissipation rate starting at the ZAMS; and (iii) set 3, in which we emulate the change from synchronous rotation during the MS to asynchronous rotation as the primary star approaches the terminal age main sequence (TAMS; defined in this paper as the time when the central hydrogen mass fraction falls below $10^{-4}$). In set 3, we multiply all of our dissipation rates by the time-dependent factor of
\begin{equation}
\mathit{fac} = \min\left[\frac{e^{(t/t_\mathrm{TAMS})^2} - 1}{(e - 1)}, 1\right] ,
\label{eq:fac}
\end{equation}
which increases sharply to unity as $t\rightarrow t_\mathrm{TAMS}$, where it becomes saturated.


It is evident from Table~\ref{tab_input_dotE} that the energy dissipation rate has a strong dependence on radius, internal structure, and rotation rate.  The aim of this paper is to simply explore the effect on the stellar structure of this type of energy input.  Thus, in order to take into account larger energy  input rates than given by the TIDES model 63,  we also computed models with the model 63  profile multiplied by a factor of 10.  These models are labeled set2-h10 and
set3-h10 for the set 2 and set 3 cases, respectively.  Such higher heating rates are obtained, for example, by increasing the $\beta_0$ value from 1.05 to 1.10, while holding the polytropic index constant or by increasing the radius from 0.99 to 1.4 R$_\odot$; this can be seen by comparing models 66 and 73 and models 4 and 12, respectively, in Table~\ref{table_TIDES_runs}.

\subsection{Results for constant injected heat}

The effect of adding heat into the stellar envelope on basic stellar physical properties is illustrated in Fig.~\ref{fig_MESA_Set2}.   In the top panels, we plot the stellar radius and luminosity as a function of time for the standard model h0, and for models set2-h1 and set2-h10, which correspond to time-constant heat injection throughout the MS, beginning at the ZAMS. For the same age, the h1 and h10 models have larger radii and luminosities, compared to the standard model, h0.

The bottom panels show the total power generated by the proton-proton chain and through the CNO cycle as a function of time. Throughout the MS, the pp chain dominates energy production but it is clear that the extra luminosity in the outer layers changes the energy transport in the envelope and has repercussions on the energy production in the stellar core, since the nuclear energy generation rate in the heated models is slightly smaller than in the standard model.

\begin{figure}
\centering
\includegraphics[width=0.495\columnwidth]{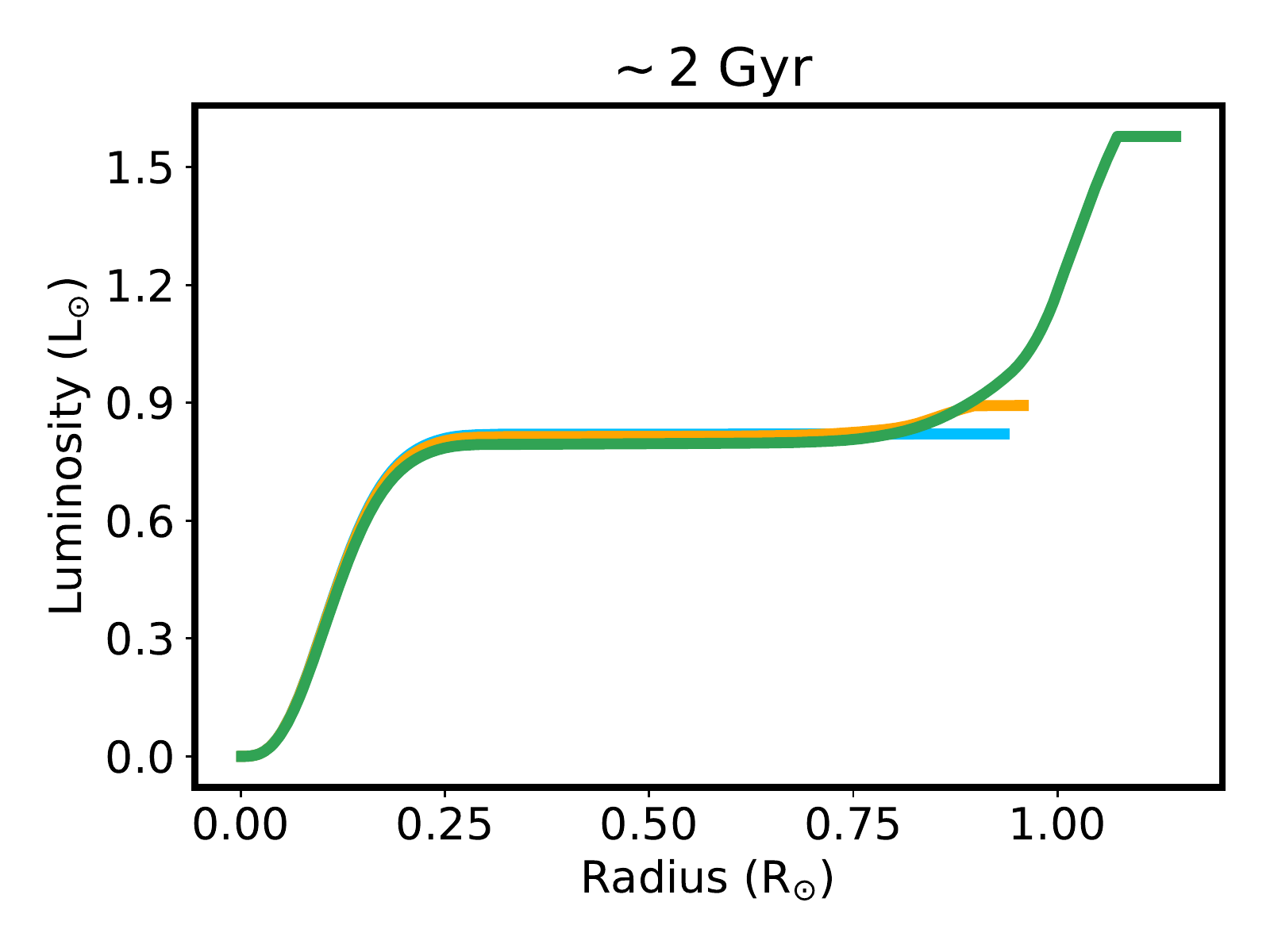}
\includegraphics[width=0.495\columnwidth]{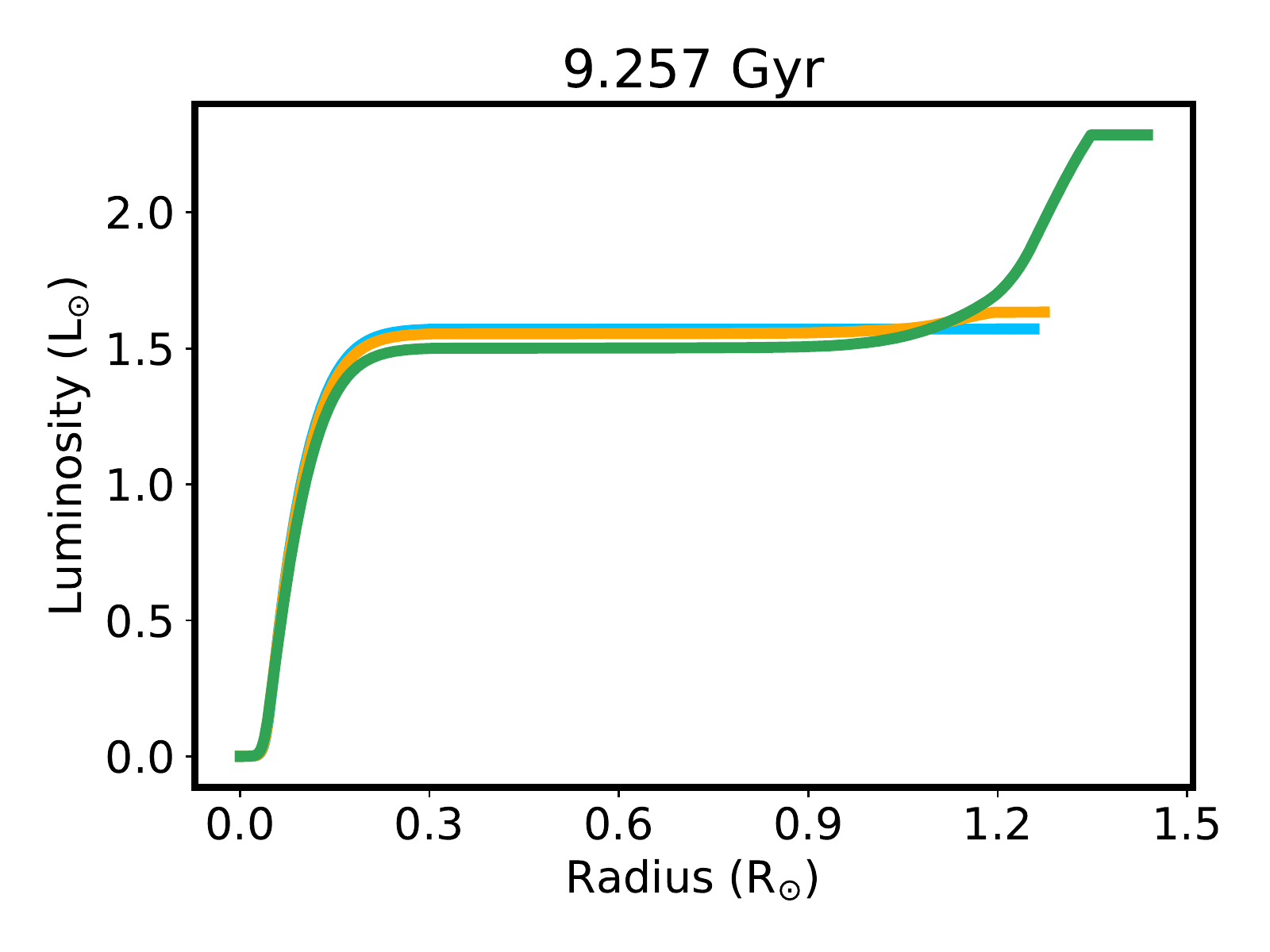}
\includegraphics[width=0.495\columnwidth]{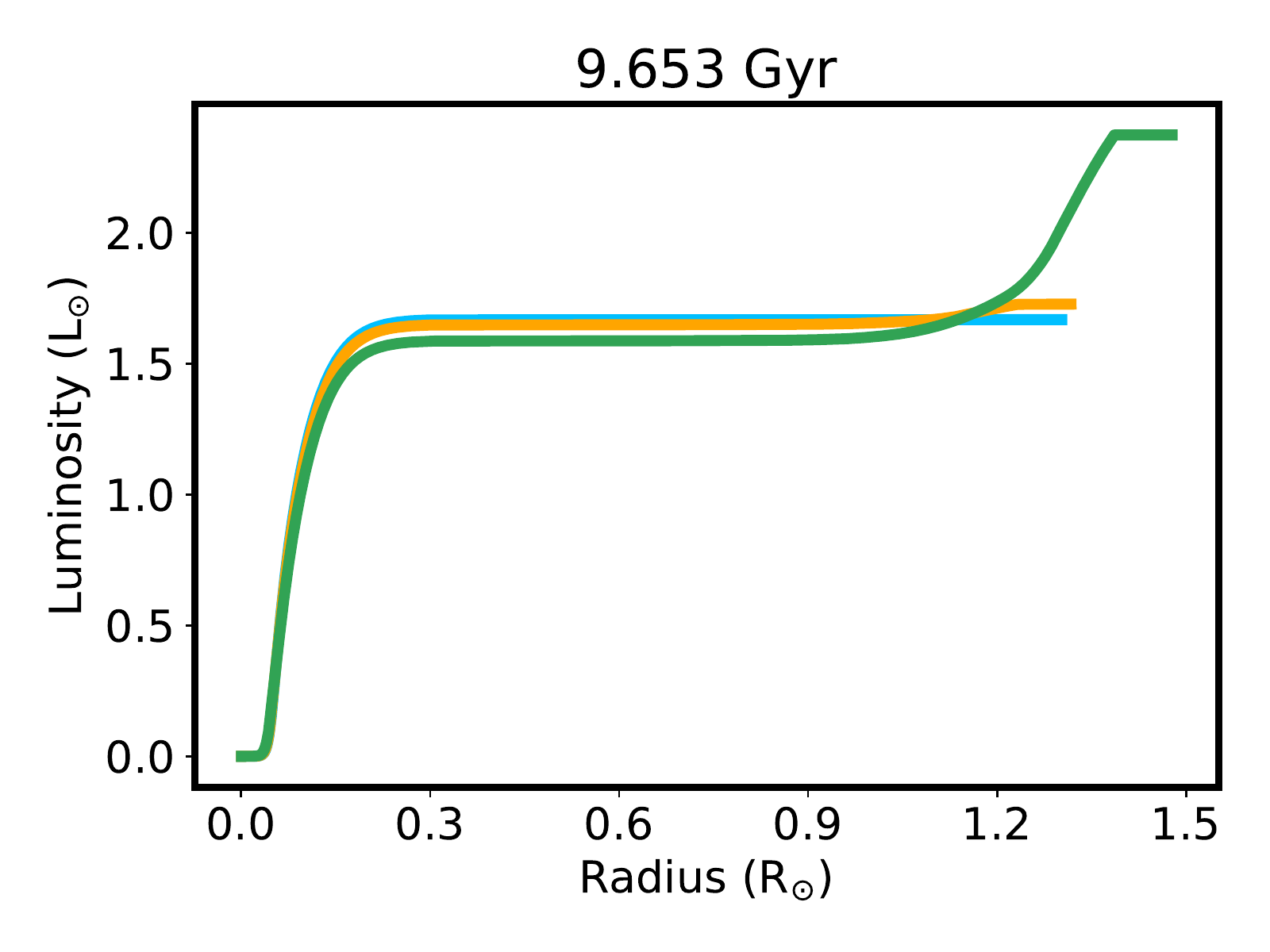}
\includegraphics[width=0.495\columnwidth]{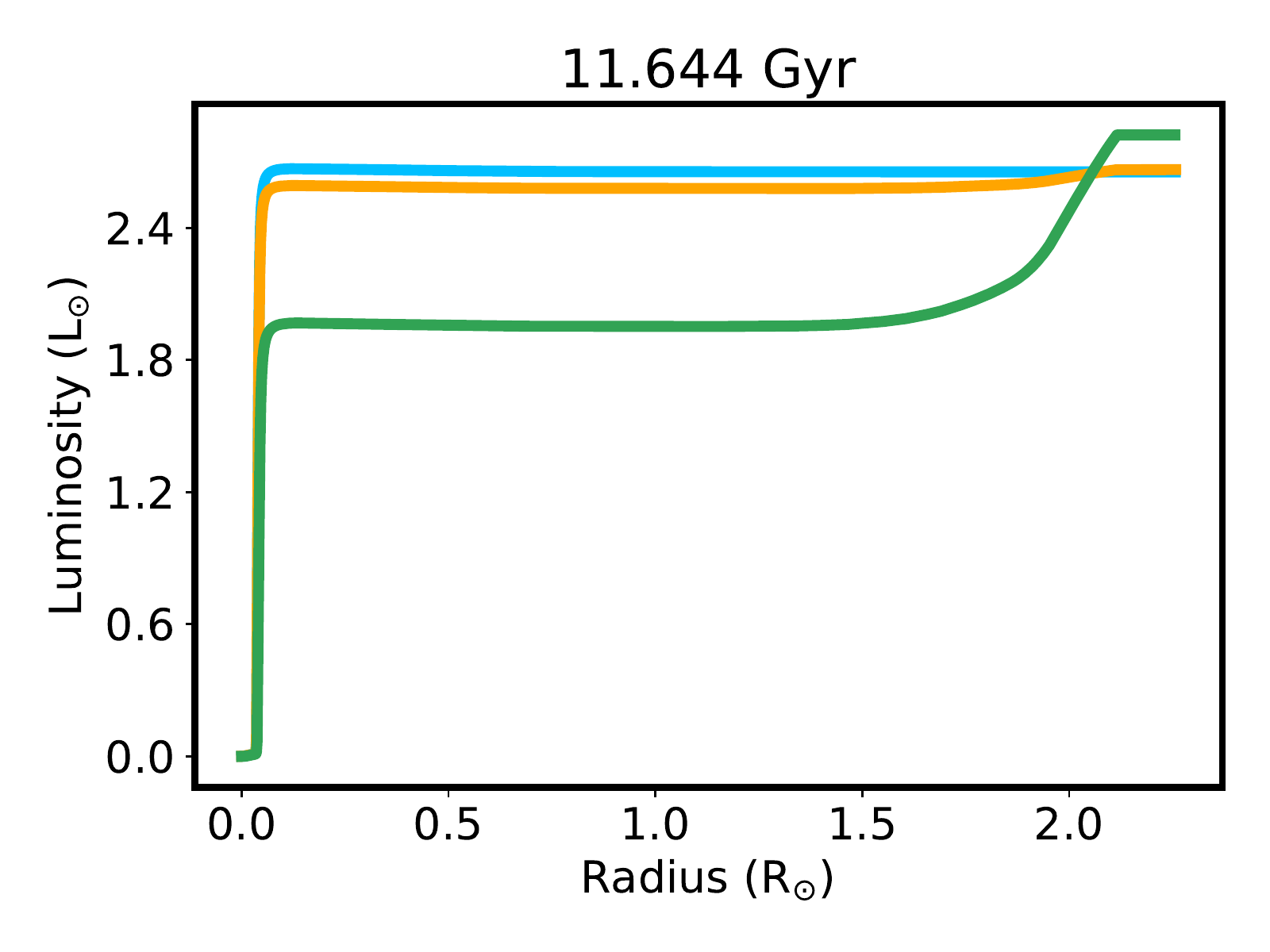}
\caption{Luminosity structure of MESA set 2 models at ages of (top left) 2 Gyr and ages between (top right) 9.257, (bottom left) 9.653 and (bottom right) 11.644~Gyr. The blue curves correspond to the models without added heating; orange to those with heating as given by model 63; green to heating by ten times that given by model 63.  The tidally heated models have surface luminosities that are larger than the nonheated models, but with lower luminosities in deeper layers. 
}
\label{fig_MESA_luminosity_Set2}
\end{figure}

We can also examine the internal structure of the model stars at any point in their evolution. The internal luminosity structure for ages of 2 Gyr and ages between 9.257 and 11.644~Gyr is shown in Fig.~\ref{fig_MESA_luminosity_Set2}. There is a significantly different structure in the heated models  compared to the standard model.  The luminosity inside $\sim$80\% of the maximum radius is smaller in the heated models than in the standard model, while the outer $\sim$20\% has a higher luminosity. Thus, the surface luminosities of the heated stars are higher than those of the standard model.

\begin{figure}
\centering
\includegraphics[width=0.98\columnwidth]{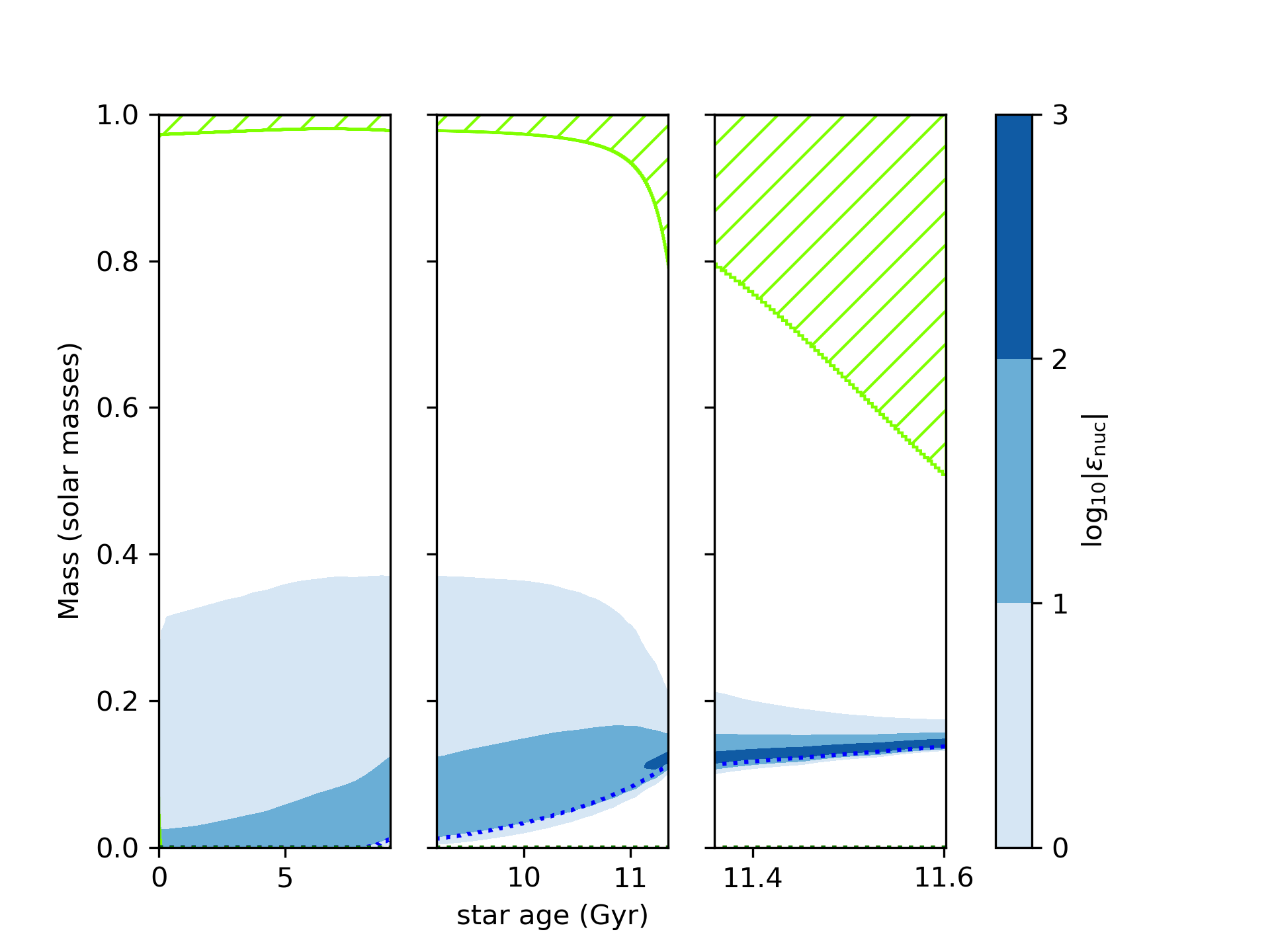}
\includegraphics[width=0.98\columnwidth]{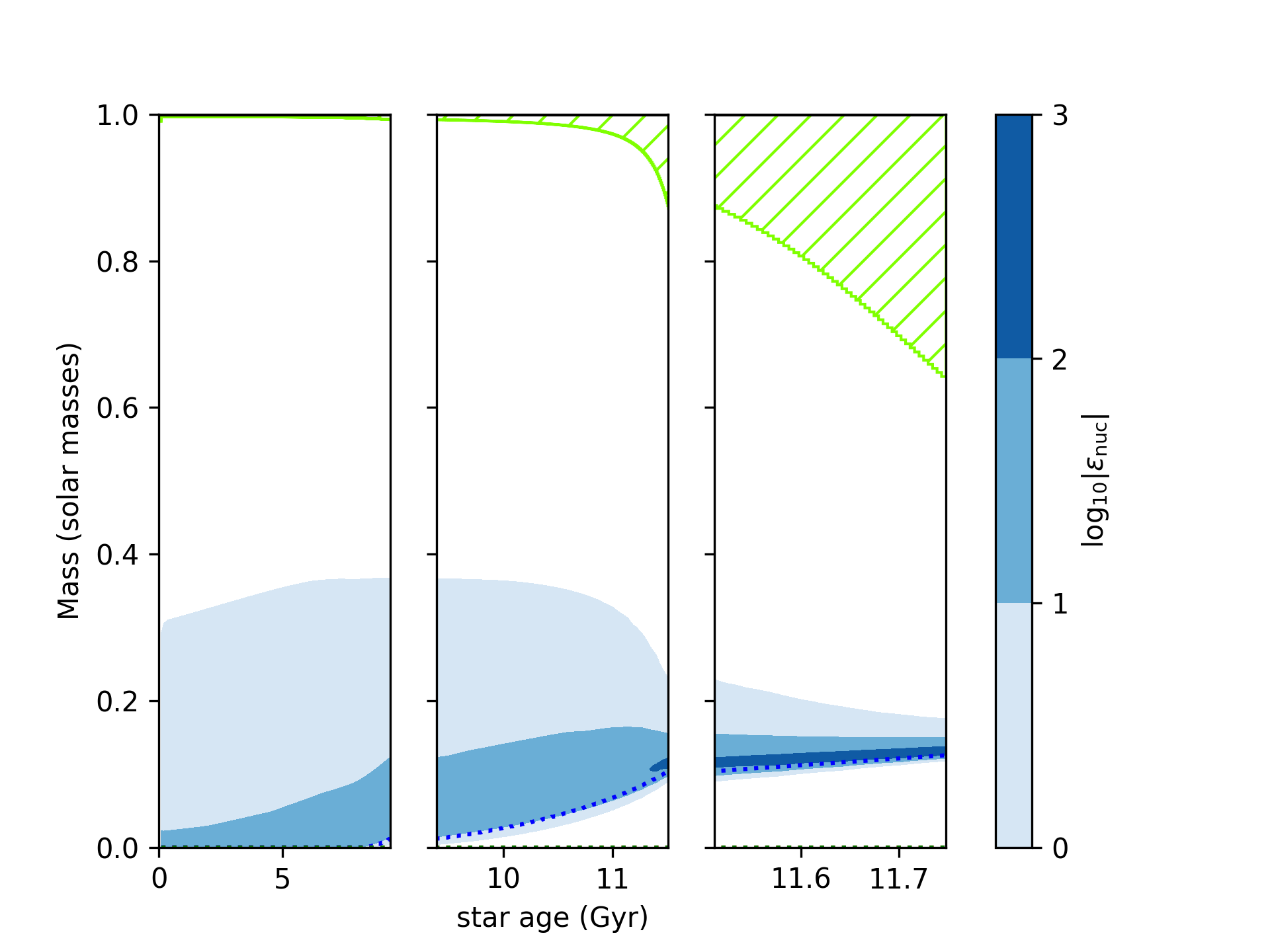}
\caption{Kippenhahn diagram for the standard model (top) and the set2-h10 model (bottom). The panels from left to right represent: \textit {Left}: Evolution from ZAMS to TAMS (defined as abundance of H first drops below 10$^{-4}$ in the stellar core). \textit {Center}: Evolution from the TAMS to 90$\%$ of the remaining time. \textit {Right}: Final 10$\%$ of the post-TAMS evolution. The ordinate is the mass coordinate that runs from the center of the star to the surface. The nuclear energy generation rates are indicated by the different shades of blue.  The green hatches indicate zones in which the energy transport is dominated by convection.} 
\label{fig_MESA_DF_Set2}
\end{figure}

In Fig.~\ref{fig_MESA_DF_Set2} we illustrate the Kippenhahn diagrams for the h0 and the h10  models.\footnote{These diagrams were made with the mkipp.py software written by Pablo Marchant \url{https://github.com/orlox/mkipp/}} These enable us to study how energy transport in different regions of the stellar interior changes as a function of time. The most notable difference is the reduced size of the convective region (green shaded region) in the h10 models compared to the unperturbed models, both on the MS and afterward.  A similar, but much less prominent reduction is present in the h1 models. This is because adding heat to the stellar envelope flattens the temperature gradient and energy can be transported by radiative diffusion almost up to the surface.

\subsection{Results for gradual injected heat}

\begin{figure}
\centering
\includegraphics[width=0.46\columnwidth]{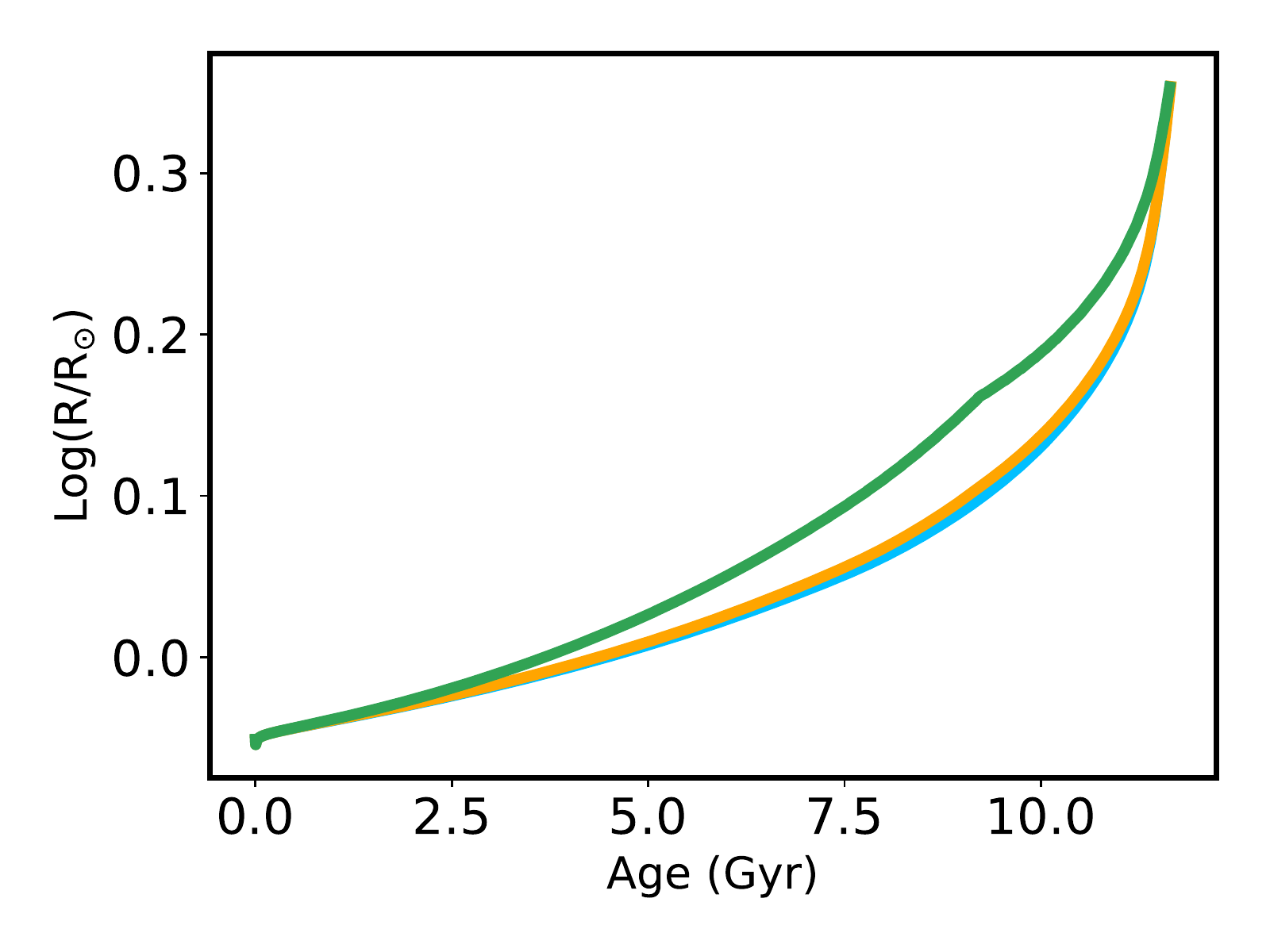}
\includegraphics[width=0.46\columnwidth]{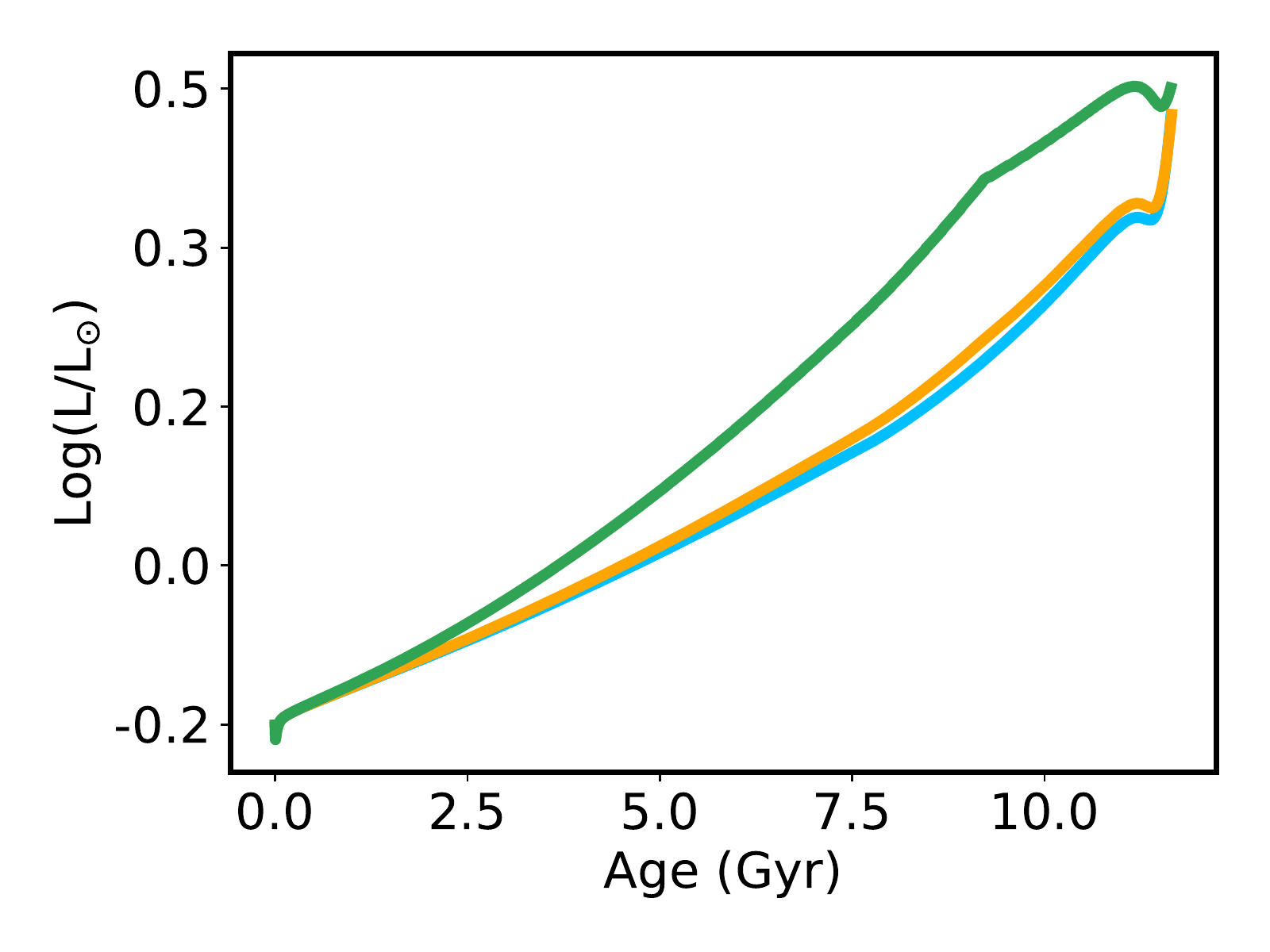}
\includegraphics[width=0.49\columnwidth]{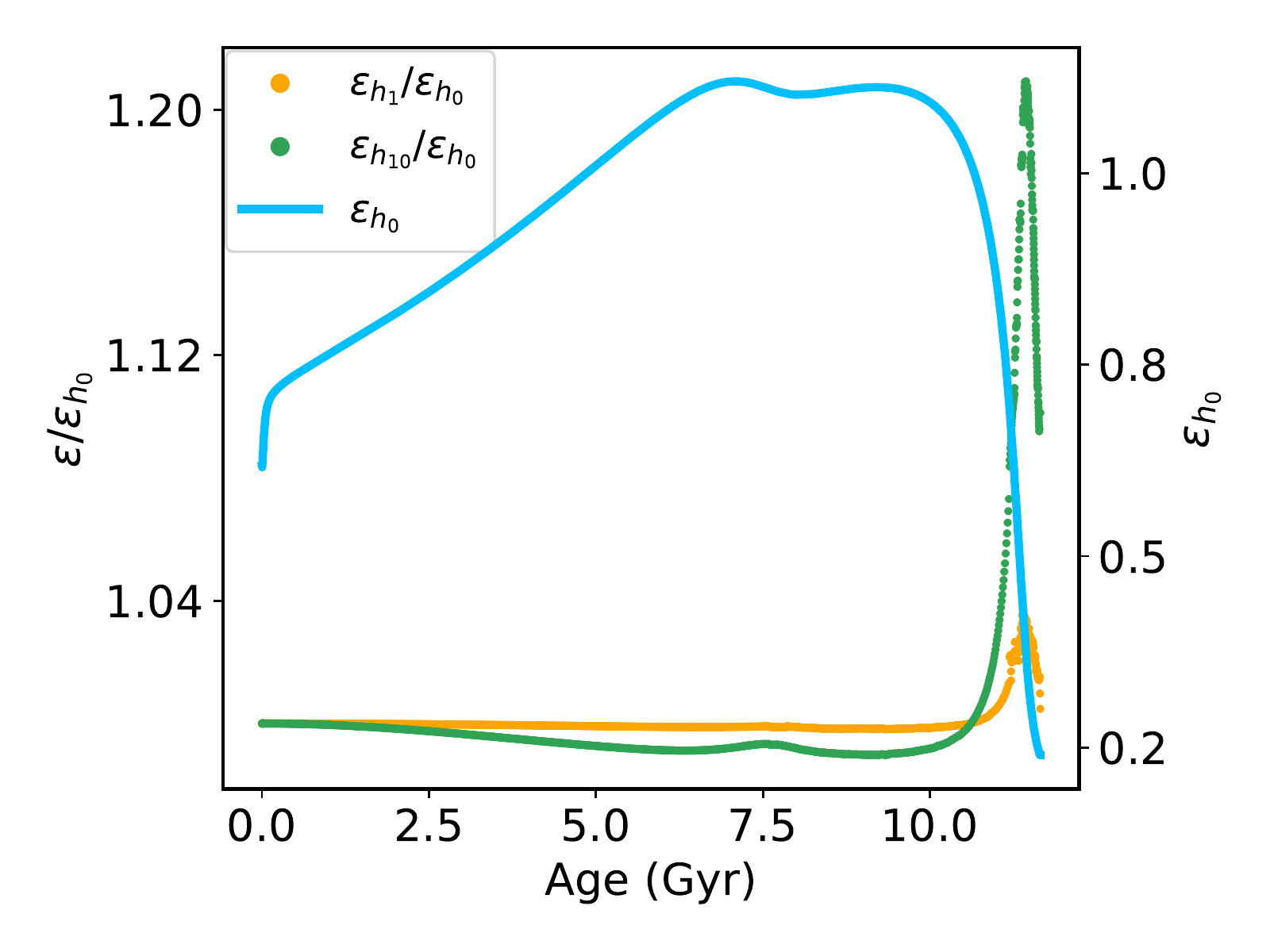}
\includegraphics[width=0.49\columnwidth]{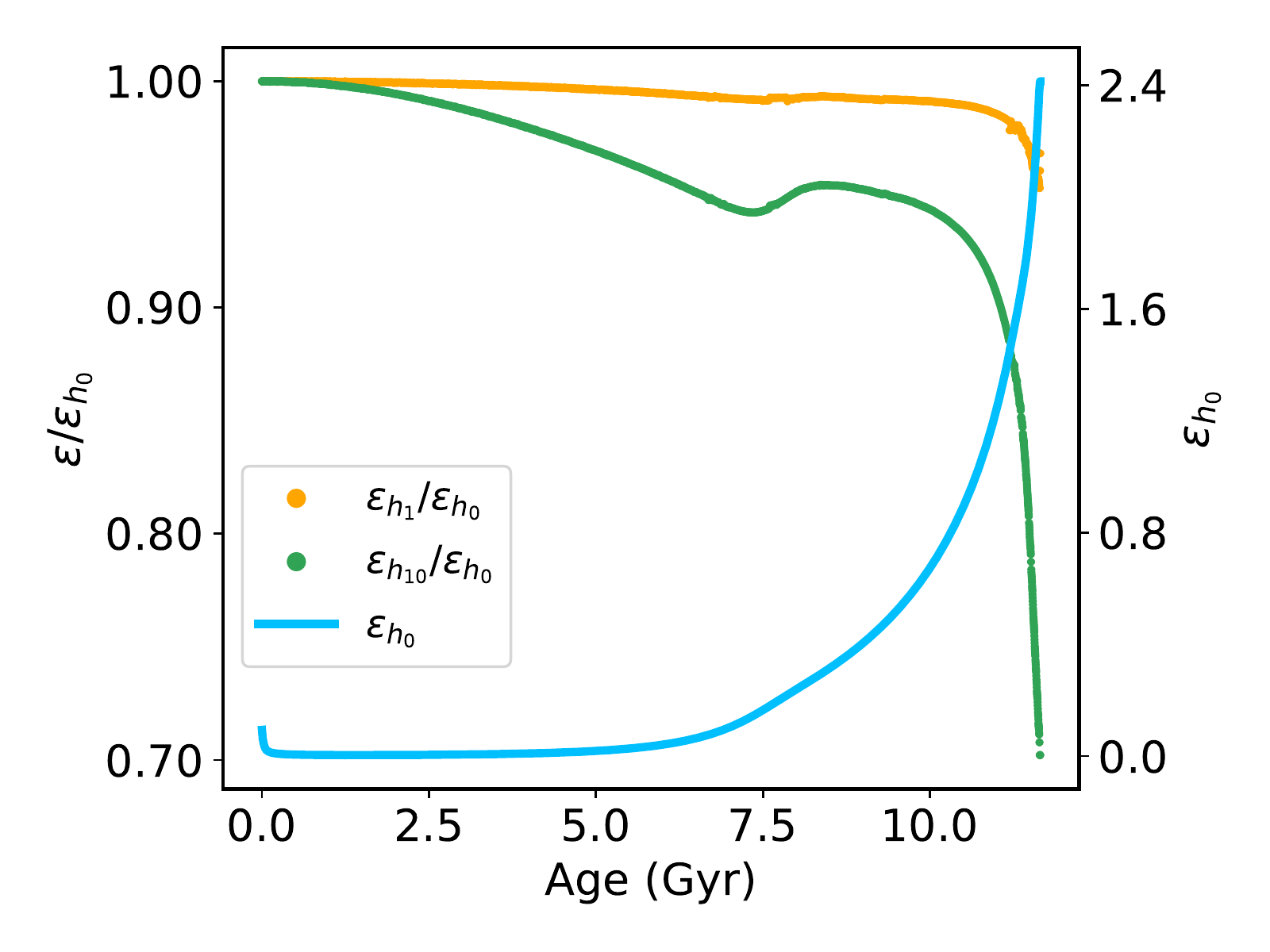}
\caption{Properties of the MESA models of set 3, in which the heating profile is introduced 
gradually, starting at the ZAMS and reaching maximum at the TAMS. {\it Top}: Stellar radius (left) and luminosity (right). The blue curve corresponds to the nonheated models.  The orange and green curves correspond, respectively, to the heating profile of model 63 and a heating profile that is ten times larger. {\it Bottom}: Proton-proton reaction rate power (left) and CNO reaction rate power (right). The blue curve corresponds to the nonheated models and its ordinate is on the right side of the plot and given in units of log(L$_\odot$). The orange and green curves correspond to the ratio of rates in the heated and nonheated models, with colors the same as in the top panels.}
\label{fig_MESA_Set3}
\end{figure}

An overview of the results for gradual time-increasing heat injection is illustrated in Fig.~\ref{fig_MESA_Set3}. In contrast to the set 2 models shown in Fig.~\ref{fig_MESA_Set2}, there is no significant difference throughout the MS between the heated and the unperturbed models; this is as expected since the injection of heat is very small until the TAMS is approached.  However, at the TAMS, the set 3 models very rapidly approach the properties of the set 2 models at similar ages. 

\begin{figure}
\centering
\includegraphics[width=0.495\columnwidth]{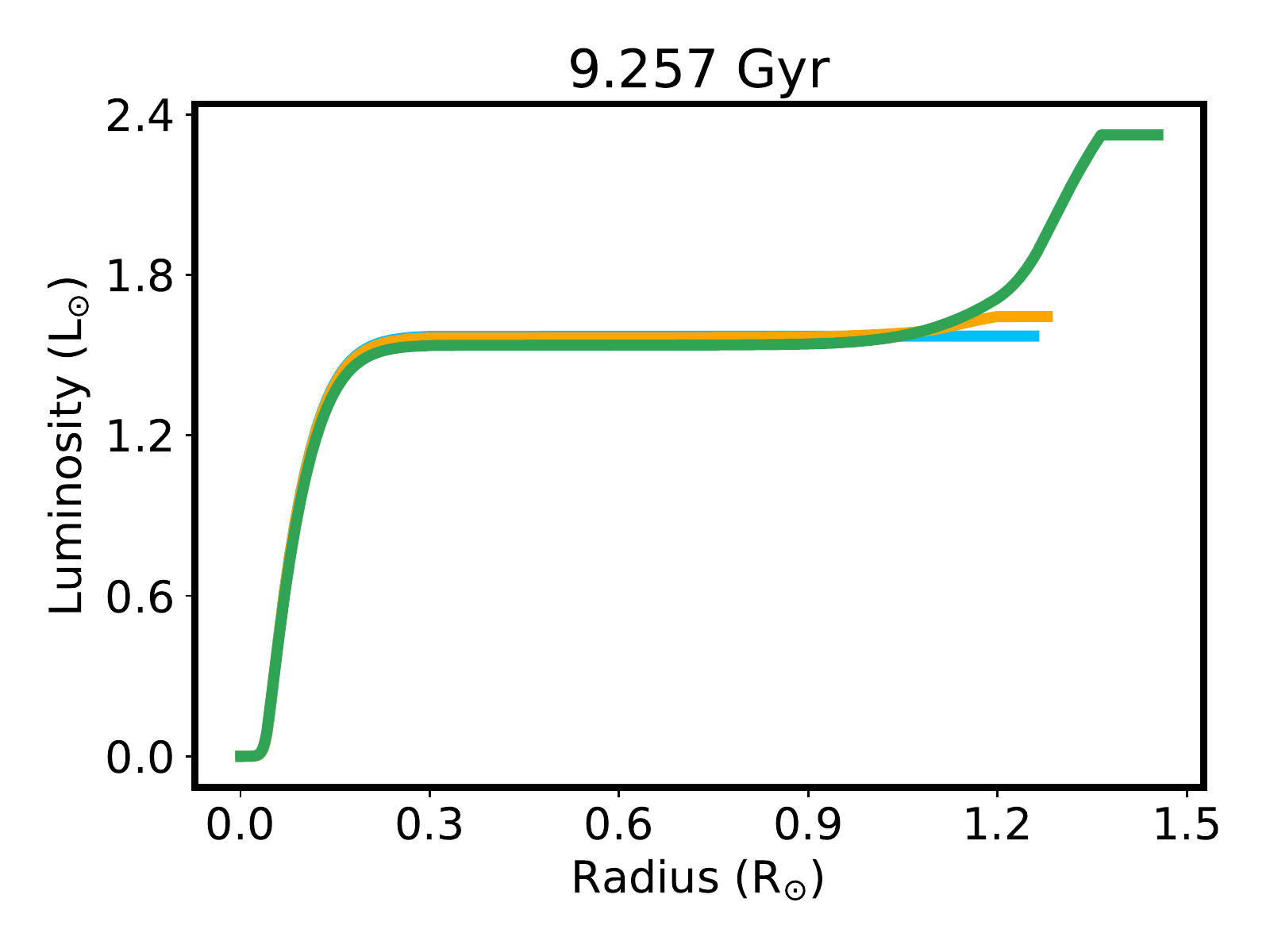}
\includegraphics[width=0.495\columnwidth]{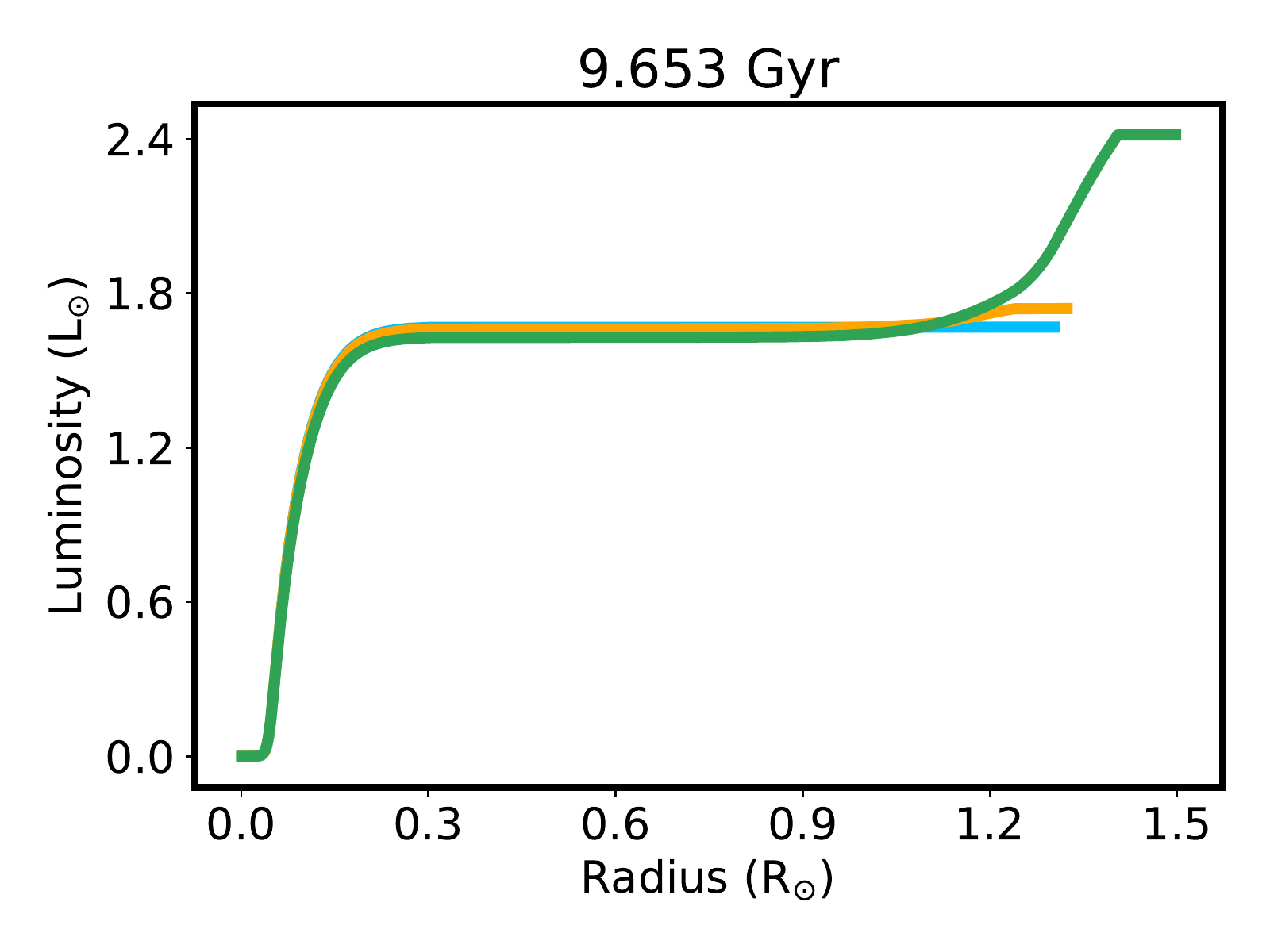}
\includegraphics[width=0.495\columnwidth]{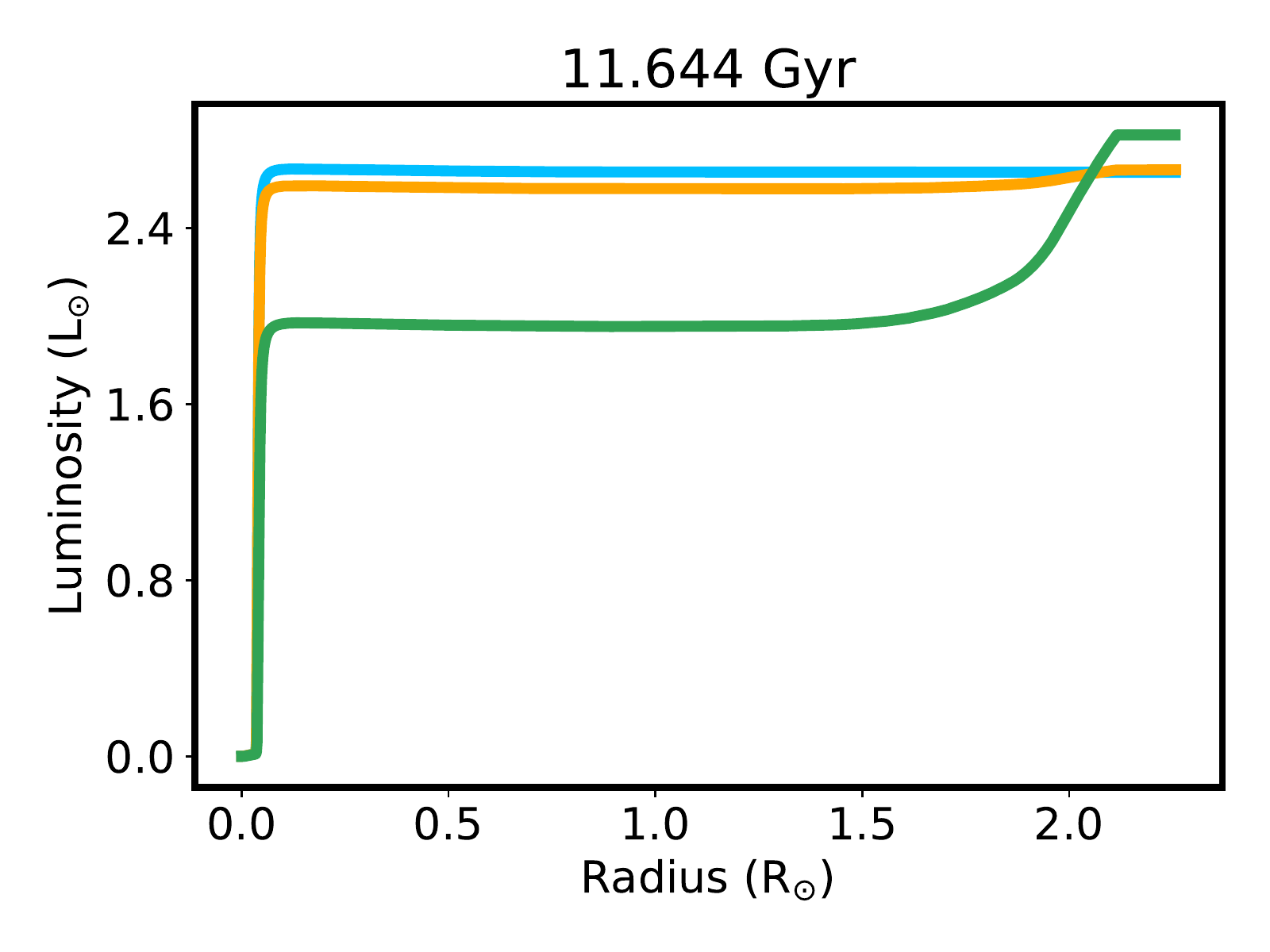}
\caption{Luminosity structure of MESA set 3 models at ages 9.257, 9.653, and 11.644 Gyr.
Blue curves correspond to the models without added heating; red to those with heating as given
by model 63; green by heating ten times that given by model 63.  The tidally heated models have a surface luminosity that is larger than the nonheated models, but lower luminosity in deeper layers.
}
\label{fig_MESA_luminosity_Set3}
\end{figure}

\begin{figure}
\centering
\includegraphics[width=0.98\columnwidth]{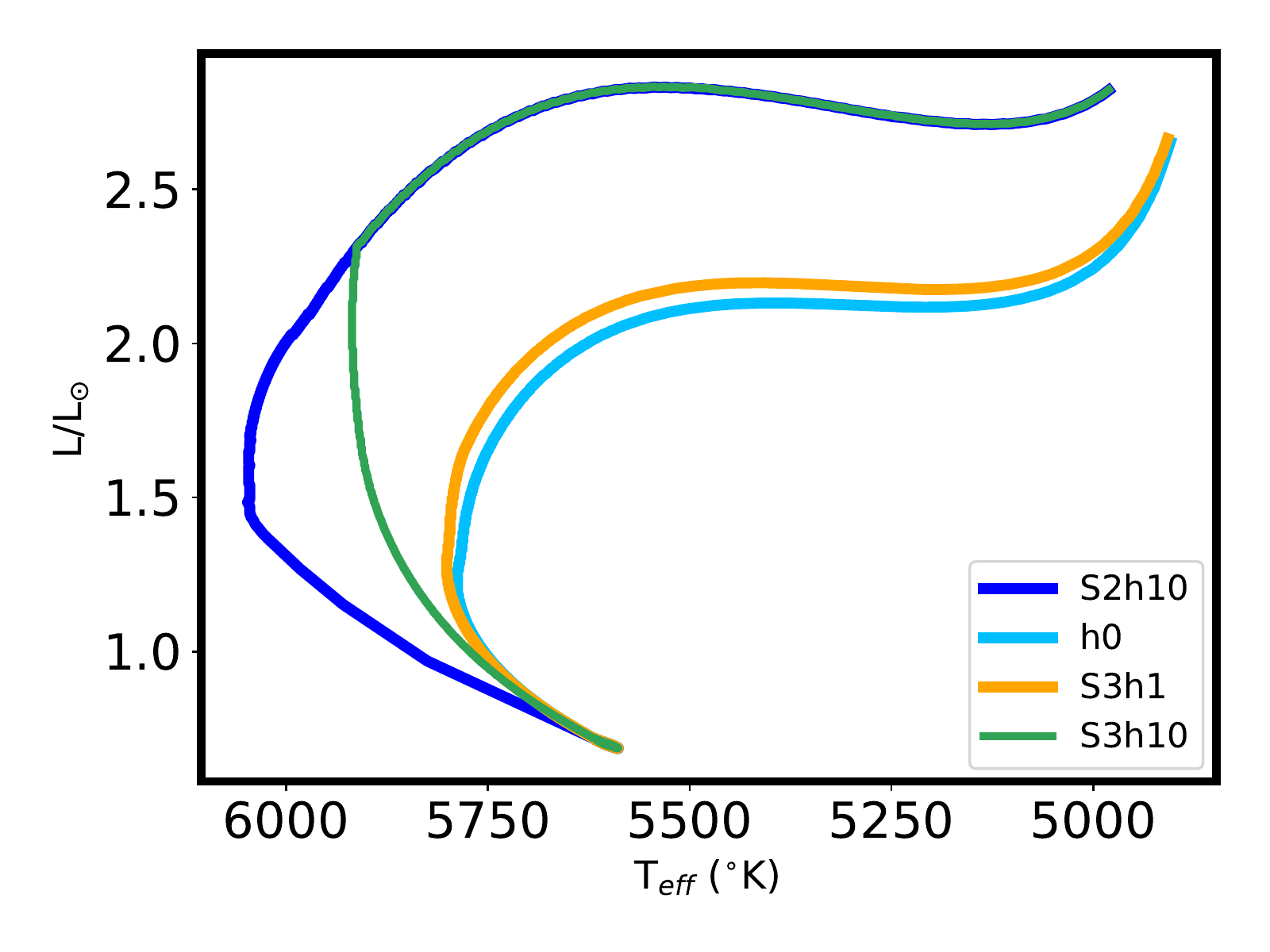}
\caption{Evolutionary tracks on the HRD of the standard model (light blue), the set3-h1  model (orange), and the set3-h10 model (green) showing the result of introducing a gradual heating near the end of the main sequence. The dark blue curve corresponds to the set2-h10  model in which there was constant heating throughout the main sequence.
}
\label{fig_MESA_evolution_Set3}
\end{figure}

The internal luminosity structure of set 3 models at ages 9.257, 9.653, and 11.644 Gyr is illustrated in Fig.~\ref{fig_MESA_luminosity_Set3}, which shows similar characteristics as those of set 2. Specifically, the heated models have a lower internal luminosity than the standard model and a significantly larger surface luminosity.
The similarity between the set 2 and set 3 heated models is best exemplified with a Hertzsprung-Russell Diagram (HRD), as shown in Fig.~\ref{fig_MESA_evolution_Set3}, where we see that the set2-h10 and the set3-h10 tracks are identical after the MS turnoff, which is when both sets are being equally heated.  

\section{Discussion}

In this paper, we explore the manner in which  extra energy that is injected into the external stellar layers affects its structure and its observational properties. The source of this extra energy is here assumed to be that which is released by the action of shearing layers whose motions are driven by the tidal interaction with a close companion.  We adopted a simple prescription for the turbulent viscosity, $\nu_\mathrm{turb} =\lambda \ell_\mathrm{t} \Delta u_\mathrm{t} $, where $\Delta u$ is the instantaneous velocity difference between two contiguous layers, $\ell_\mathrm{t}$ is a characteristic distance   taken to be the separation of the layers, and $\lambda$ is a proportionality factor.  The instantaneous values of $\Delta u$ are computed using the TIDES code, which solves the equations of motion in the rotating reference frame of the binary, taking into account the gravitational, centrifugal, Coriolis, gas pressure, and viscous accelerations.  The stellar structure and evolution is computed with MESA.

The nominal calculation performed to probe the effects of tidal shear energy dissipation on the stellar structure is for a 1\,M$_\odot$ star with an initial radius of 0.97\,R$_\odot$ in a 1.44\,d orbit and a 0.8\,M$_\odot$ companion. The TIDES computational grid consists of $\sim$14300 volume elements (one hemisphere,  north-south symmetry is assumed) covering the outer $\sim$54\% in radius of the star.  The synchronicity parameter $\beta_0$=1.05 is chosen such that the core of the star rotates at a rate that is 5\% faster than co-rotation.  Additional models are computed for stellar radii up to 2.25\,R$_\odot$, polytropic indices 1.5$\leq n \leq$3.8, and synchronicity parameters $0.20 \leq \beta_0 \leq 1.1$.


\subsection{Viscosity and tidal shear energy dissipation}

For the $\sim$1\,R$_\odot$ models, we find maximum turbulent viscosity values near the stellar surface at the equator in the range 5$\times$10$^{-4}$ to 10$^{-3}$\,R$_\odot^2/$d in calculations with  $\lambda$=1. These values correspond to 3--6$\times$10$^{13}$\,cm$^2$ s$^{-1}$, respectively.   The corresponding total energy dissipation rates integrated over the entire star $\dot{E}_\mathrm{tot}$ are in the range 2--20$\times$10$^{30}$\,erg\,s$^{-1}$.  These values decrease in proportion to the value of $\lambda$. 

As a star evolves and becomes larger,  the turbulent viscosity grows and, assuming that the density structure does not significantly change,  the energy dissipation rate also increases.
The maximum value of $\nu_\mathrm{turb}$ obtained with $\lambda$=1  in this paper is not very different from the values used in Koenigsberger \& Moreno (2016, henceforth KM2016), except for the largest radii listed in block 1 of Table 2, where we see a difference of a factor of $\sim$3.  The difference arises because the value of $\nu$ in the KM2016 is an input parameter which, broadly speaking, is unconstrained. To avoid using a completely arbitrary $\nu$ value, it was estimated assuming it to scale with the characteristic spatial and velocity dimensions. This was not a very precise estimate.  In our current formulation, $\nu$ is computed internally, allowing for a more consistent estimate.


Our current $\dot{E}_\mathrm{tot}$ values are significantly smaller  compared to those listed in KM2016 (their Table 3). For example, they list $\dot{E}_{n=1.5}\sim$10$^{34}$ ergs/s for the $R_1$=0.99 R$_\odot$ model, while we obtain $\dot{E}$=2$\times$10$^{30}$ ergs/s for our model 4.  The dominant source of this difference is that KM2016 assumes a $n$=1.5 polytropic index, while all the models in  block 1 of our Table 2 were computed with $n$=3.  The density at a radius of 0.965\,R$_\odot$ is $\sim$3 orders of magnitude higher for $n$=1.5 than for $n$=3. We performed an analogous one-layer computation to that of KM2016 using their same model, but this time with $n$=3 and obtained $\dot{E}_{n=3}$=6$\times$10$^{30}$ ergs/s, which is only a factor $\sim$3 larger than what we obtain in the current calculations for model 4. This remaining difference is due to a combination of  factors. The first is that in the KM2016 model, the viscosity is constant over the entire surface layer, while in our current model its value significantly decreases at various locations along the azimuthal coordinate and, especially, in the polar direction. The second is that the KM2016 model includes radial motions which, although they are approximately ten times smaller than the azimuthal motions, also contribute toward the energy dissipation rates.

The lower energy dissipation rates that we find in this paper could impact the interpretation of the V1309 Sco  merger phenomenon in terms of a tidally-induced runaway process, as discussed in KM2016.  However, the higher energy dissipation rates needed for a rapid orbital evolution timescale can still be obtained in our current model by increasing the departure from synchronicity. While it is beyond the scope of this paper,  a re-analysis using our current model, but relaxing the conditions on $\beta_0$ and the polytropic index, is warranted before abandoning the tidal runaway scenario.

\subsection{Implications for stellar structure} 

The  energy dissipation rates computed by TIDES for each layer were injected into MESA stellar structure and evolution calculations.  With even a small ($\sim$5\%) departure from synchronicity,  the radius and luminosity values of the tidally heated stars are larger than those of the equivalent unperturbed model at all evolutionary times.  The star also has a smaller surface convective region and lower nuclear processing rates, the latter allowing the tidally perturbed star to live longer. The differences  between an asynchronous binary and its unperturbed counterpart depend  on the amount of injected energy which, for a fixed set of stellar and orbital parameters, depends on how much the stellar rotation departs from synchronicity and the value of turbulent viscosity.

From an observational perspective, determining whether a star is truly in synchronous rotation is a challenging problem as the only available information is the projected surface equatorial speed.  Because the synchronization time scales as the radius of the layer and its viscosity, $\tau_\mathrm{visc} \sim r^2$/$\nu_\mathrm{turb}$, the star tends to synchronize from the surface inward \citep{1989ApJ...342.1079G, 2021A&A...653A.127K}. Thus, stars in circular orbits that are thought to be synchronized may actually retain an internal angular velocity gradient upon which tidally excited oscillations are superposed.   Furthermore, all eccentric binary systems are asynchronously rotating during most of their orbital trajectory; hence, they suffer from tidal perturbations regardless of their age.  This would be particularly true of recently-formed binaries in dense regions of stellar clusters where close encounters of single stars are believed to frequently occur.

In this context, it is interesting to note that radii of low-mass binary stars in short-period orbits have been determined to be as much as  10\% larger than their counterparts in long-period orbits \citep{1973A&A....26..437H,1977ApJS...34..479L,1997AJ....114.1195P,1999ASPC..173..265C,2006ApJ...640.1018T},   consistent with the radius increase that we find in the tidally heated stars explored in this paper. 
Many low-mass stars are associated with significant surface activity, as evidenced by their light curves and the emission cores of \ion{Ca}{II} H and K lines \citep{2006ApJ...640.1018T,2007ApJ...660..732L,2008A&A...478..507M}.  This activity is  explained in terms of the presence of strong surface magnetic fields and  these fields are thought to inhibit efficient convection \citep{2006ApJ...640.1018T,2007A&A...472L..17C,2009A&A...502..253C,2016A&A...593A..99F}. Tidal flows and magnetic fields are not mutually exclusive, but the manner in which these two physical processes might interact is an open question.

Our scenario for tidal shear energy dissipation may also have a bearing on the mass discrepancy problem  in massive stars that has been known for several decades, but that has still eluded explanation.  The discrepancy, first noted by \citet{1992LNP...401...21H}, consists of the fact that the masses derived from spectroscopic analysis are systematically lower than those found from evolutionary models; alternatively, they are more luminous than predicted by the evolutionary models. In their analysis of a set of eclipsing binary stars, \citet{2012ApJ...748...96M} found them to be on average 11\% less massive or conversely, 0.2 dex more luminous, as compared to stellar structure models.  Because it is now generally accepted that a large majority of massive stars are in binary systems, it is tempting to suggest that a possible solution to the mass-discrepancy problem may reside in the phenomena we discuss in this paper.

\subsection{Implications for stellar evolution and population synthesis}

The effects of binary interactions on stellar evolution have,  until now, focused mainly on the effects of mass-loss and mass transfer between the components in late evolutionary stages.  In the case of massive stars, binary interactions during the main sequence  have been incorporated only indirectly in the sense that tides are invoked to maintain short-period binary stars in rapid rotation, allowing them to be treated as  rapid rotators throughout their main sequence lifetime.  The possible modification in the stellar structure that results from the tidal interactions, however, is generally neglected. These effects  include tidal shear energy dissipation and turbulent viscosity, which, in turn, have an impact on  the internal energy budget as well as  the rates of angular momentum and chemical transport.  

The presence of tidal perturbations does not depend on the age of a star, but it may be most easily detected in very close binaries or those in which the radius of one of the stars is significant compared to the orbital separation -- specifically, stars at the end of the main sequence.   We find that  evolutionary tracks of our tidally heated  stars extend further to the blue during the end stages of the main sequence than does the track for the standard model.  This effect is reminiscent of the extended main sequence turnoff (eMSTO) phenomenon that was first detected by \citet{2003AJ....125..770B}  and \citet{2007MNRAS.379..151M}  in the Large Magellanic Cloud clusters NGC\,2173 and NGC\,1846; this is now considered an ubiquitous feature of  Magellanic Cloud Clusters with ages between $\sim$20 Myr and $\sim$2 Gyr \citep{2018MNRAS.477.2640M,2014ApJ...797...35G}. The eMSTO has been interpreted as the result of a prolonged star formation \citep{2008ApJ...681L..17M,2008AJ....136.1703G,2011ApJ...737....4G,2011ApJ...731...22K} or as a result of stellar rotation \citep{2009MNRAS.398L..11B}.  We suggest that tidal heating may be an additional potential explanation. In globular clusters, the presence of  blue straggler stars not showing evidence of mass-transfer \citep{2006ApJ...647L..53F} could also potentially be a manifestation of the effectiveness of this process.  

Another interesting application of our model refers to the stability of a binary star's outer layers as it leaves the main sequence and heads up the giant branch.  As the stellar radius increases, so do the tidal amplitudes (Section~\ref{sec:plots_block1}).  We can speculate that the growing surface velocities could attain the escape velocity  before  the Roche Lobe radius is reached.  Because the tidal amplitudes are largest around the equator,  this is where the star would become most bloated and where mass loss might be expected to  occur first and take the form of an excretion disk.  Observational evidence for asymmetrical mass-loss episodes is found in planetary nebulae, many of which display bipolar morphologies  \citep{2001ApJ...557..256S,2009PASP..121..316D,2011PhDT........15J,2017NatAs...1E.117J,2019ibfe.book.....B}, as well as the presence of structures such as jets, rings, and halos \citep{1997ApJ...487..809H,2002AAS...201.2206K,2009MNRAS.399.1126P}. Therefore, taking into account the perturbations caused by tidal forces on the progenitor star during its expansion stages may contribute to improving our understanding of the processes that give rise to the wide variety of morphologies of such interstellar structures. Finally, we note that although our focus  is the potential effects on the stellar structure and evolution due to tidal perturbations, any nonstandard process that can heat sub-surface stellar layers would produce the same effects we describe in this paper.

\subsection{Caveats}
 
There are several important caveats to our results.  The first is that our simplified prescription for the turbulent viscosity depends on a $\lambda$ parameter, which can be associated with the fraction of kinetic energy in the shearing flows that is transformed into turbulent eddies. Our calculations were performed for $\lambda$=0.1 and 1, and we find that the values of $\dot{E}$ scale approximately linearly with $\lambda$. Significant differences between the heated and the standard MESA models appear mainly for the larger $\lambda$ value, which is likely to be unrealistic.  However,  $\dot{E}$ values large enough  to produce significant differences can also be obtained by increasing the synchronicity parameter, as we showed for the cases with $\beta_0$=1.05 and 1.1, or decreasing it as illustrated for $\beta_0$=0.2.  This means that even if $\lambda$ is small, stars having significant departures from synchronous rotation could be observed to display the tidal heating effects that we have described. 

An underlying assumption in our treatment is that the criteria for triggering shear instabilities are met, an issue that depends on the hydrodynamical properties and the microphysics of the fluid \citep[][and references therein]{2016ApJ...821...49G} processes that our approach cannot compute. In the case of solar-type stars,  turbulent viscosity is associated with the convective eddies, so in principle, the criteria for triggering the shear instability are met.   However, the nature of the interaction between convective motions and the tidal flows remains to be resolved \citep{2021MNRAS.503.5789T, 2020ApJ...888L..31V, 2020MNRAS.497.3400D}.

Finally, it is important to keep in mind that  $\dot{E}$  is highly nonisotropic in 3D space, attaining maximum values near the equator and decreasing toward the poles, as well as having a dependence on azimuth. Although the TIDES model captures the 3D tidal perturbation structure, the MESA models are currently only 1D.  The  $\dot{E}$ radial profile that was injected into the h1 and h10 MESA models corresponds  to the total energy dissipation in each layer and, hence, the heating is strongly concentrated around the equator. This would lead to equatorial bloating unless the horizontal energy transport processes are sufficiently efficient to re-distribute the added energy.

\section{Conclusions}

Stars in a binary system can interact in different ways, depending on both their evolutionary stage and their orbital parameters \citep{1997ApJS..112..487S}, and these interactions have an impact on the stellar evolution of the components.  In this paper, we explore the potential role of tidal shear energy dissipation in altering not only the evolutionary path of a star in its post-main sequence stages, but also its internal structure during the main sequence. Tidal heating offers a possible alternative for describing discrepancies between observations and the standard stellar structure models.  Examples include phenomena such as the eMSTO in clusters, bloated or overluminous binary components, age discrepancies, and aspherical mass ejection.  However, establishing the actual role of tidal heating requires incorporating the nonspherically symmetric properties of the tidal perturbations into stellar structure models. It also requires a hydrodynamical approach to determining the turbulent viscosity of the stellar fluid.

\begin{acknowledgements}
We acknowledge support from CONACYT project 252499 and DGAPA/PAPIIT projects IN103619 and IN105723. We thank an anonymous referee for the very insightful and helpful comments.  GK thanks Catherine Pilachowski and Constantine Deliyannis for enlightening discussions.
\end{acknowledgements}

\bibliographystyle{aa} 
\bibliography{TIDES_2022} 

\begin{appendix}

\section{Polytropic structures, angular velocity, and viscosity for different values of TIDES input parameters}  \label{sec:plots_block1}

\begin{figure*}
\centering
\includegraphics[width=0.65\columnwidth]{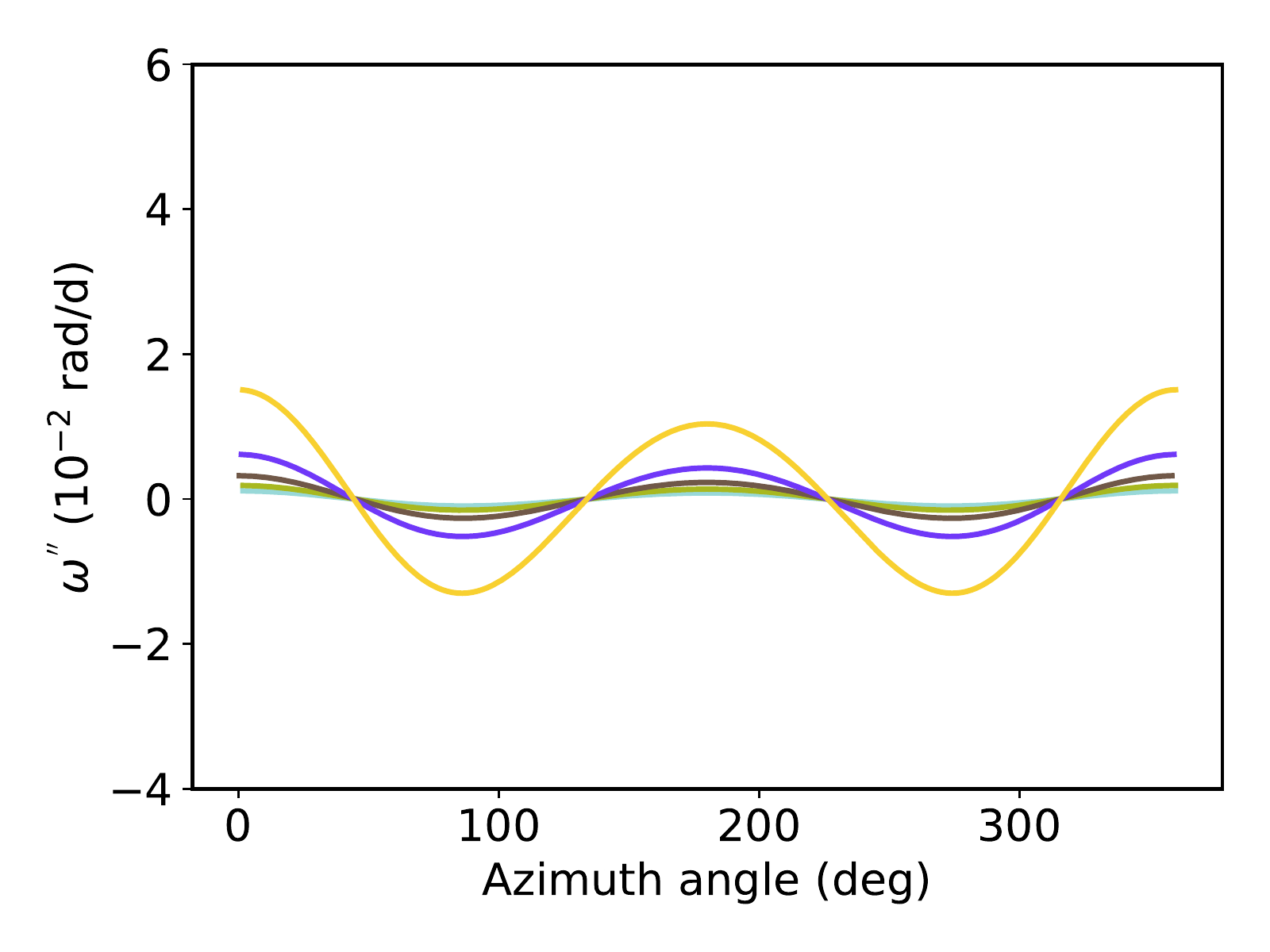}
\includegraphics[width=0.65\columnwidth]{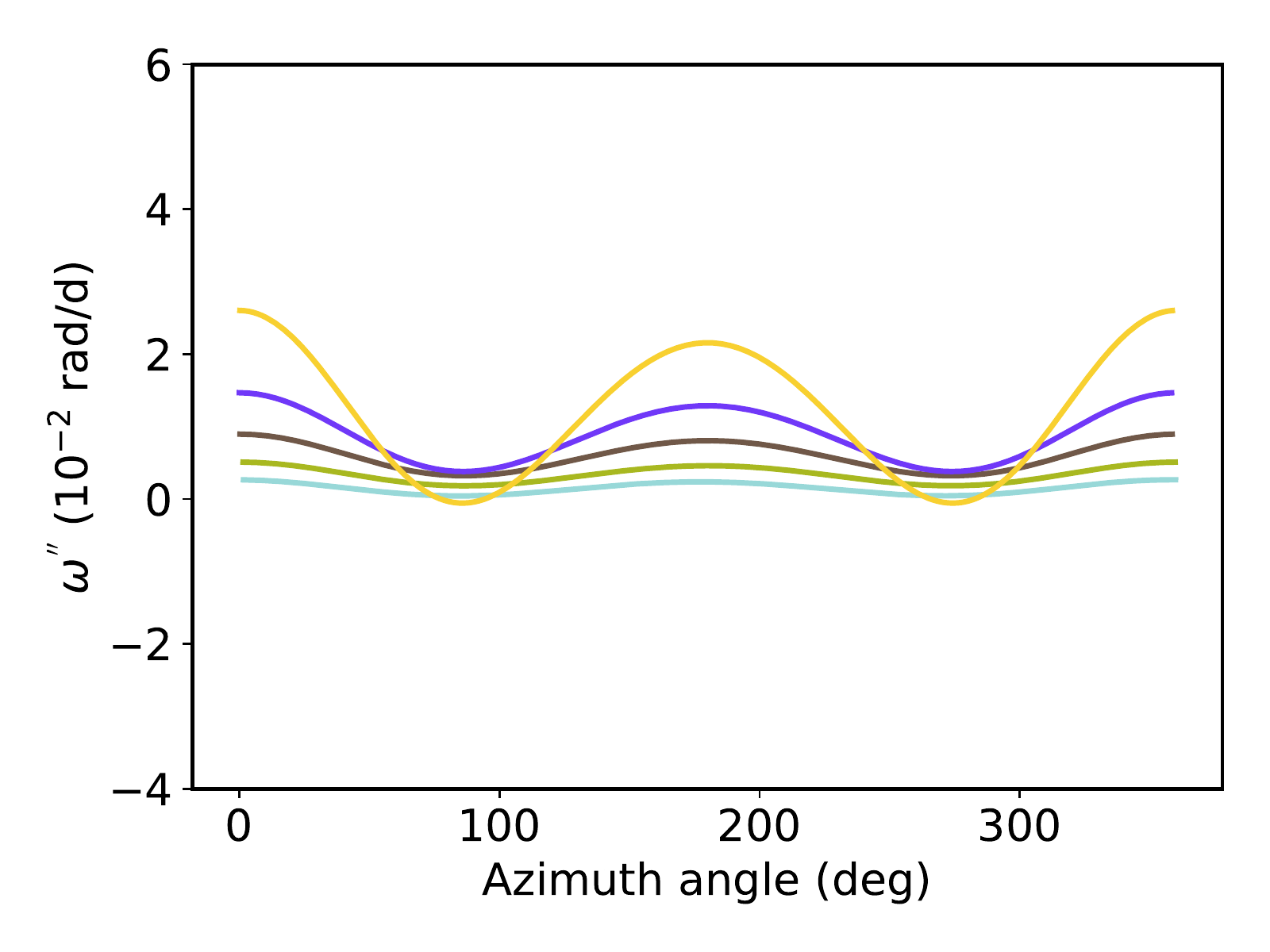}
\includegraphics[width=0.65\columnwidth]{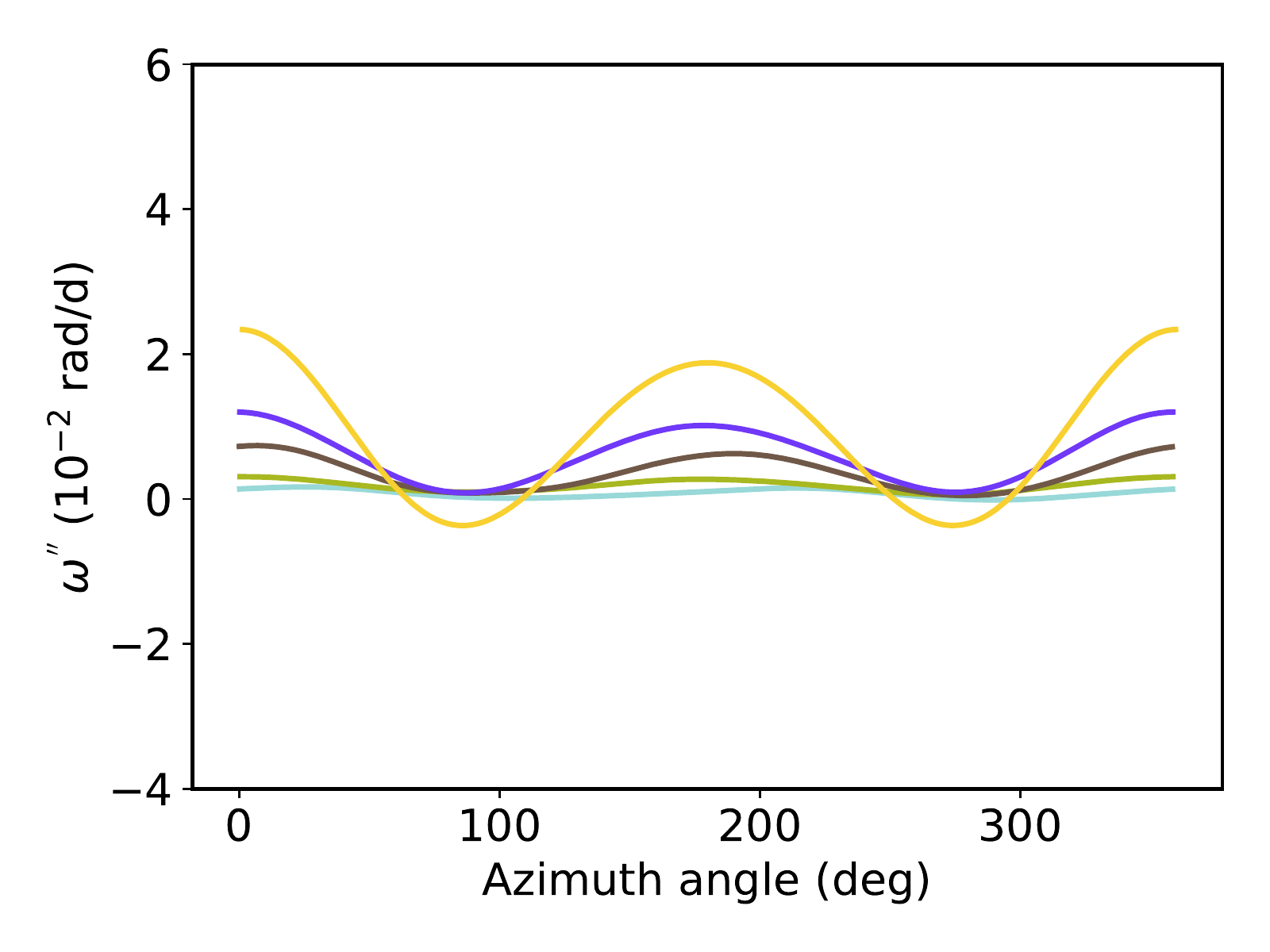}
\caption{Angular velocity at the equator as a function of azimuth angle in five layers 
in the rest frame of the star. Left: Model 2, constant $\nu$; Middle: Model 4, $\lambda$=1; 
Right: Model 6, $\lambda$=0.1.}
\label{fig_models246_omdp}
\end{figure*}

\begin{figure*}
\centering
\includegraphics[width=0.65\columnwidth]{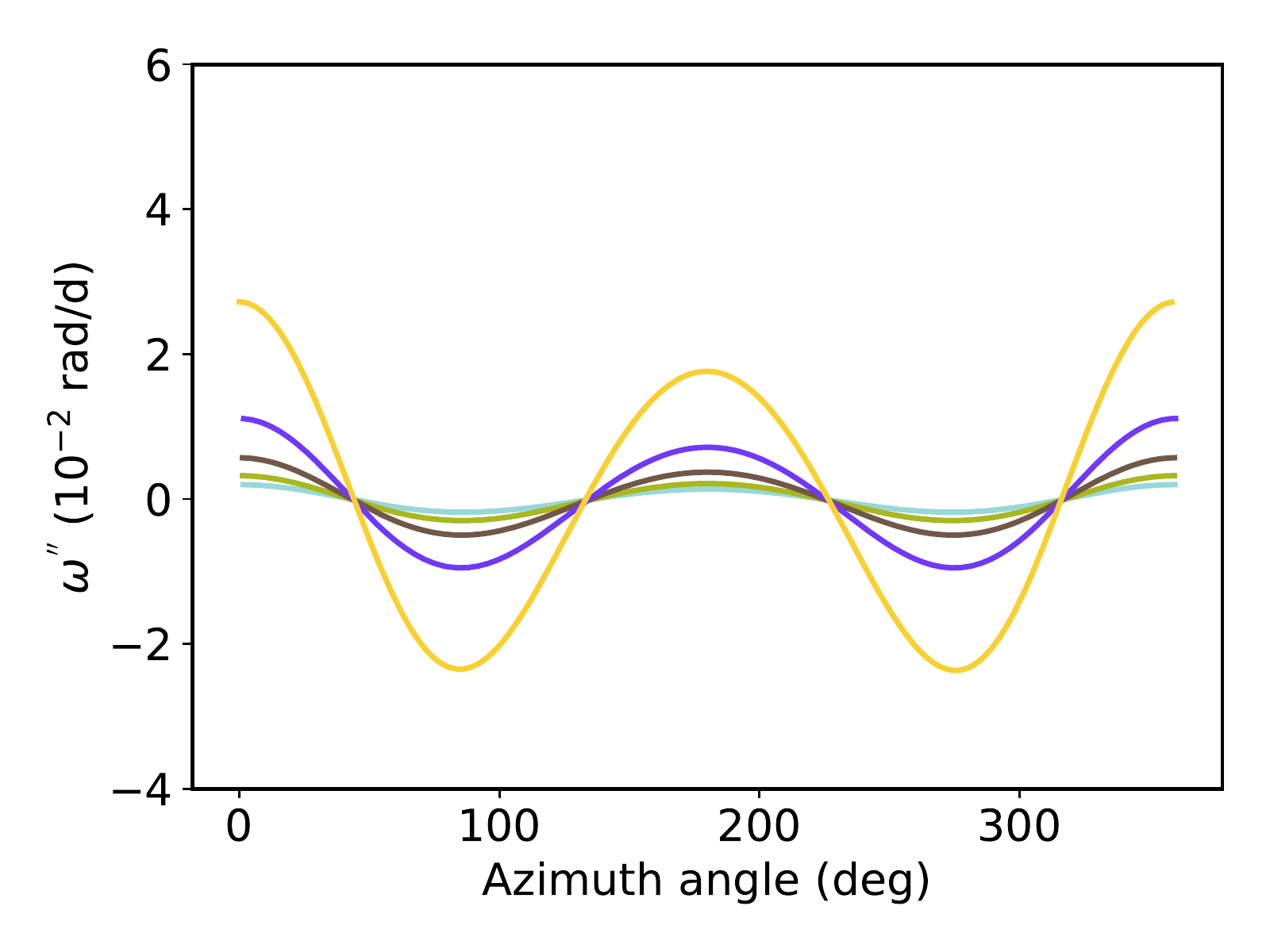}
\includegraphics[width=0.65\columnwidth]{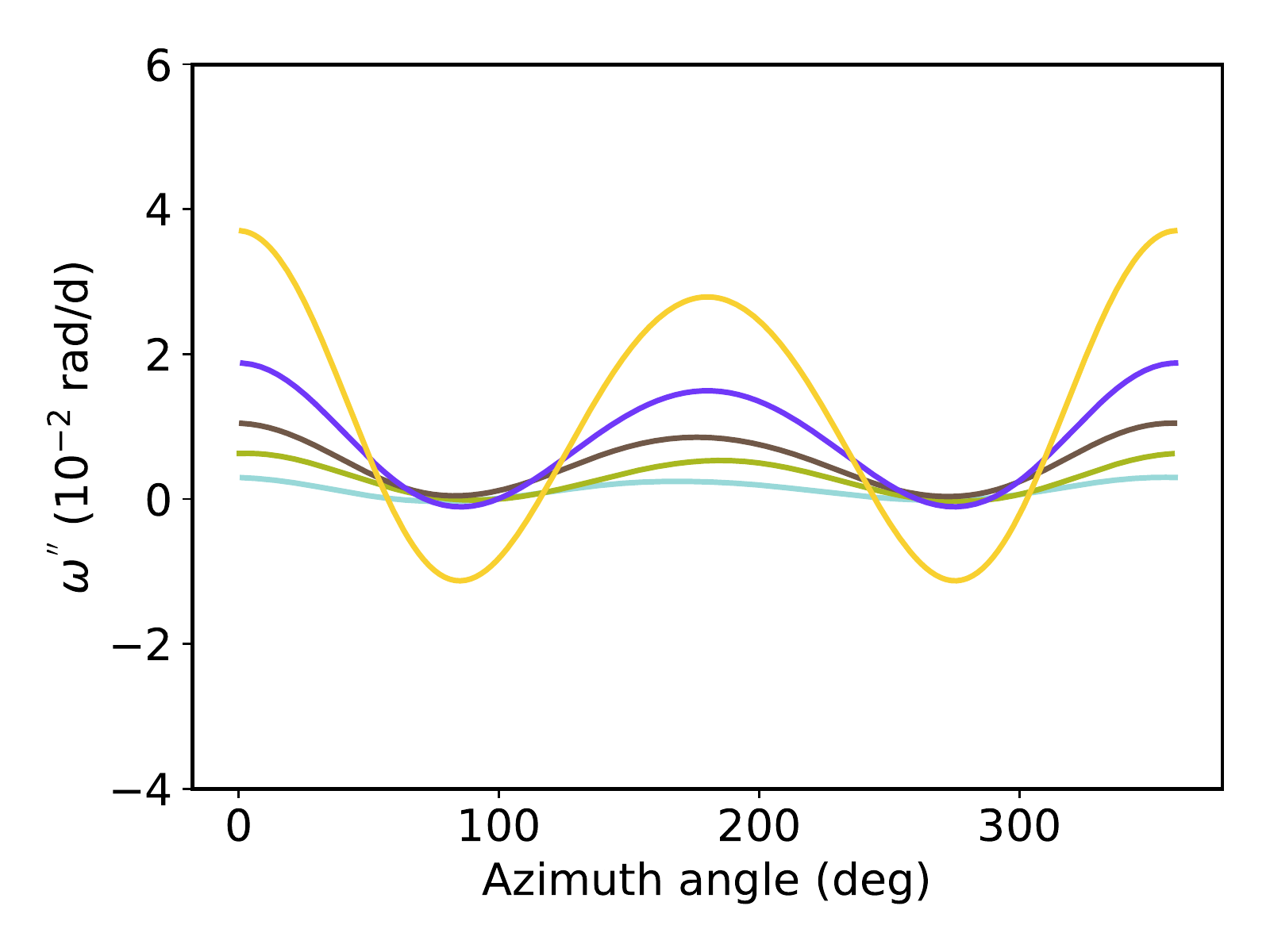}
\includegraphics[width=0.65\columnwidth]{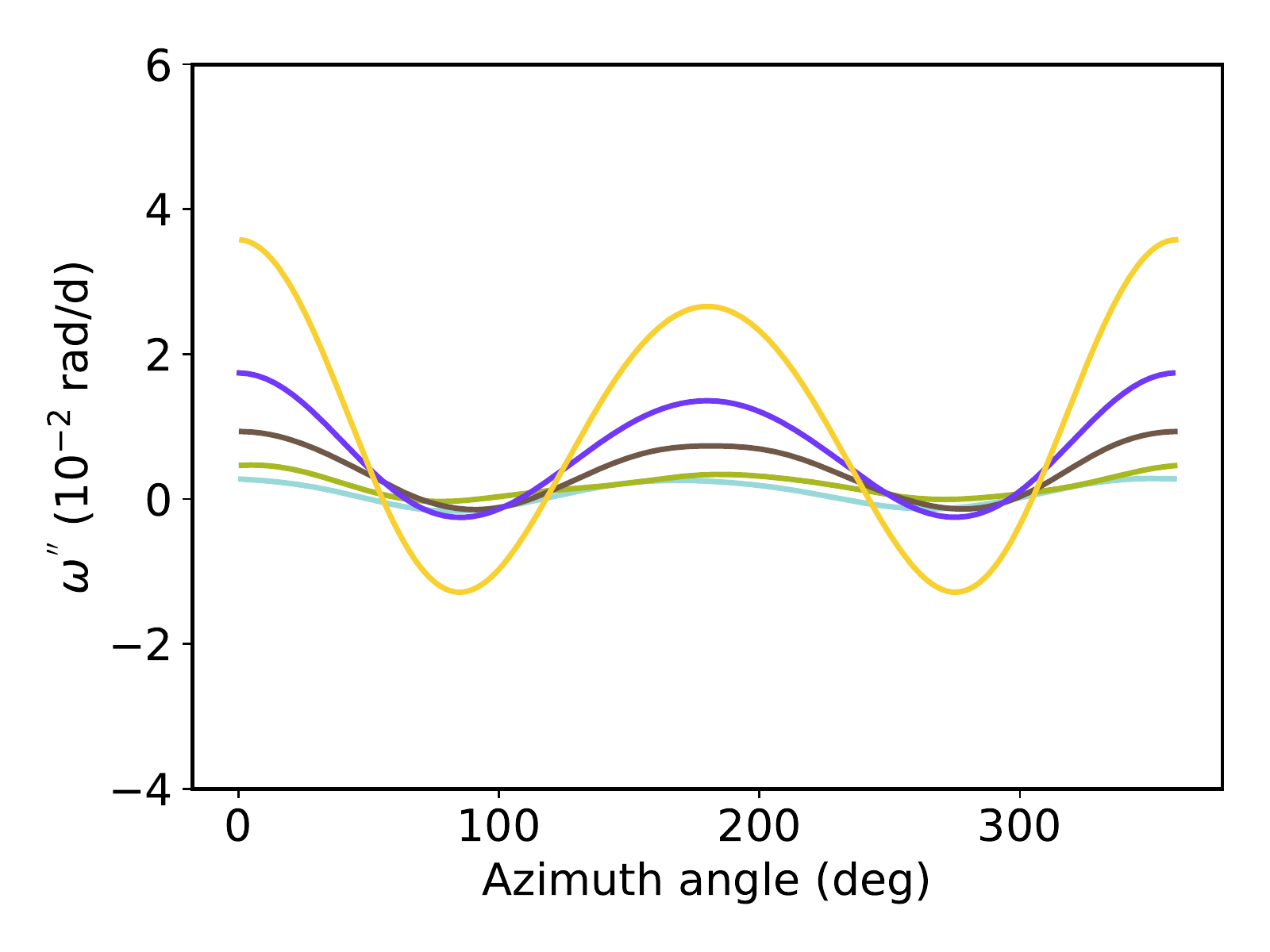}
\caption{Angular velocity at the equator as a function of azimuth angle in five layers
in the rest frame of the star. Left: Model 7, constant $\nu$; Middle: Model 8, $\lambda$=1;           
Right: Model 10, $\lambda$=0.1.
}
\label{fig_models7810_omdp}
\end{figure*}

\begin{figure*}
\centering
\includegraphics[width=0.65\columnwidth]{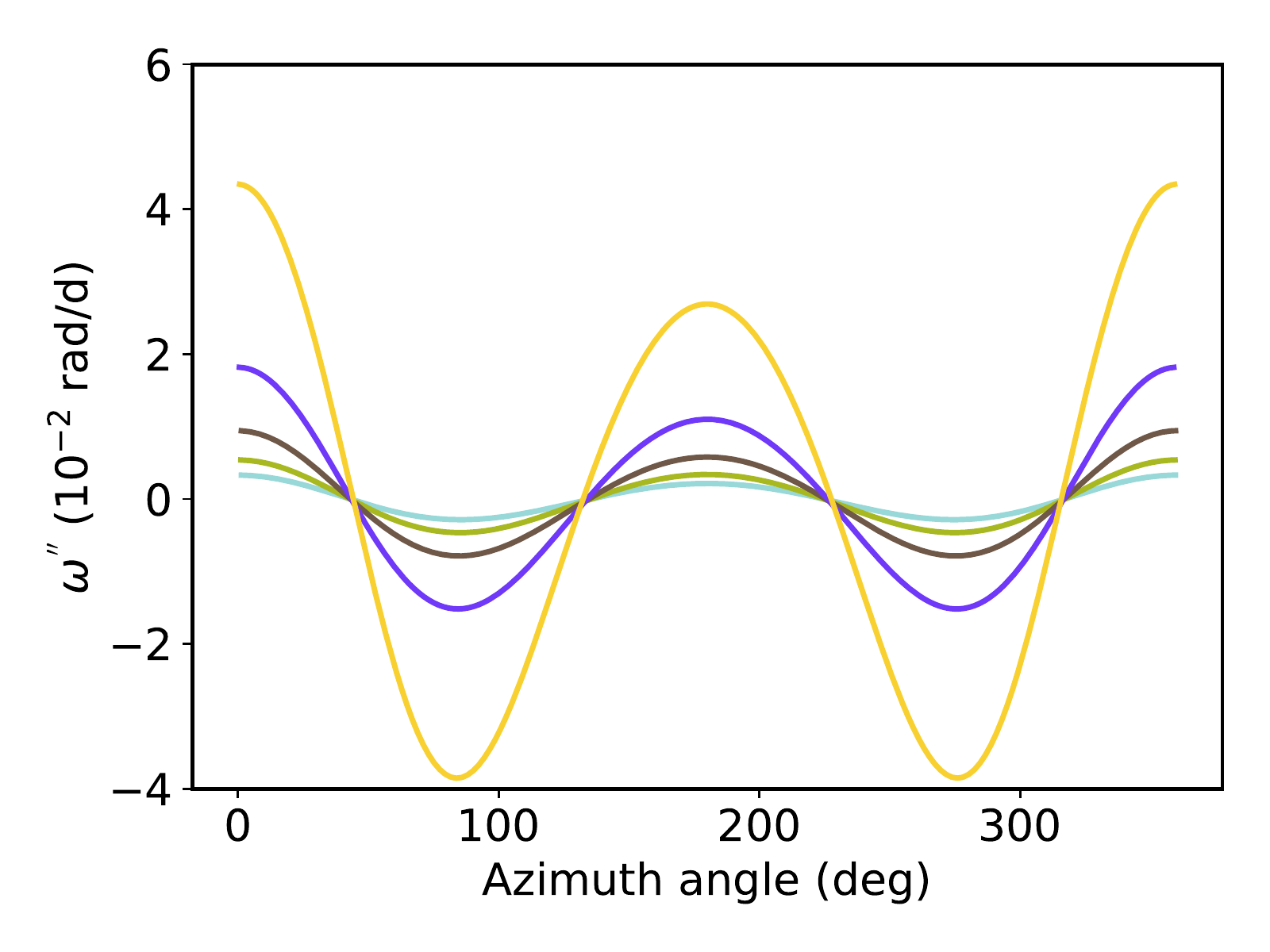}
\includegraphics[width=0.65\columnwidth]{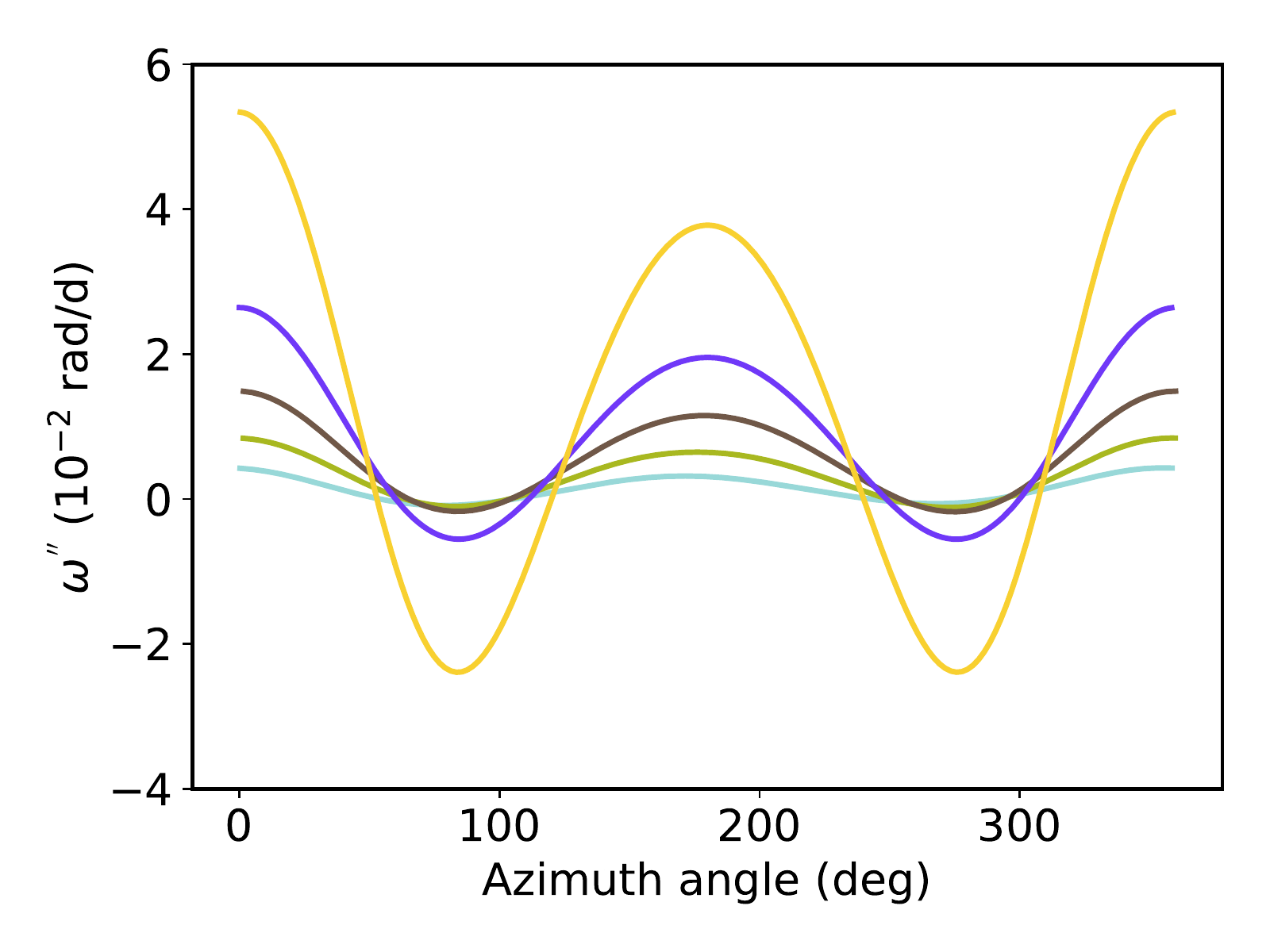}
\includegraphics[width=0.65\columnwidth]{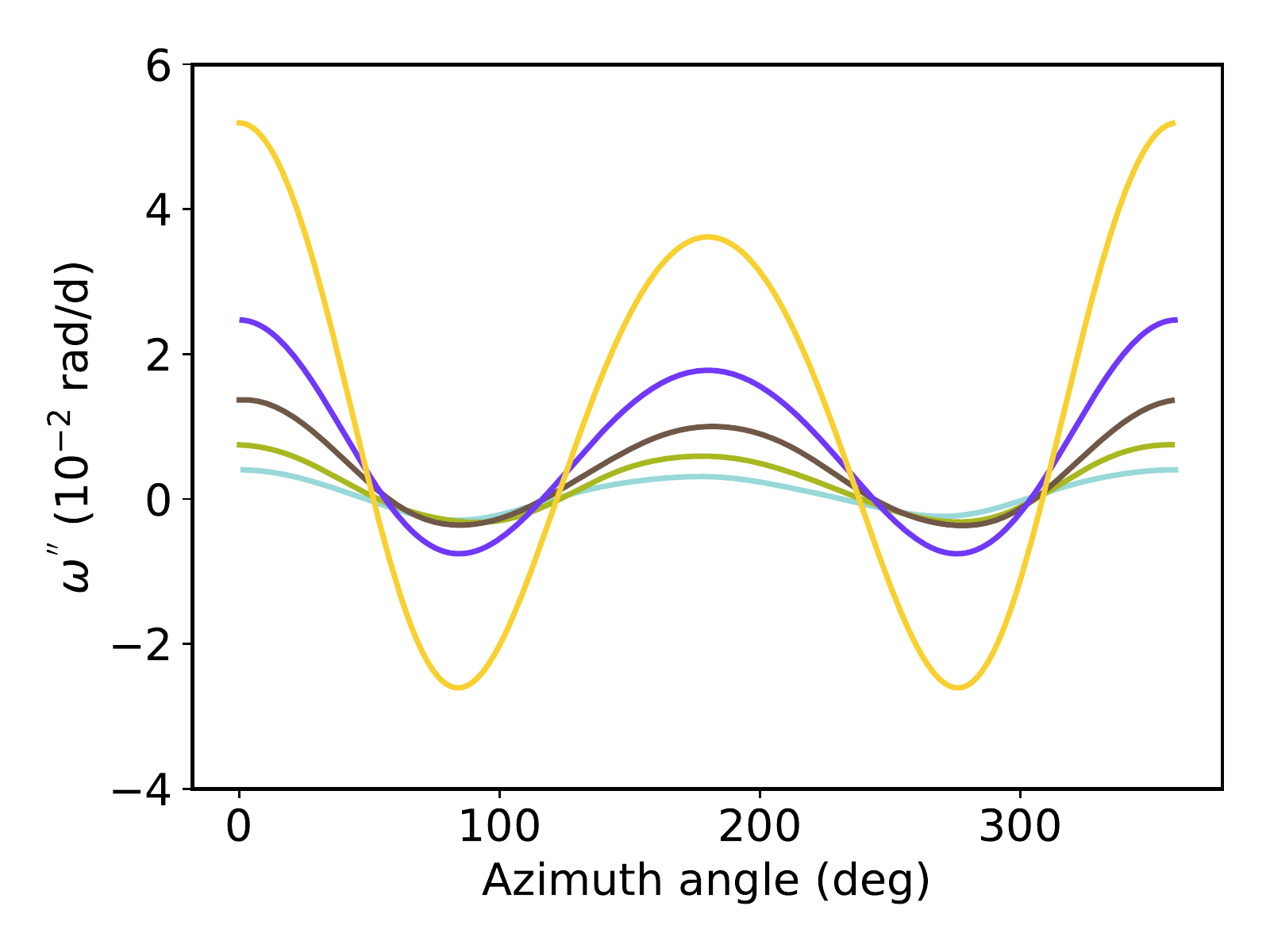}
\caption{Angular velocity at the equator as a function of azimuth angle in five layers
in the rest frame of the star. Left: Model 11, constant $\nu$; Middle: Model 12, $\lambda$=1;
Right: Model 14, $\lambda$=0.1.
}
\label{fig_models111213_omdp}
\end{figure*}

\begin{figure*}
\centering
\includegraphics[width=0.65\columnwidth]{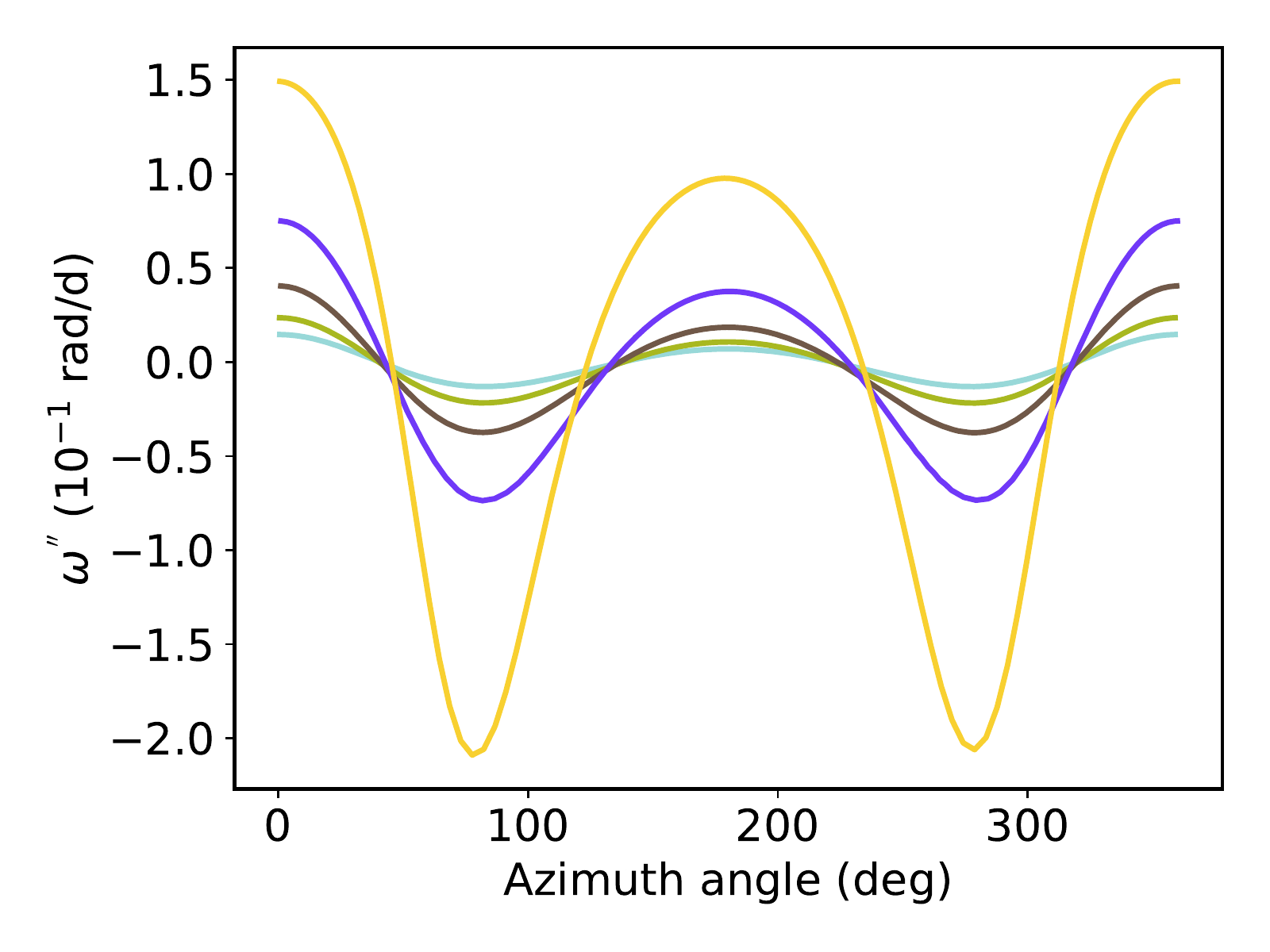}
\includegraphics[width=0.65\columnwidth]{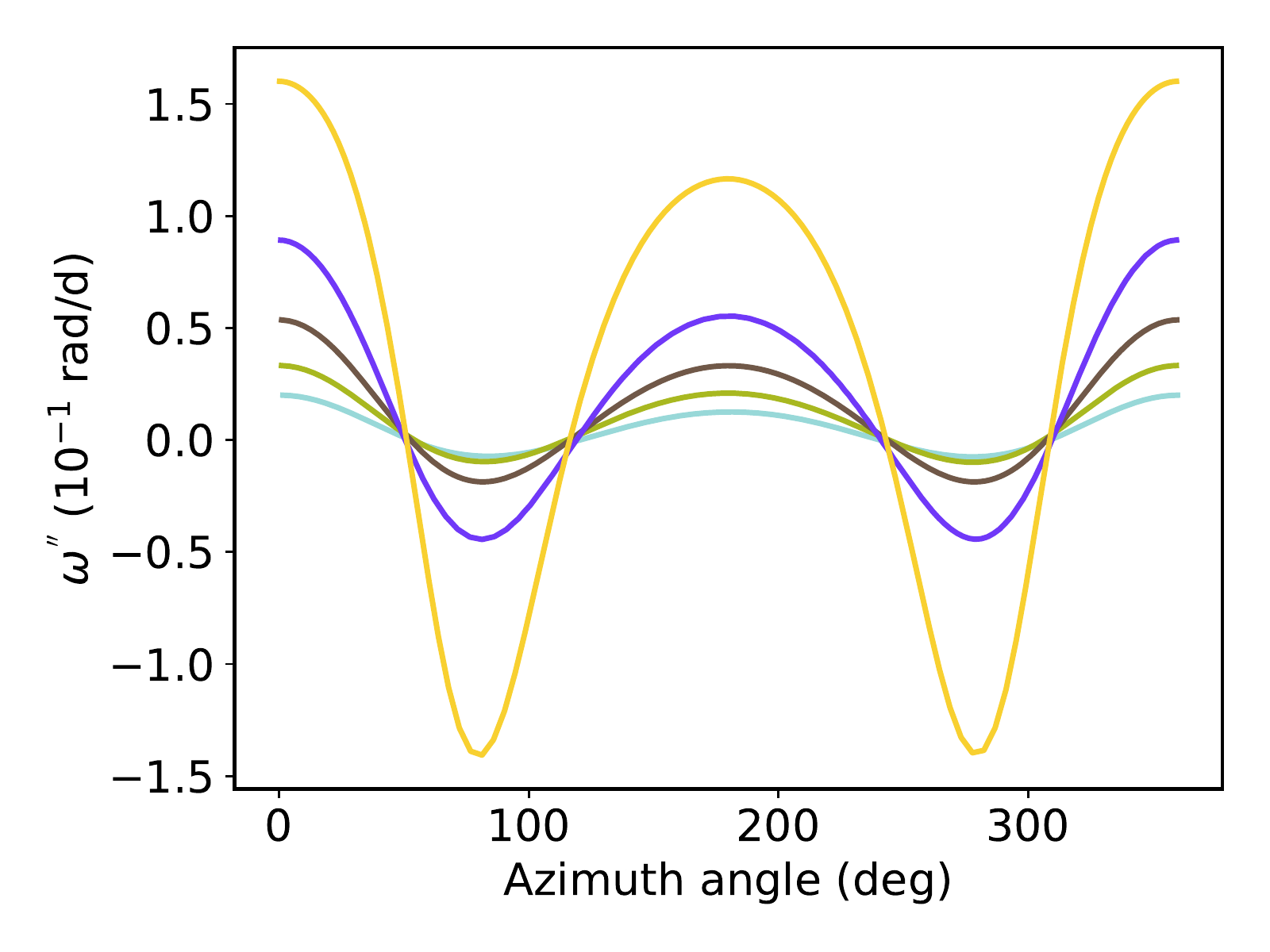}
\includegraphics[width=0.65\columnwidth]{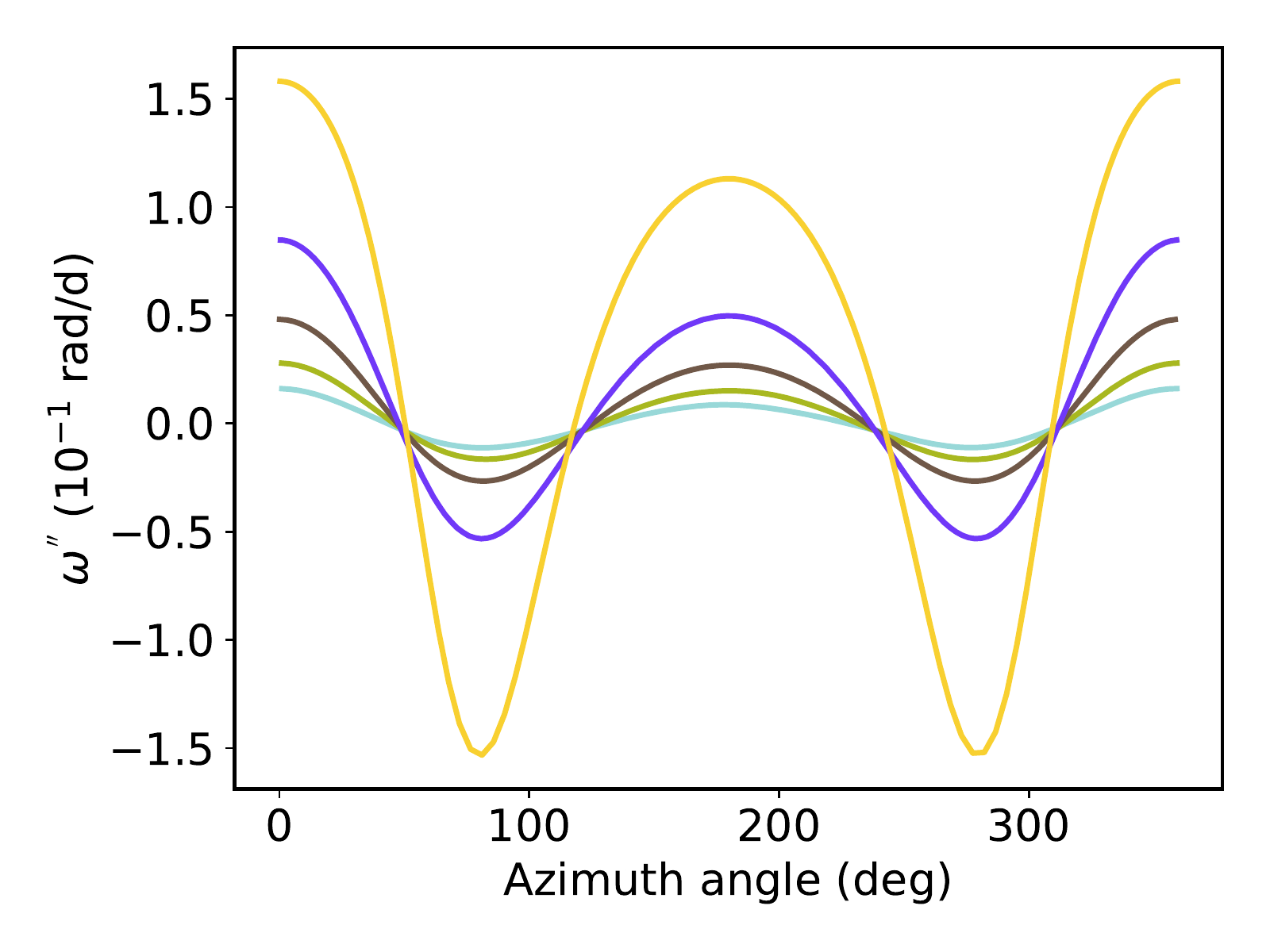}
\caption{Angular velocity at the equator as a function of azimuth angle in five layers
in the rest frame of the star. Left: Model 23, constant $\nu$; Middle: Model 24, $\lambda$=1;
Right: Model 26, $\lambda$=0.1.
}
\label{fig_models232426_omdp}
\end{figure*}

\begin{figure*}
\centering
\includegraphics[width=0.95\columnwidth]{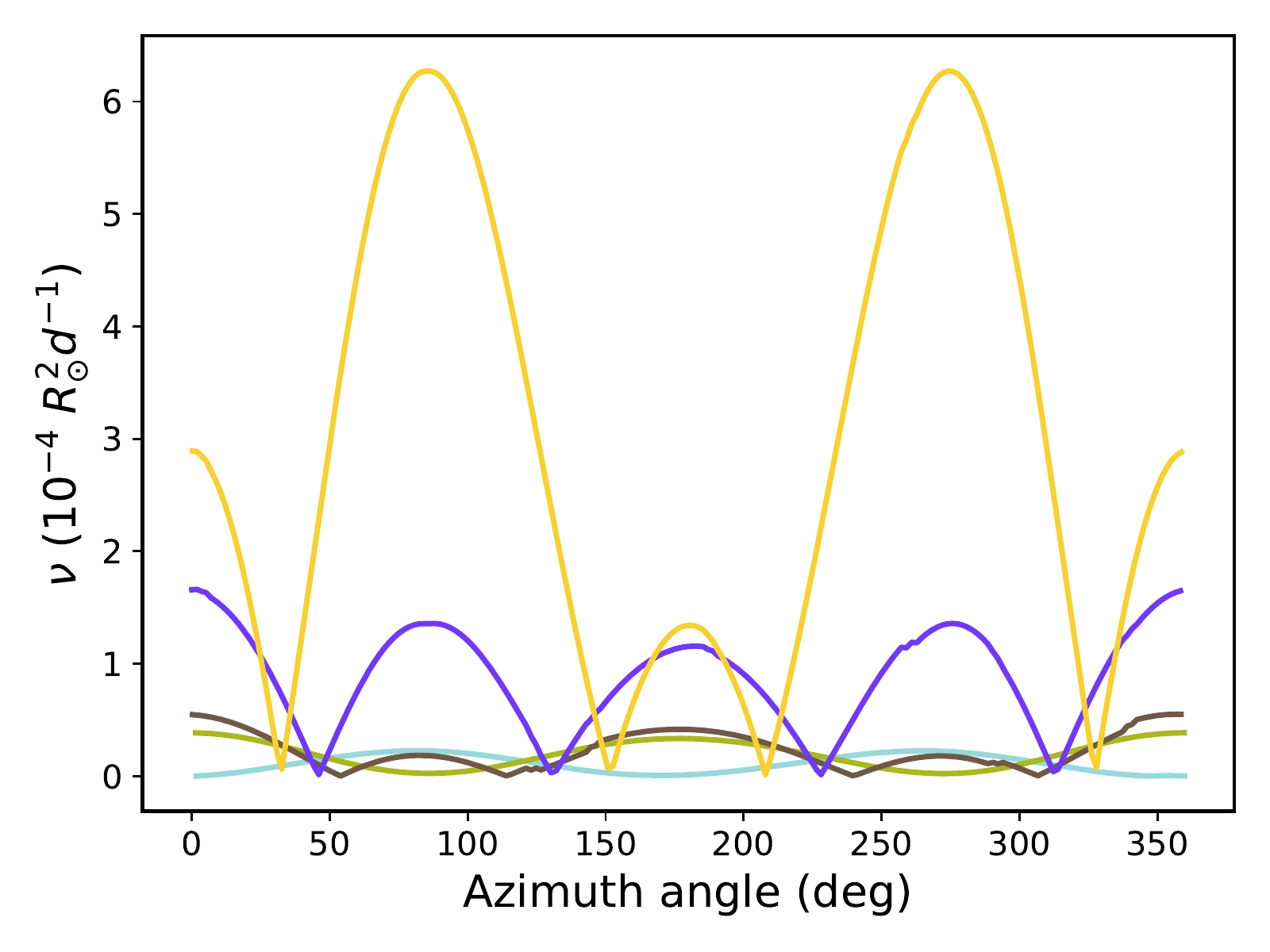}
\includegraphics[width=0.95\columnwidth]{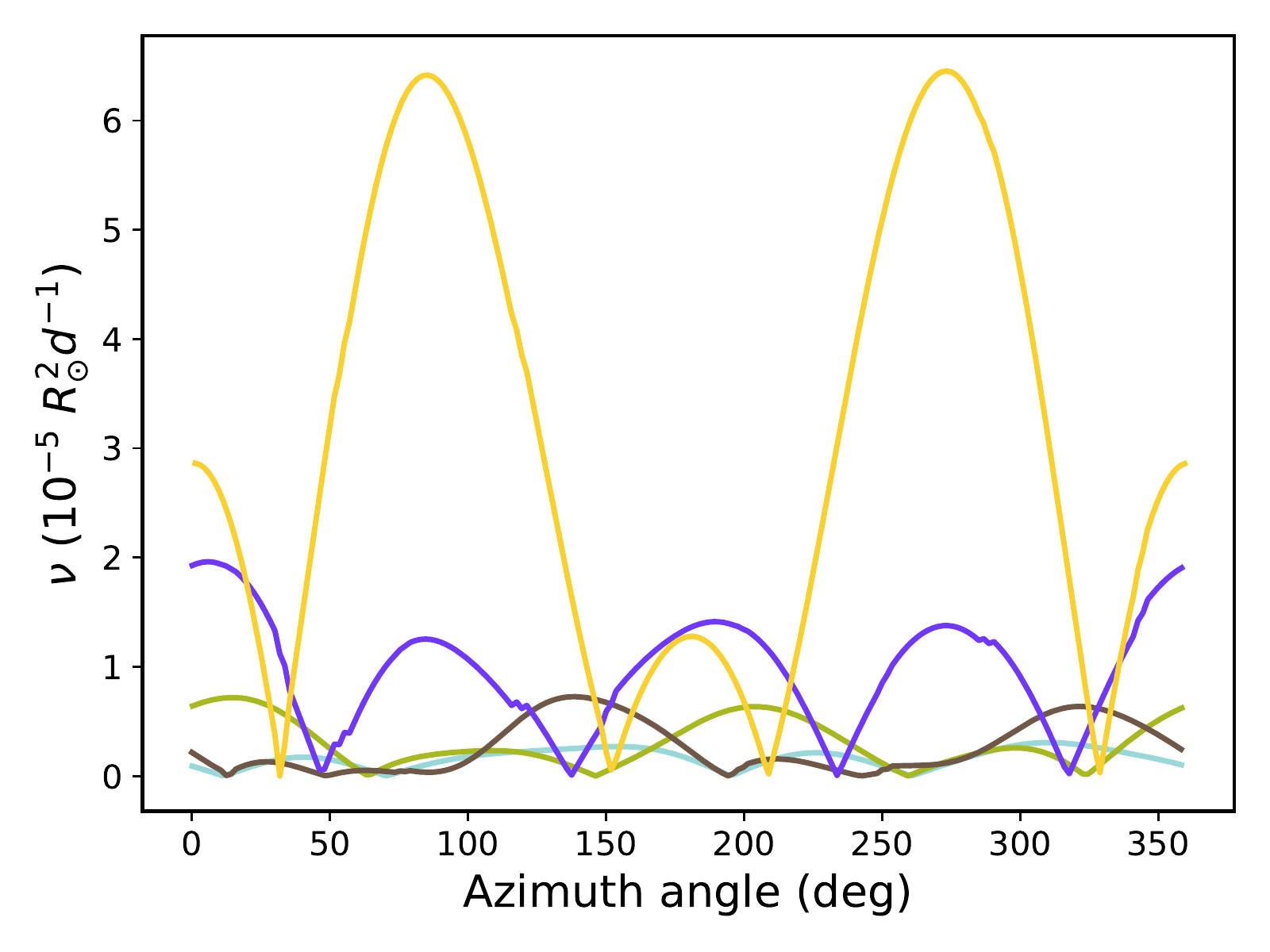}
\caption{Viscosity at the equator as a function of azimuth angle in five layers
in the rest frame of the star. Left:  Model 4, $\lambda$=1;
Right: Model 6, $\lambda$=0.1.  
}
\label{fig_models246_visc}
\end{figure*}

\begin{figure*}
\centering
\includegraphics[width=0.95\columnwidth]{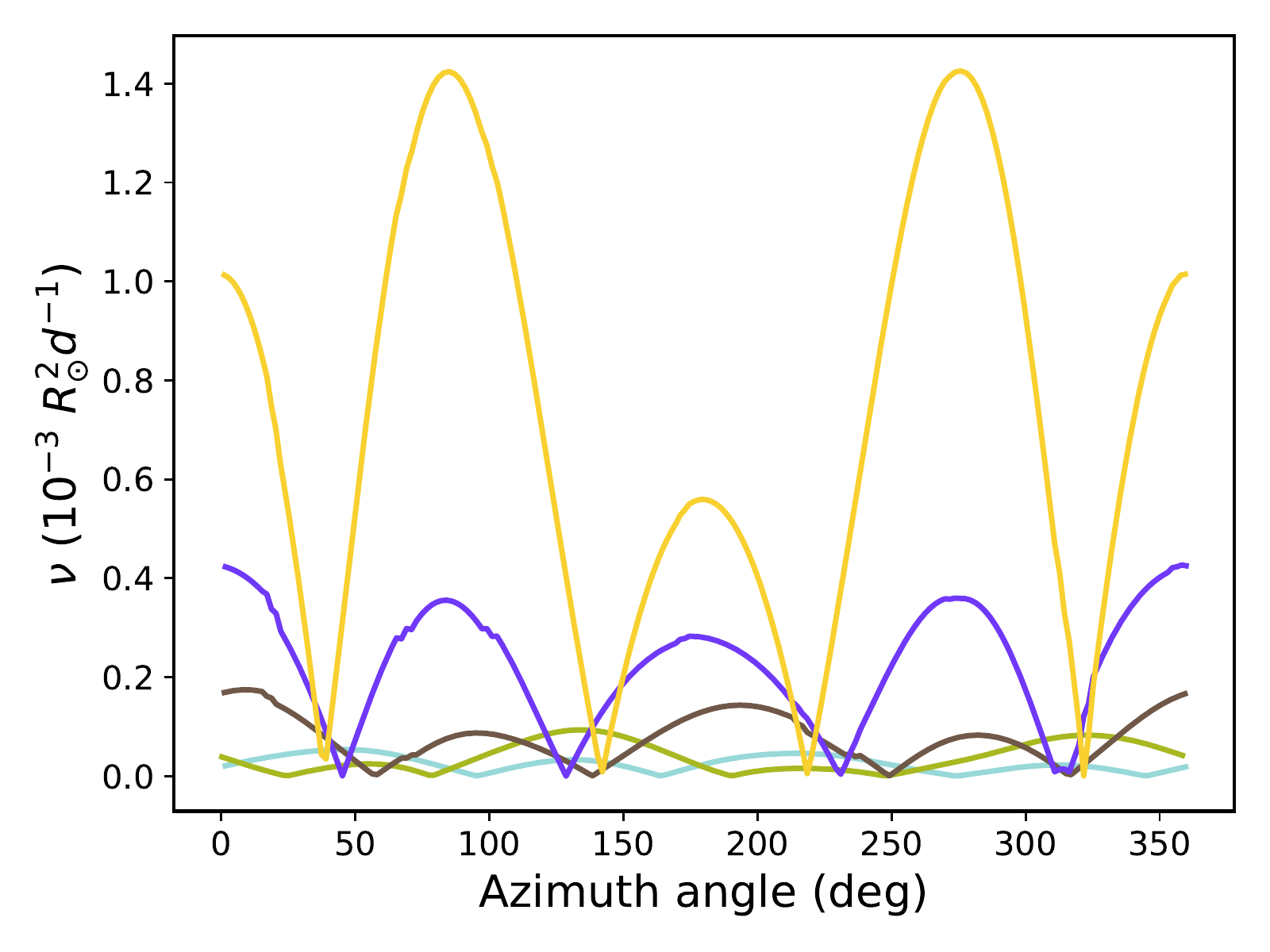}
\includegraphics[width=0.95\columnwidth]{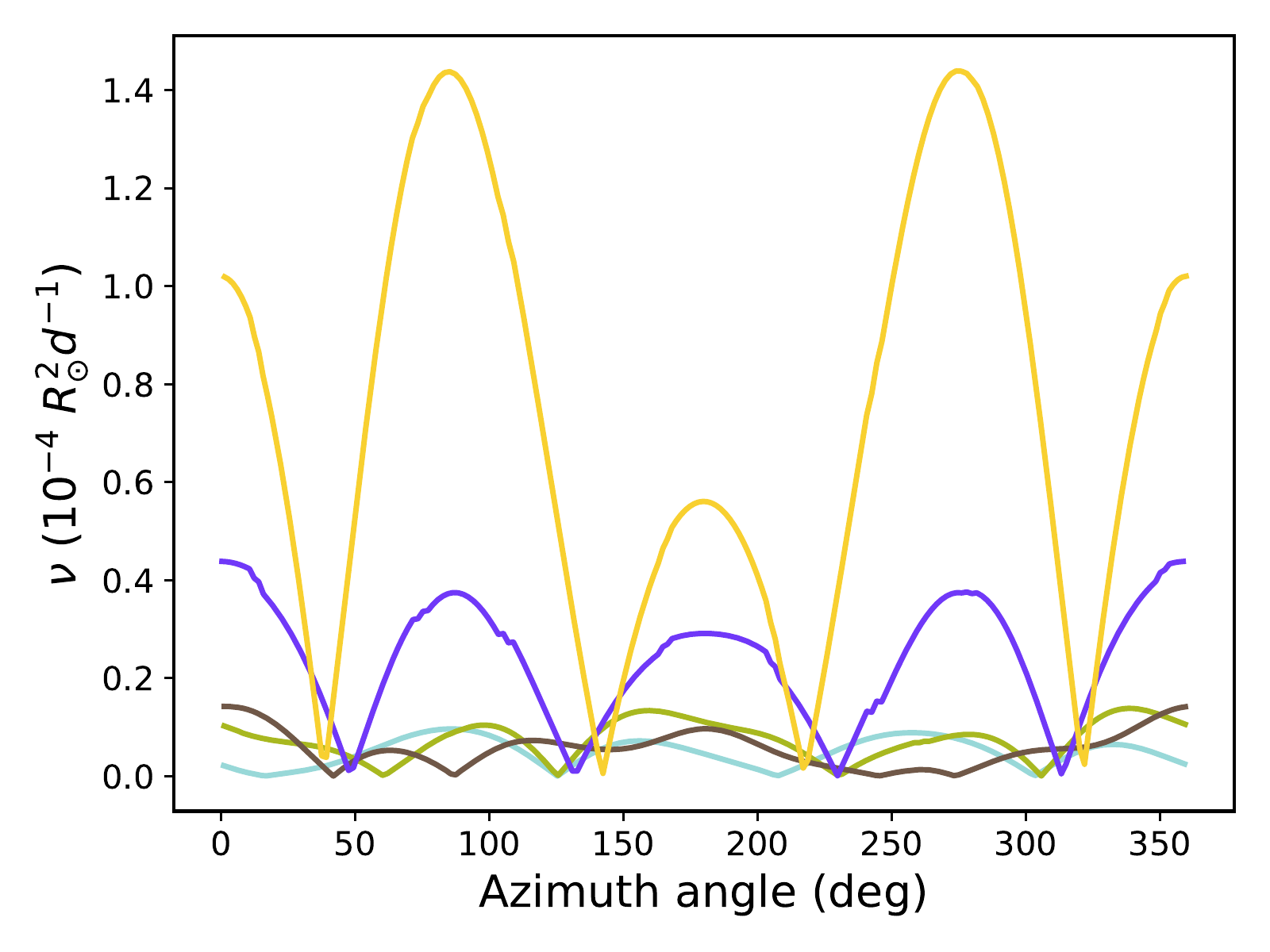}
\caption{Viscosity at the equator as a function of azimuth angle in five layers
in the rest frame of the star. Left: Model 8, $\lambda$=1;
Right: Model 10, $\lambda$=0.1.
}
\label{fig_models7810_visc}
\end{figure*}

\begin{figure*}
\centering
\includegraphics[width=0.95\columnwidth]{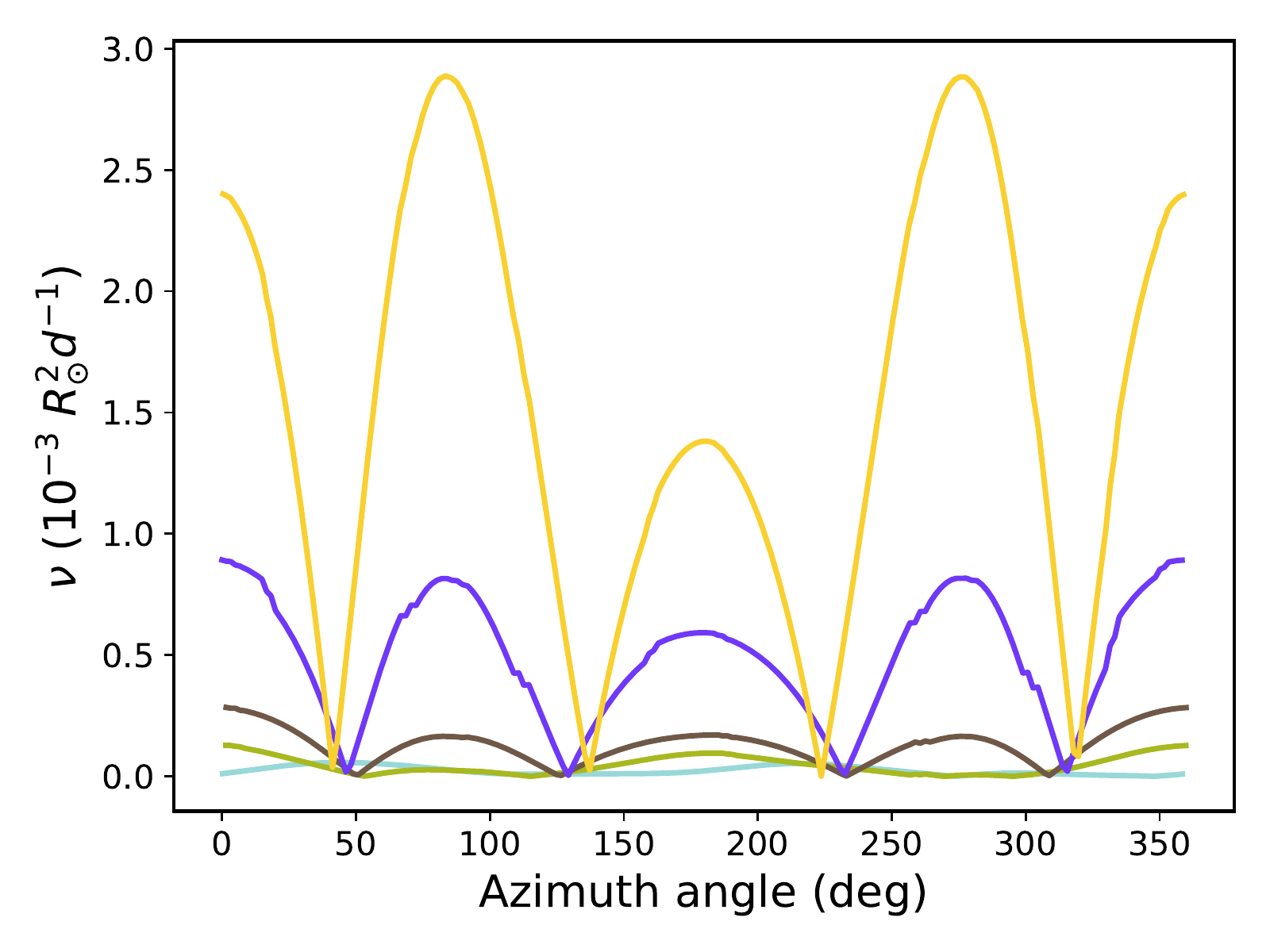}
\includegraphics[width=0.95\columnwidth]{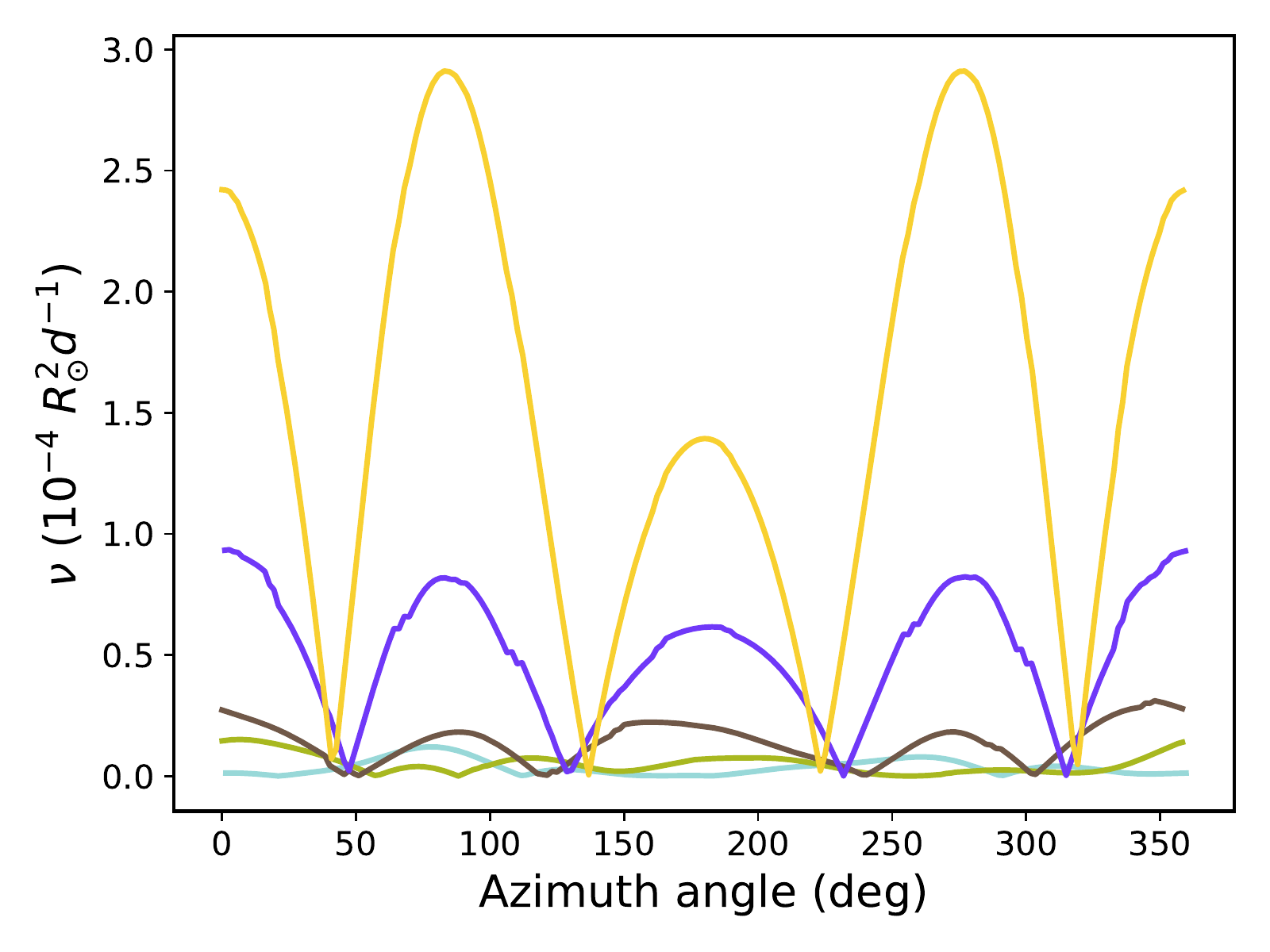}
\caption{Viscosity at the equator as a function of azimuth angle in five layers
in the rest frame of the star. Left:  Model 12, $\lambda$=1;
Right: Model 14, $\lambda$=0.1.
}
\label{fig_models111213_visc}
\end{figure*}

\begin{figure*}
\centering
\includegraphics[width=0.95\columnwidth]{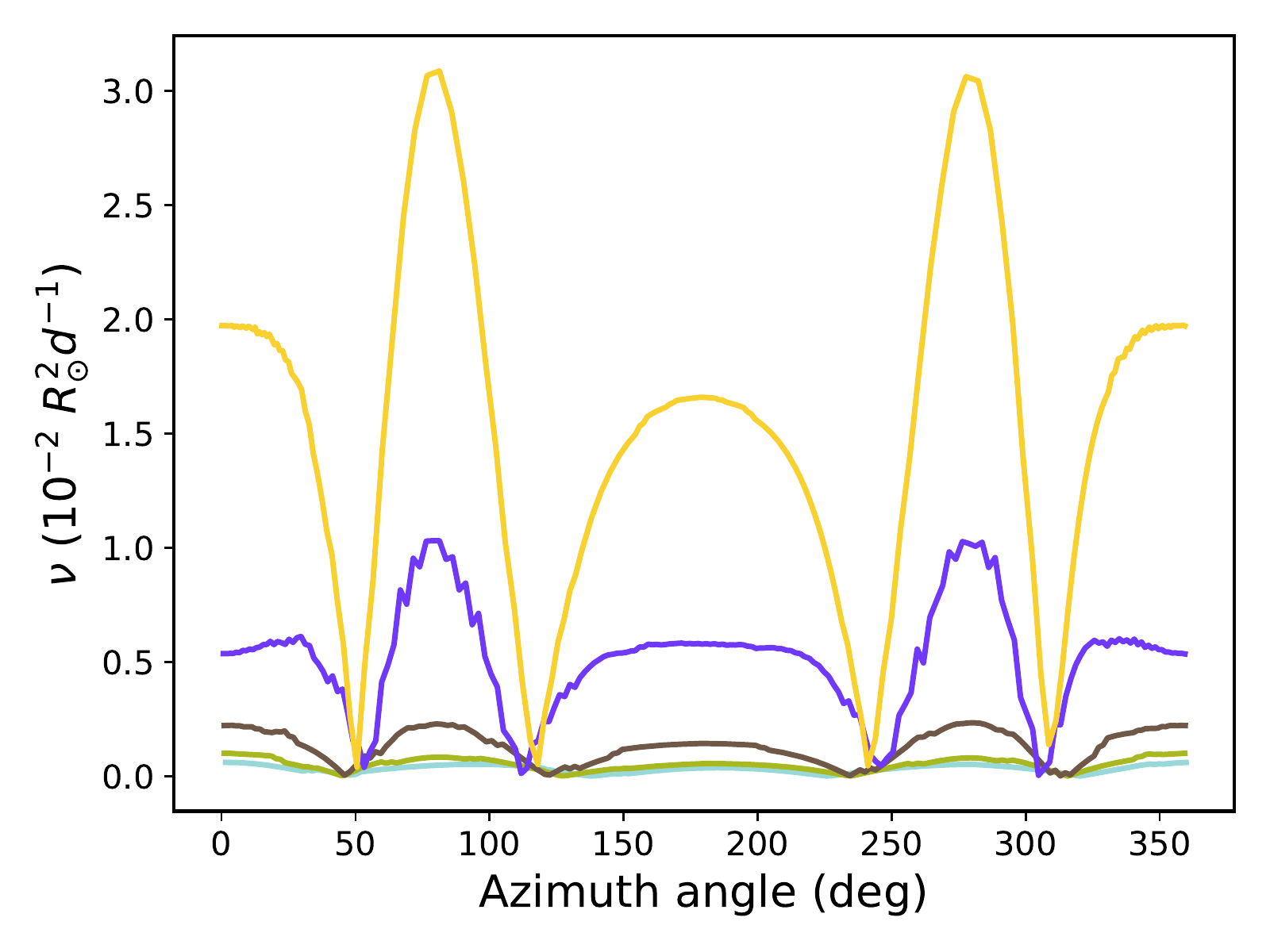}
\includegraphics[width=0.95\columnwidth]{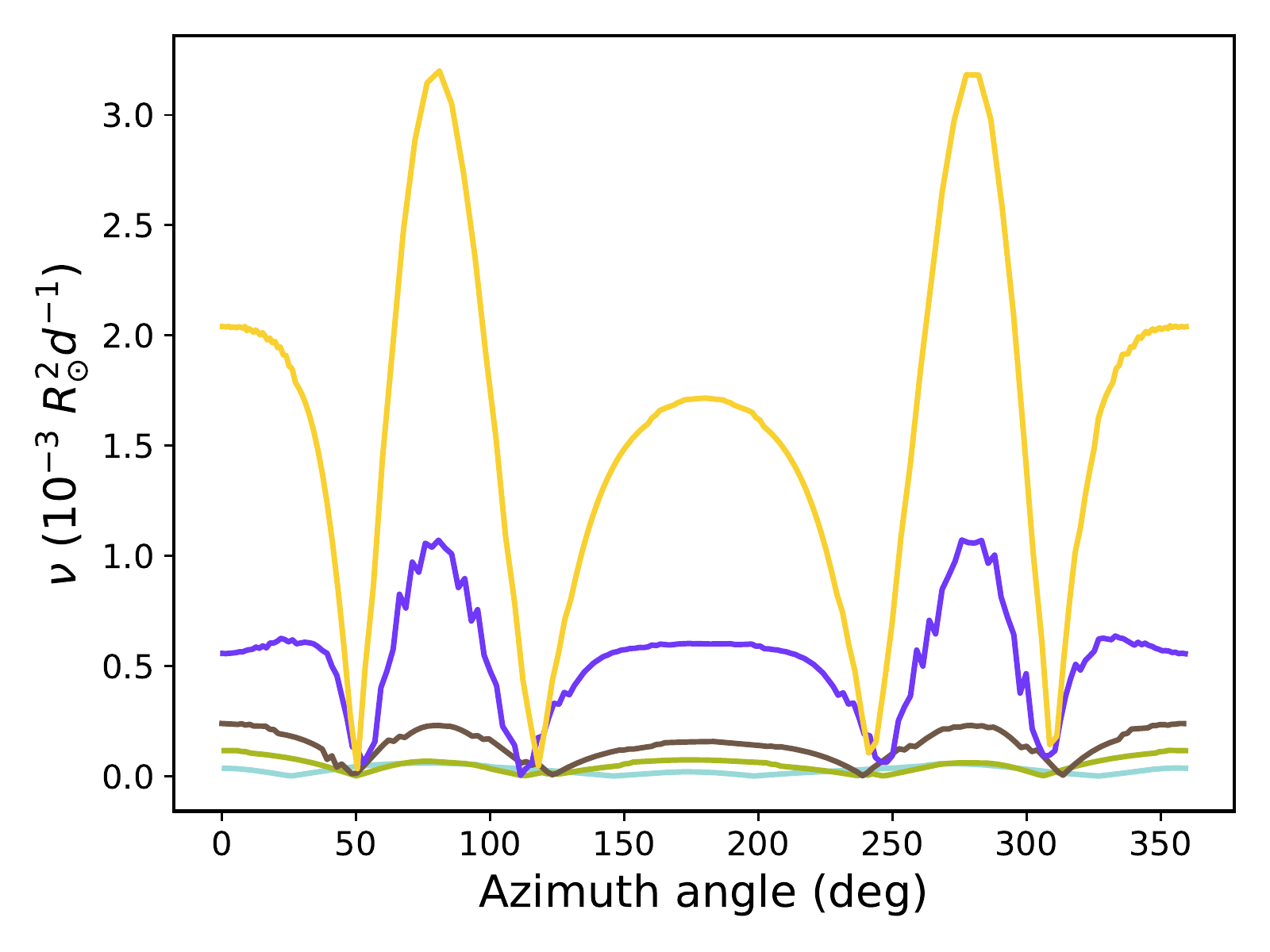}
\caption{Viscosity at the equator as a function of azimuth angle in five layers
in the rest frame of the star. Left:  Model 24, $\lambda$=1;
Right: Model 26, $\lambda$=0.1.
}
\label{fig_models2426_visc}
\end{figure*}

We present the behavior of angular velocity as a function of azimuth angle for the models of the block 1 in the Table \ref{table_TIDES_runs}, both with constant viscosity and the cases with variable viscosity for different radii (Figs. \ref{fig_models246_omdp}, \ref{fig_models7810_omdp}, \ref{fig_models111213_omdp}, and \ref{fig_models232426_omdp}). The aim is to observe more clearly how the variation of the radius as well as of $\lambda$ modifies the behavior of $\omega''$. Analogous graphs of the viscosity as a function of azimuth angle calculated for $\lambda$=0.1 and 1 are illustrated in Figs.~\ref{fig_models246_visc}, \ref{fig_models7810_visc}, \ref{fig_models111213_visc}, and \ref{fig_models2426_visc}.
The shape of these functions reveals four maxima per 360$^\circ$ in azimuth angle, each maximum corresponding to the extrema in the angular velocity curves.  The maximum value of these computed viscosities\footnote{$\nu_\mathrm{max}$
is generally found in the surface layer or the one just below it.} is listed in column 7 of
Table~\ref{table_TIDES_runs} for comparison with the constant viscosity values used in KM2016 and shows that they are of similar order of magnitude.  However, there is a major difference for the inner layers as mentioned above, where the smaller velocity gradients lead to significantly smaller viscosity values.\\
Table \ref{table_compare_layers} shows the comparison of the value of the energy dissipation rate in different orbital cycles for the M4 and M72 models whose parameters are the same, but with five and ten layers, respectively. The values shown correspond to layers 1 to 5 for M4 and layers six to ten (the outermost) for M72. Table \ref{table_compare_cycles} presents the values of the rate of energy dissipation in different orbital cycles for all the layers of M72.


\begin{figure*}
\centering
\includegraphics[width=0.65\columnwidth]{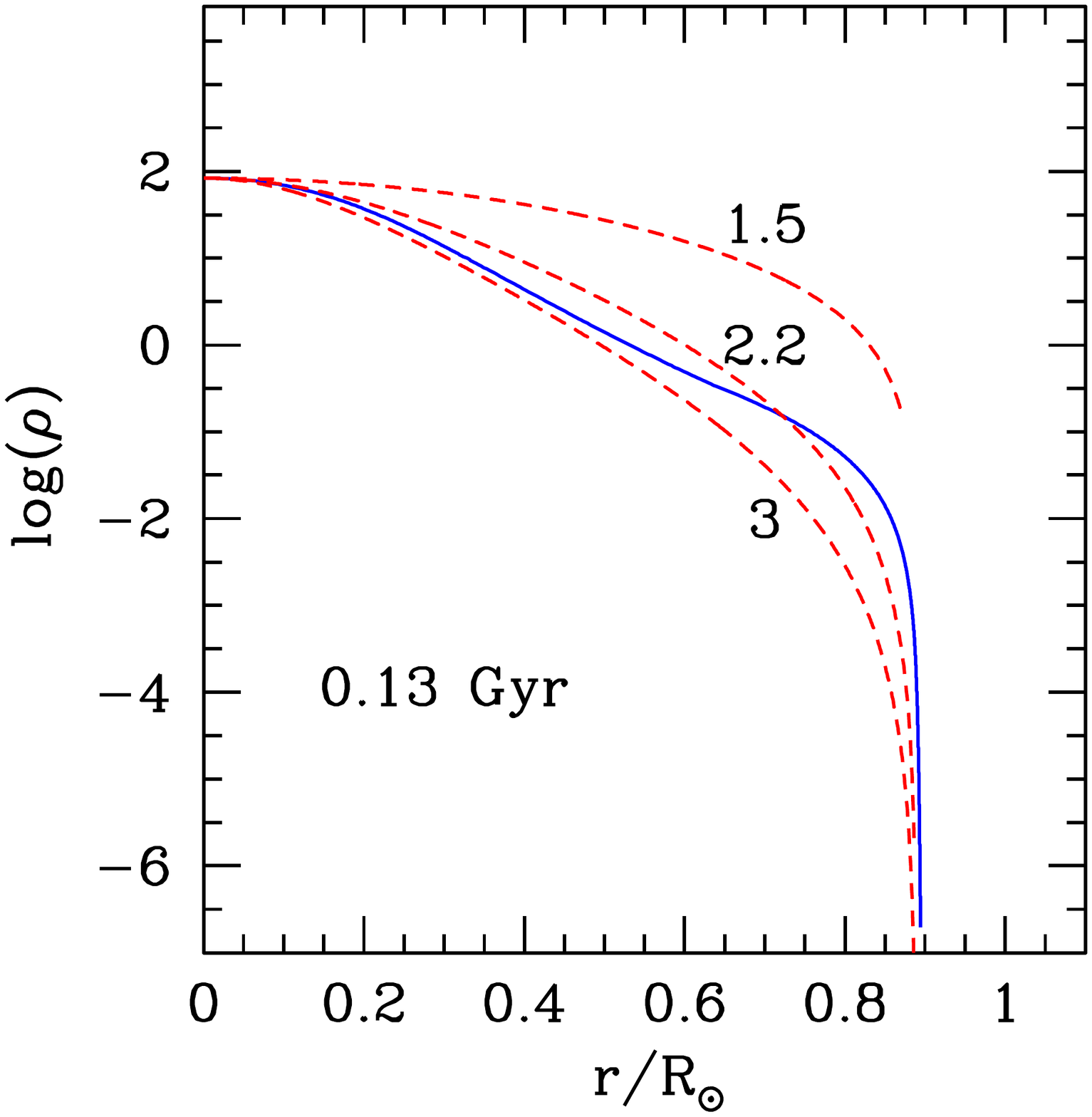}
\includegraphics[width=0.65\columnwidth]{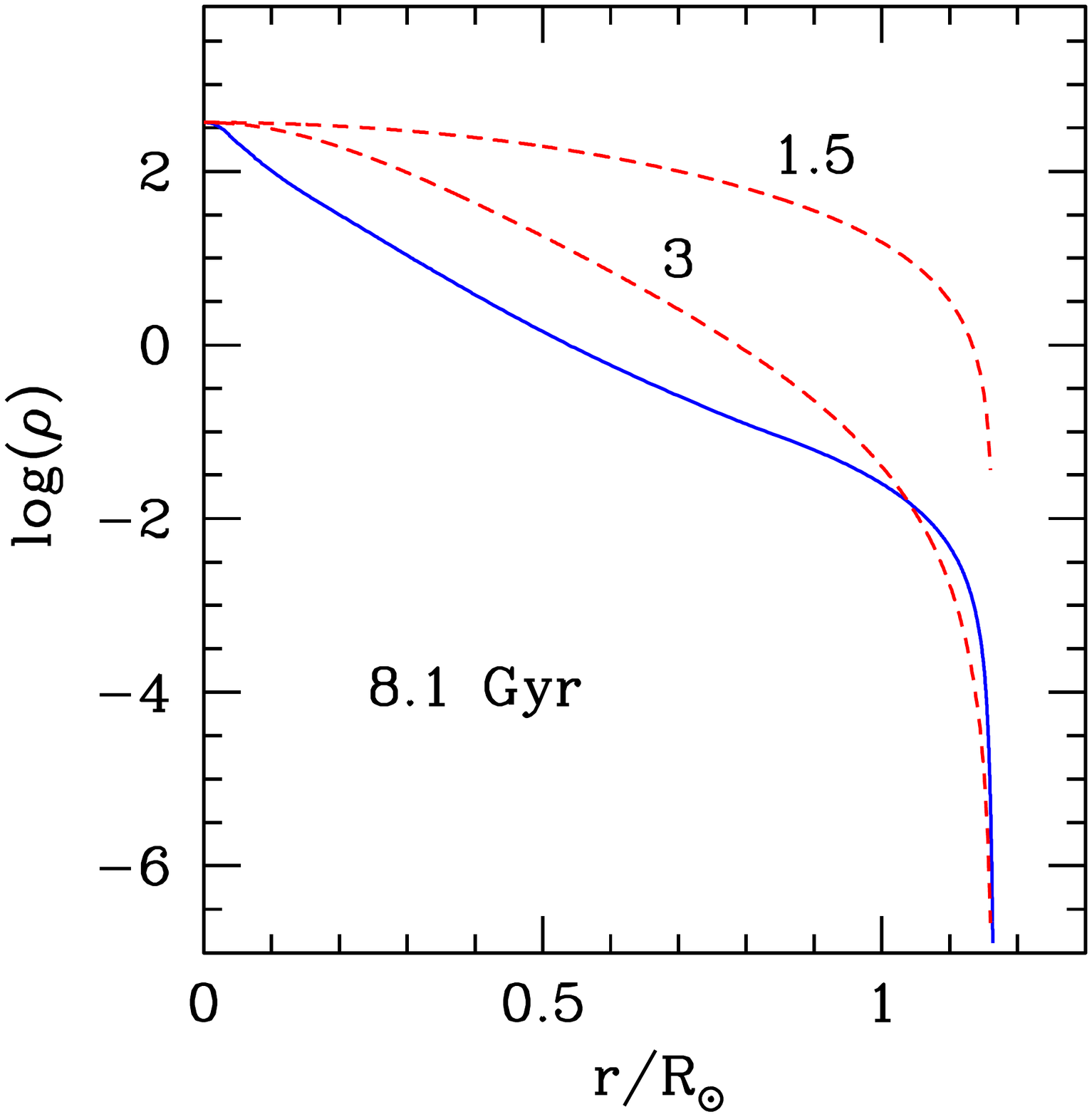}
\includegraphics[width=0.65\columnwidth]{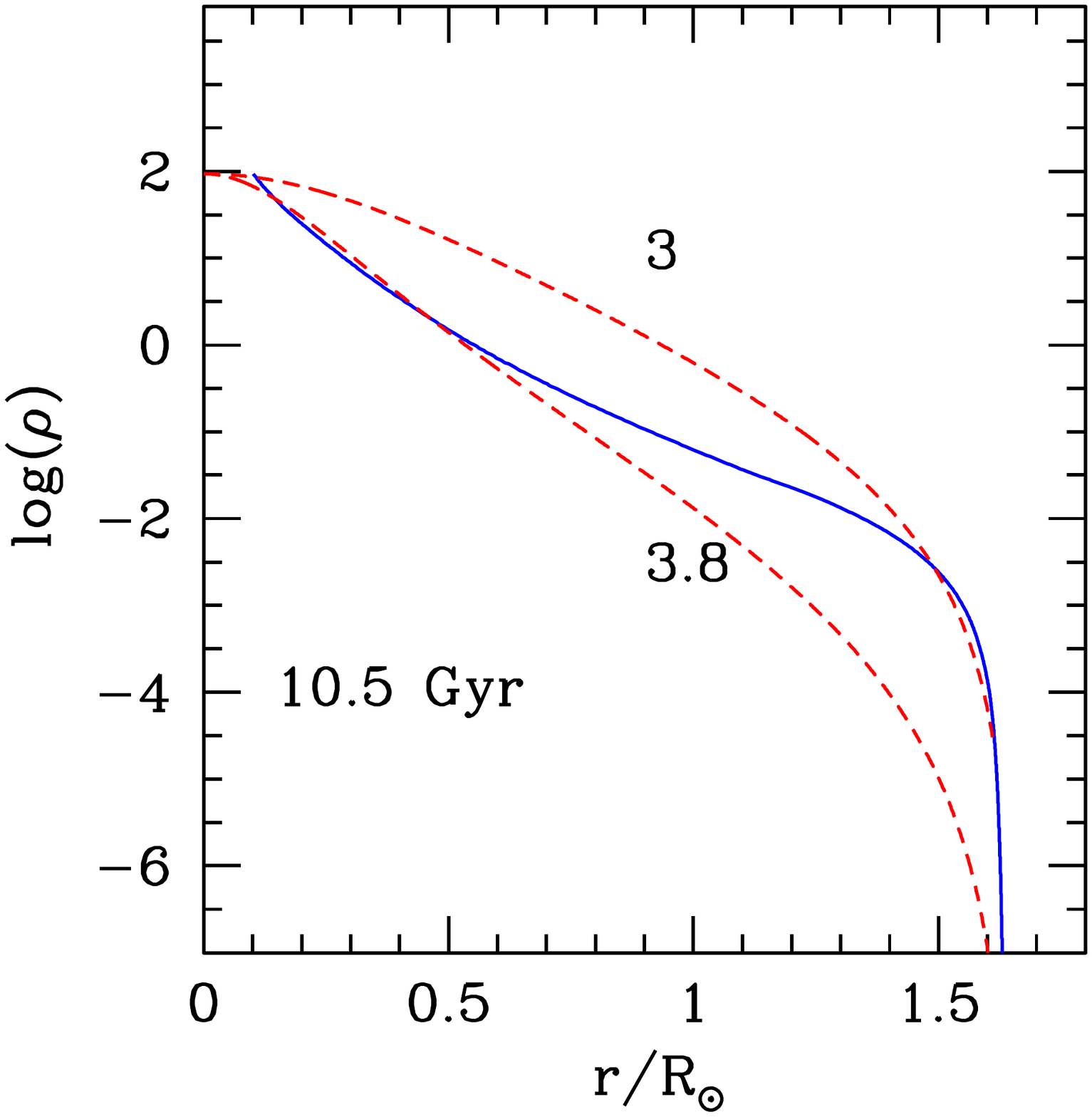}
\caption{Density structure of the MESA models (blue) compared to a polytropic structure (red dash). The MESA models are: Standard model at 0.13 Gyr (h0-3, top left) and at 8.1 Gyr (h0-11, top right) as well as the set 2 10x heated model at 10.5 Gyr (h10-22, bottom). The polytropes are labeled with their corresponding index, and their central density was chosen to coincide with that of the MESA models for illustration purposes. The density is given in units of g\,cm$^{-3}$.
}
\label{compare_logrho_polytropes}
\end{figure*}

The TIDES calculations of Moreno \& Koenigsberger (2016) were performed under the assumption of a $n$=1.5 polytropic structure which resulted in densities that are significantly larger than those obtained in this paper and which are obtained under the assumption that $n$=3 or larger.  In Fig.~\ref{compare_logrho_polytropes}, we show a comparison of the polytropic structures that we explored with the density structure given by the MESA models. 

\clearpage
\section{Comparison of model runs with five and with ten layers}

In Table~\ref{table_compare_layers}, we compare the energy dissipation rates obtained in the five-layer computation (model 4) with those obtained in the top five layers of a ten-layer computation (model 72), demonstrating that they are comparable.
In Table~\ref{table_compare_cycles}, we show that the energy dissipation rates in the inner layers have larger cycle-to-cycle variation than those of outer layers.  The orbital cycles listed are 150, 200, and 250.

\begin{table}[H]
\caption{Comparison of a five-layer and a ten-layer computation}
\label{table_compare_layers}
\centering
\begin{tabular}{ccccc}     
\hline\hline
&M4&   & M72&\\        
Cycle&Layer&$\dot{E}_\mathrm{k}$&Layer&$\dot{E}_\mathrm{k}$\\
\hline
&1&6.2424$\times$10$^{29}$&6&5.4581$\times$10$^{29}$\\
&2&5.9248$\times$10$^{29}$&7&5.7266$\times$10$^{29}$\\
20&3&5.6616$\times$10$^{29}$&8&5.6139$\times$10$^{29}$\\
&4&7.9975$\times$10$^{29}$&9&7.9735$\times$10$^{29}$\\
&5&1.7331$\times$10$^{29}$&10&1.7291$\times$10$^{29}$\\  
\hline
&1&1.4770$\times$10$^{29}$&6&1.6312$\times$10$^{29}$\\
&2&3.5352$\times$10$^{29}$&7&3.4355$\times$10$^{29}$\\
100&3&4.2569$\times$10$^{29}$&8&4.2623$\times$10$^{29}$\\
&4&7.0774$\times$10$^{29}$&9&7.0618$\times$10$^{29}$\\
&5&1.5233$\times$10$^{29}$&10&1.5204$\times$10$^{29}$\\ 
\hline
&1&1.7544$\times$10$^{29}$&6&2.1969$\times$10$^{29}$\\
&2&4.1049$\times$10$^{29}$&7&4.0143$\times$10$^{29}$\\
150&3&4.2719$\times$10$^{29}$&8&4.3463$\times$10$^{29}$\\
&4&6.9268$\times$10$^{29}$&9&6.9052$\times$10$^{29}$\\
&5&1.5124$\times$10$^{29}$&10&1.5079$\times$10$^{29}$\\  
\hline
\hline
\end{tabular}
\end{table}

\begin{table}[H]
\caption{Comparison of different orbital cycles for model 72}
\label{table_compare_cycles}
\centering
\begin{tabular}{cccc}     
\hline\hline
Cycle&150&200&250\\
\hline
Layer&&$\dot{E}_\mathrm{k}$&\\
\hline
1&4.3829$\times$10$^{27}$&1.7895$\times$10$^{27}$
&1.0173$\times$10$^{27}$\\
2&4.3547$\times$10$^{27}$&2.3716$\times$10$^{27}$
&1.7739$\times$10$^{27}$\\
3&7.2450$\times$10$^{27}$&5.9248$\times$10$^{27}$
&5.8900$\times$10$^{27}$\\
4&1.9956$\times$10$^{28}$&2.3581$\times$10$^{28}$
&3.2818$\times$10$^{28}$\\
5&7.2835$\times$10$^{28}$&1.0214$\times$10$^{29}$
&1.3587$\times$10$^{29}$\\  
6&2.1969$\times$10$^{29}$&2.7278$\times$10$^{29}$
&3.1836$\times$10$^{29}$\\
7&4.0143$\times$10$^{29}$&4.3273$\times$10$^{29}$
&4.4988$\times$10$^{29}$\\
8&4.3463$\times$10$^{29}$&4.2546$\times$10$^{29}$
&4.1610$\times$10$^{29}$\\
9&6.9052$\times$10$^{29}$&6.9448$\times$10$^{29}$
&6.8349$\times$10$^{29}$\\
10&1.5080$\times$10$^{29}$&1.5043$\times$10$^{29}$
&1.4943$\times$10$^{29}$\\ 
\hline
\hline
\end{tabular}
\end{table}
\end{appendix}
\end{document}